\documentclass[11pt,british]{article}

\usepackage{setspace}
\usepackage{jheppub}                   
\usepackage{mathrsfs}
\usepackage[utf8]{inputenc}
\usepackage{babel}
\usepackage{booktabs}
\usepackage{microtype} 								
	\SetTracking{encoding={*}, shape=sc}{0}			

\usepackage{amsmath}
\usepackage{amssymb}
\usepackage{amsthm}
\usepackage{mathtools}
\usepackage{esint}
\usepackage{tikz-cd}    							
\usepackage{stmaryrd}								
	\SetSymbolFont{stmry}{bold}{U}{stmry}{m}{n}		
\usepackage{slashed}								

\allowdisplaybreaks					
\numberwithin{equation}{section}	

\makeatletter
\gdef\@fpheader{\ }
\makeatother

\makeatletter 
\g@addto@macro\bfseries{\boldmath}
\makeatother

\DeclareMathOperator{\vol}{vol}					
\DeclareMathOperator{\tr}{tr}					
\DeclareMathOperator{\End}{End}					
\DeclareMathOperator{\re}{Re}					
\DeclareMathOperator{\im}{Im}					
\DeclareMathOperator{\ad}{ad}					
\DeclareMathOperator{\image}{im}				
\DeclareMathOperator{\Lie}{Lie}					

\newcommand{\quotient}{/}							
\newcommand{\qquotient}{/\!\!/}				
\newcommand{\qqquotient}{/\!\!/\!\!/}	
\newcommand{\eqspace}{\mathrel{\phantom{=}}{}} 				
\newcommand{\ext}{\mbox{\large $\wedge$}} 					
\newcommand{\dd}{\mathrm{d}} 								
\newcommand{\ee}{\mathrm{e}} 								
\newcommand{\ii}{\mathrm{i}} 								
\newcommand{\Dorf}{L} 
\newcommand{\Bgen}[2]{\left\llbracket#1,#2\right\rrbracket} 
\newcommand{\Tint}{W_{\text{int}}}							
\newcommand{\TTint}{T_{\text{int}}}							
\newcommand{\XX}{\mathcal{X}}
\newcommand{\proj}[1]{\times_{#1}}							
\newcommand{\oadj}{\proj{\text{ad}}}						
\newcommand{\CC}{\text{c.c.}}								
\newcommand{\rep}[1]{\boldsymbol{#1}} 						
\newcommand{\repp}[2]{(\rep{#1}, \rep{#2})} 				
\newcommand{\id}{\boldsymbol{1}} 							
\newcommand{\bbZ}{\mathbb{Z}} 								
\newcommand{\bbR}{\mathbb{R}} 								
\newcommand{\bbC}{\mathbb{C}} 								
\newcommand{\bbH}{\mathbb{H}} 								
\newcommand\qqq{\qquad\quad} 								
\newcommand\qqqq{\qquad\qquad} 								

\newcommand{\GL}[1]{\mathrm{GL}(#1)}
\newcommand{\SL}[1]{\mathrm{SL}(#1)}
\newcommand{\SU}[1]{\mathrm{SU}(#1)}
\newcommand{\Uni}[1]{\mathrm{U}(#1)}
\newcommand{\SUstar}[1]{\mathrm{SU}^{*}(#1)}
\newcommand{\Spin}[1]{\mathrm{Spin}(#1)}
\newcommand{\Spinstar}[1]{\mathrm{Spin}^{*}(#1)}
\newcommand{\Cliff}[1]{\mathrm{Cliff}(#1)}
\newcommand{\Symp}[1]{\mathrm{Sp}(#1)}
\newcommand{\USp}[1]{\mathrm{USp}(#1)}
\newcommand{\Orth}[1]{\mathrm{O}(#1)}
\newcommand{\SO}[1]{\mathrm{SO}(#1)}

\newcommand{\Ex}[1]{\mathrm{E}_{#1}}
\newcommand{\Fx}[1]{\mathrm{F}_{#1}}
\newcommand{\Gx}[1]{\mathrm{G}_{#1}}
\newcommand{\GDiff}{\mathrm{GDiff}}
\newcommand{\Hd}[1]{H_{#1}}

\newcommand{\so}[1]{\mathfrak{so}_{#1}}
\newcommand{\su}[1]{\mathfrak{su}_{#1}}

\newcommand{\spinstar}[1]{\mathfrak{spin}^{*}_{#1}}
\newcommand{\symp}[1]{\mathfrak{sp}_{#1}}

\newcommand{\ex}[1]{\mathfrak{e}_{#1}}

\newcommand{\gl}[1]{\mathfrak{gl}_{#1}}

\newcommand{\gdiff}{\mathfrak{gdiff}}

\theoremstyle{definition}
\newtheorem*{defn}{Definition}

\newcommand{\Mv}{\mathcal{A}_\text{V}}						%
\newcommand{\MV}{\mathcal{A}_\text{V}}					
\newcommand{\MH}{\mathcal{A}_\text{H}}					
\newcommand{\MT}{\mathcal{A}_\text{T}}					
\newcommand{\MHV}{\mathcal{A}}								
\newcommand{\ModH}{\mathcal{M}_\text{H}}	
\newcommand{\ModV}{\mathcal{M}_\text{V}}	
\newcommand{\ModT}{\mathcal{M}_\text{T}}	
\newcommand{\ModHV}{\mathcal{M}}		

\newcommand{\ZH}{Z_\text{H}}							
\newcommand{\ZV}{Z_\text{V}}							
\newcommand{\ZT}{Z_\text{T}}							

\newcommand{\GS}[1]{\tilde{P}_{#1}}

\newcommand{\Gf}{G'}
\newcommand{\gf}{\mathfrak{g}'}

\newcommand{\HM}{H}
\newcommand{\VM}{V}
\newcommand{\TM}{T}
\newcommand{\HV}{ECY}										

\title{Exceptional Calabi--Yau spaces: the geometry of $\mathcal{N}=2$ backgrounds with flux}

\author[a]{Anthony Ashmore}
\emailAdd{a.ashmore12@imperial.ac.uk}
\author[a,b]{and Daniel Waldram}
\emailAdd{d.waldram@imperial.ac.uk}

\affiliation[a]{Department of Physics,
   Imperial College London, \\
   Prince Consort Road, London, SW7 2AZ, UK} 

\affiliation[b]{Berkeley Center for Theoretical Physics,
   LeConte Hall MC 7300, \\
   University of California, Berkeley, CA 94720, U.S.A.}

\subheader{\textrm{Imperial/TP/15/DW/02}\\ \textrm{UCB-PTH-15/09}}

\abstract{In this paper we define the analogue of Calabi--Yau geometry for generic $D=4$, $\mathcal{N}=2$ flux backgrounds in type II supergravity and M-theory. We show that solutions of the Killing spinor equations are in one-to-one correspondence with integrable, globally defined structures in $\Ex{7(7)}\times\mathbb{R}^{+}$ generalised geometry. Such ``exceptional Calabi--Yau'' geometries are determined by two generalised objects that parametrise hyper- and vector-multiplet degrees of freedom and generalise conventional complex, symplectic and hyper-K\"{a}hler geometries. The integrability conditions for both hyper- and vector-multiplet structures are given by the vanishing of moment maps for the ``generalised diffeomorphism group'' of diffeomorphisms combined with gauge transformations. We give a number of explicit examples and discuss the structure of the moduli spaces of solutions. We then extend our construction to $D=5$ and $D=6$ flux backgrounds preserving eight supercharges, where similar structures appear, and finally discuss the analogous structures in $\Orth{d,d}\times\bbR^+$ generalised geometry.}

\makeatother

\begin{document}
\maketitle


\section{Introduction}

Supersymmetric backgrounds are a key ingredient in both string phenomenology and the AdS/CFT correspondence. For backgrounds without flux, supersymmetry requires the internal manifold $M$ to have special holonomy. The archetypal example is a six-dimensional Calabi--Yau manifold~\cite{CHSW85}, where the geometry is characterised by a holomorphic three-form $\Omega$ and a symplectic two-form $\omega$ satisfying the integrability conditions $\dd\Omega=\dd\omega=0$. The integrability of $\Omega$ implies the manifold is complex, and the tools of complex and algebraic geometry can then be used to construct examples and calculate many important physical  properties, such as moduli spaces, particle spectra and couplings, as first discussed for example in~\cite{Strominger85,SW85,CL91}. 

A natural question is whether there is an analogous description of generic supersymmetric flux compactifications preserving eight supercharges. In the context of type II reductions to four dimensions, this defines the natural string-theory generalisation of the notion of a Calabi--Yau manifold to backgrounds including both NS-NS and R-R flux.\footnote{Note that there are general ``no-go'' theorems~\cite{CR84,Candelas85,MN01,GMPW04,GMW03,IP00} that, in the absence of sources, exclude reductions on a compact space to a Minkowski background when fluxes are present. Thus the backgrounds in this paper are generically either non-compact or have boundaries corresponding to the excision of the sources.}  In this paper we show that exceptional generalised geometry~\cite{Hull07,PW08,CSW11,CSW14} gives precisely such a reformulation: the supersymmetric background defines an \emph{integrable} generalised structure, which we call an ``exceptional Calabi--Yau'' (\HV{}) geometry. In fact, the tensors $\omega$ and $\Omega$ are replaced by a pair of generalised structures that interpolate between complex, symplectic and hyper-K\"ahler geometries. With respect to the $\mathcal{N}=2$ supersymmetry, one structure is naturally associated to hypermultiplets and the other to vector multiplets, and the integrability conditions, defined using generalised intrinsic torsion~\cite{CSW14b}, have an elegant interpretation in terms of moment maps. 

A description in terms of integrable structures is important since it should provide new approaches for tackling problems such as analysing the deformations and moduli spaces of arbitrary flux backgrounds, as well as potentially constructing new examples. We also note that it not only gives generalisations of the classical target-space theories of the topological string A and B models to include R-R modes, but also defines a corresponding pair of theories in M-theory.  

The conventional notion of a $G$-structure has already provided a very useful way of analysing flux backgrounds~\cite{GMPW04,GGHPR03,GP03}. While the manifold $M$ no longer has special holonomy, the Killing spinor bilinears still define a set of tensors invariant under $G\subset\SO{d}\subset\GL{d;\bbR}$ where $d$ is the dimension of $M$. The fluxes then describe the lack of integrability of this $G$-structure. Formally this is encoded in the intrinsic torsion, and only when this vanishes does the background have special holonomy. For generic backgrounds, the structure is only locally defined since there can be points where the stabiliser group of the Killing spinors changes. The basic point of this paper, as in~\cite{CSW14b}, is that there is actually a natural extended geometry in which supersymmetry for a generic flux background again corresponds to integrable, globally defined $G$-structures. 

Focussing for the moment on $\mathcal{N}=2$, $D=4$ backgrounds, in the case of NS-NS flux such a reformulation has already appeared under the guise of generalised complex geometry~\cite{Hitchin02,Gualtieri04,GMPT04b,GMPT05}. One considers a generalised tangent bundle $E\simeq TM\oplus T^*M$, admitting a natural $\Orth{d,d}$ metric. For a large class of supersymmetric backgrounds with non-trivial two-form $B$-field and dilaton $\phi$, the holomorphic and symplectic forms generalise to a pair of $\Orth{6,6}$ pure spinors $\Phi^{\pm}\in\Gamma(\ext^\pm T^*M)$, each defining an $\SU{3,3}\subset\Orth{6,6}$ structure. They satisfy compatibility conditions, analogues of $\omega\wedge\Omega=0$ and $\omega^3\propto\Omega\wedge\bar{\Omega}$, that imply that together they define an $\SU{3}\times\SU{3}$ structure. Furthermore, the $\mathcal{N}=2$ Killing spinor equations imply $\dd\Phi^{\pm}=0$, and one says the $\SU{3}\times\SU{3}$ structure is integrable~\cite{GMPT05}. Each such integrable $\Phi^\pm$ defines a generalised complex structure~\cite{Hitchin02} and the integrable $\SU3\times\SU3$ structure is known as a generalised Calabi--Yau metric structure~\cite{Gualtieri04}. This language has been useful for a whole range of applications including addressing deformations~\cite{Gualtieri04,Tomasiello07,CG11,TY14}, topological strings~\cite{Kapustin04,KL07,PW05}, steps towards classifying flux backgrounds~\cite{GMPT07,LT09b,Andriot08} and the AdS/CFT
correspondence~\cite{MPZ06,BFMMPZ08,GGPSW10}. 

To include R-R fluxes and also M-theory compactifications, we need to consider $\Ex{d(d)}\times\mathbb{R}^{+}$ or exceptional generalised geometry~\cite{Hull07,PW08}. The generalised tangent space is further extended to include the R-R gauge symmetries, such that it admits a natural action of $\Ex{d(d)}\times\bbR^+$. It gives a unified geometrical description of type II and M-theory restricted to a $(d-1)$- or $d$-dimensional manifold~\cite{CSW11,CSW14}, invariant under local transformations by the maximal compact subgroup $\Hd{d}$ of $\Ex{d(d)}\times\bbR^+$. The bosonic symmetries combine in a generalised Lie derivative, there is a generalised metric, invariant under $\Hd{d}$, that encodes all the bosonic degrees of freedom, and, defining a generalised connection $D$ that is the analogue of the Levi-Civita connection, the full bosonic action is equal to the corresponding generalised Ricci scalar, and the fermion equations of motion and supersymmetry variations can all be written in terms of $D$. We note that there is a long history of considering exceptional groups in supergravity, in many cases by positing the existence of extra coordinates, for example see~\cite{DN86,West00,West01,West03,KNS00,DHN02,DHN03,Hillmann09} and more recently~\cite{BP11a,BGP11,BGPW12,AGMR13,HS13}. Although in general defining such putative extensions of spacetime is problematic~\cite{BCP14,Cederwall14,Hull14,LSW15}, we note that all the constructions here are equally applicable as local descriptions to any situation where a suitable spacetime can be defined. 

The two generalised structures defining generic $\mathcal{N}=2$, $D=4$ backgrounds are invariant under $\Spinstar{12}$ and $\Ex{6(2)}$ subgroups of the $\Ex{7(7)}\times\bbR^+$ acting on the generalised tangent space. We refer to them as  \HM{} and \VM{} structures respectively, standing for``hyper-'' and ``vector-multiplet''. If compatible, together they define an exceptional Calabi--Yau or \HV{} structure that is invariant under $\SU6$.\footnote{Strictly it is more natural to define ECY as an \emph{integrable} $\SU6$ structure. For simplicity we are not always careful in making this distinction.} Such structures were first introduced in the context of type II theories in~\cite{GLSW09}. Since the supersymmetry parameters transform under $\Hd{7}=\SU8$ in the exceptional generalised geometry, the $\SU6$ structure appears as $\SU6$ is the stabiliser group of a pair of Killing spinors. Some steps towards rephrasing supersymmetry in terms of integrable generalised structures in the $\mathcal{N}=1$ case, where the structure is $\SU7$, were taken in~\cite{PW08} in M-theory and in~\cite{GLSW09} in type II. The full set of $\mathcal{N}=1$ conditions, written using a particular unique generalised connection, were given in~\cite{GO11}, and this was extended to $\mathcal{N}=2$ in~\cite{GO12}. The four-dimensional effective theories in both $N=1$ and $N=2$ have been considered in~\cite{PW08,GLSW09,GT13}. 

As first noted in~\cite{GLSW09}, the infinite-dimensional spaces of hypermultiplet and vector-multiplet structures admit hyper-K\"ahler and special K\"ahler metrics respectively. Strikingly, we find that the integrability conditions for each can be formulated as the vanishing of the corresponding moment maps for the action of the generalised diffeomorphism group. The moduli spaces of structures are then given by a hyper-K\"ahler or symplectic quotient. For \HV{} structures there is an additional integrability condition that involves both structures. That differential conditions appear as moment maps on infinite-dimensional spaces is a ubiquitous phenomenon~\cite{Kirwan98,Thomas06}. Examples include the Atiyah--Bott description of flat gauge connections on a Riemann surface~\cite{AB83}, the Donaldson--Uhlenbeck--Yau equations~\cite{Donaldson85,UY86,UY89}, the Hitchin equations~\cite{Hitchin87}, and even the equations for K\"ahler--Einstein metrics~\cite{Fujiki90,Donaldson97}. In our case we see that there are also moment maps for geometries defining generalisations of complex and symplectic structures that, in addition, use the full (generalised) diffeomorphism group. 

For each structure, we show that the integrability conditions correspond to the existence of a torsion-free $G$-compatible generalised connection. This follows the analysis of~\cite{CSW14b} where it was shown that there is a natural definition of intrinsic torsion for generalised $G$-structures, and one can define generalised special holonomy as structures with $G\subset \Hd{d}$ and vanishing generalised intrinsic torsion. The minimally supersymmetric backgrounds of type II supergravity and M-theory in various dimensions are constrained to have generalised special holonomy in both the Minkowski~\cite{CSW14b} and AdS~\cite{CS15} case. Here, we use the same notion of generalised intrinsic torsion to prove that our integrability conditions are equivalent to the Killing spinor equations. 

Physically the appearance of moment maps is very natural. It is possible to reformulate the full ten- or eleven-dimensional supergravity as a four-dimensional $\mathcal{N}=2$ theory~\cite{GLW06,GLW07,GLSW09}. The $\Spinstar{12}$ structures then naturally parametrise an infinite-dimensional space of hypermultiplets, while the $\Ex{6(2)}$ structures encode an infinite-dimensional space of vector multiplets. This is the origin of our names for the two types of structures. The $\mathcal{N}=2$ theory will be gauged, and supersymmetry implies that the gauging defines a triplet of moment maps on the hypermultiplets and a single moment map on the vector multiplets (see for example~\cite{ABCDFFM97}). This structure was already noted in~\cite{GLSW09}, where it was pointed out that the gauged symmetry was simply the R-R gauge transformations. However for generic backgrounds, as we show here, not only the R-R gauge transformations but actually the whole set of generalised diffeomorphisms are gauged, including NS-NS gauge transformations and conventional diffeomorphisms. The integrability conditions can then be directly translated into
the vanishing of the gaugino, hyperino and gravitino variations, following a similar analysis for $\mathcal{N}=1$ backgrounds in~\cite{KM07,PW08,GLSW09,GO11}. In making this translation we partly rephrase the standard conditions, as given in~\cite{HLV09,LST12,ALMTW14}, showing that the gaugino variation generically implies a vanishing of the vector-multiplet moment map. 

This paper is rather long, primarily because of the inclusion of a number of examples which we hope will clarify some of the more abstract constructions. However, it does fall into two fairly distinct parts. The first, sections~\ref{sec:Calabi--Yau} to~\ref{sec:int-tor}, discusses $\mathcal{N}=2$, $D=4$ backgrounds in type II and M-theory. The first three sections cover some examples of $D=4$ supersymmetric flux backgrounds, an introduction to generalised geometry and generalised structures, integrability of the structures and finally some example calculations. More technical aspects, such as the equivalence of integrability with torsion-free $G$-structures, the origin of the integrability conditions from gauged supergravity and the moduli space of supersymmetric compactifications, are all in section~\ref{sec:int-tor}. 

Our formalism also applies to both type II and M-theory backgrounds in $D=5$ and $D=6$ preserving eight supercharges, and to $\Orth{d,d}\times\bbR^+$ generalised geometry. Discussion of these cases, along with some simple examples, forms the second part of the paper, sections~\ref{sec:Ed_structures} to~\ref{sec:0dd}. The hypermultiplet structure is always of the same form, but the second generalised structure that is compatible with it is dependent on the case in hand. The $\Orth{d,d}\times\bbR^+$ case is of physical interest because these structures capture those supersymmetric NS-NS backgrounds that do not have a description in generalised complex geometry, most notably the NS5-brane solution. We note that AdS backgrounds can also be described in this formalism but, in the interest of avoiding an even longer paper, we leave these for a companion work~\cite{AW15b}. 


\section{\texorpdfstring{$\mathcal{N}=2$}{N=2}, \texorpdfstring{$D=4$}{D=4} flux backgrounds}\label{sec:Calabi--Yau}

Our aim is to reformulate generic supersymmetric flux backgrounds with eight supercharges as integrable structures within generalised geometry. We will focus first on the case of type II and M-theory backgrounds giving $\mathcal{N}=2$ in four dimensions. In this section, we briefly summarise the Killing spinor equations, along with the standard $\mathcal{N}=2$ examples of Calabi--Yau and generalised Calabi--Yau metrics in type II theories, and discuss how the notion of a Calabi--Yau structure extends when one includes fluxes. In appendix~\ref{sec:examples}, we give a number of other backgrounds that will provide useful examples when we come to discuss generalised structures. 

 
\subsection{Supersymmetric backgrounds in type II and M-theory}

We consider type II and M-theory spacetimes of the form $\mathbb{R}^{D,1}\times M$, with a warped product metric 
\begin{equation}
\label{eq:warp-metric}
\dd s^2 = \ee^{2\Delta} \dd s^2 (\mathbb{R}^{D-1,1}) + \dd s^2 (M),
\end{equation}
where $\Delta$ is a scalar function on $M$. Initially we will assume $D=4$ and hence $M$ is six-dimensional for type II and seven-dimensional for M-theory. For the type II theories we use the string frame metric so that the warp factors for type II and M-theory are related by $\Delta_{\text{II}}=\Delta_{\text{M}}+\frac{1}{3}\phi$, where $\phi$ is the dilaton.  We allow generic fluxes compatible with the Lorentz symmetry of $\bbR^{3,1}$. Thus for M-theory, of the eleven-dimensional four-form flux $\mathcal{F}$ we keep the components 
\begin{equation}
\label{eq:FtF}
  F_{m_1 \ldots m_4} = \mathcal{F}_{m_1 \ldots m_4} , \qqq
  \tilde{F}_{m_1 \ldots m_7} = (\star\mathcal{F})_{m_1 \ldots m_7} ,
\end{equation}
where $m=1,\ldots,7$ are indices on $M$, while for type II we use the democratic formalism~\cite{BKORP01} and keep only the flux components that lie entirely on $M$. 

In M-theory, the eleven-dimensional spinors $\varepsilon$ can be decomposed into four- and seven-dimensional spinors $\eta^{\pm}$ and $\epsilon$ respectively according to
\begin{equation}
\label{eq:mtheoryspinor}
   \varepsilon = \eta^+ \otimes \epsilon + \eta^- \otimes \epsilon^* ,
\end{equation}
where $\pm$ denotes the chirality of $\eta^\pm$. The internal spinor $\epsilon$ is complex, and can be thought of as a pair of real $\Spin7$ spinors $\epsilon = \re\epsilon + \ii \im\epsilon$. The Killing spinor equations read~\cite{KMT03,BJ03,DP04,LS05}
\begin{equation}
\label{eq:elevendsusy}
\begin{split}
   \nabla_m \epsilon + \tfrac{1}{288} F_{n_1 \dots n_4} (
      \gamma_m{}^{n_1 \dots n_4}-8\delta_{m}{}^{n_1}\gamma^{n_2 n_3 n_4}
      ) \epsilon
      - \tfrac{1}{12} \tfrac{1}{6!} \tilde{F}_{mn_1 \dots n_6}
         \gamma^{n_1 \dots n_6} \epsilon &=0, \\
   \gamma^m\nabla_m \epsilon + (\partial_m \Delta) \gamma^m \epsilon 
      - \tfrac{1}{96} F_{m_1 \dots m_4}\gamma^{m_1 \dots m_4}  \epsilon 
      - \tfrac{1}{4} \tfrac{1}{7!} \tilde{F}_{m_1 \dots m_7}
         \gamma^{m_1 \dots m_7} \epsilon &=0,
\end{split}
\end{equation}
where $\nabla$ is the Levi-Civita connection for the metric on $M$ and $\gamma^m$ are the $\Cliff{7;\bbR}$ gamma matrices. These imply that $\tilde{F}$ vanishes for Minkowski backgrounds~\cite{KMT03}, since it can only be supported by a cosmological constant.

There are similar expressions for the Killing spinor equations in type II (see for example~\cite{GMPT05}). In this case, there are a pair of real ten-dimensional spinors $\{\varepsilon_1,\varepsilon_2\}$ which decompose under $\Spin{3,1}\times\Spin6$ as 
\begin{equation}
\label{eq:typeIIspinor}
\begin{split}
   \varepsilon_1 &= \eta^+_1\otimes \chi^-_1 
       +  \eta^-_1\otimes \chi^+_1 , \\
   \varepsilon_2 &= \eta^+_2 \otimes \chi^\pm_2 
       +  \eta^-_2\otimes \chi^\mp_2 , 
\end{split}
\end{equation}
where the $\pm$ superscripts denote chiralities, $\chi_i^-$ and $\eta^-_i$ are the charge conjugates of $\chi_i^+$ and $\eta^+_i$ respectively, and the upper and lower signs in the second line refer to type IIA and IIB respectively. The two internal spinors can be combined into a single, complex, eight-component object
\begin{equation}
\label{eq:IIspinor}
   \epsilon = \begin{pmatrix} \chi^-_1 \\ \chi^\pm_2 \end{pmatrix} ,
\end{equation}
which for type IIA is simply the lift to the $d=7$ complex spinor of the M-theory reduction. 

In both type II and M-theory, the gamma matrices generate an action of $\SU8$ on the eight-component spinors $\epsilon$. Using the $\SU8$ norm, given in type II by $\bar{\epsilon}\epsilon=\bar{\chi}^+_1\chi_1^+ + \bar{\chi}^+_2\chi^+_2$, supersymmetry implies $\bar{\epsilon}\epsilon = \text{const.} \times \ee^{\Delta}$~\cite{KMT03,BJ03,DP04,LS05,GMPT05}. For $\mathcal{N}=2$ backgrounds we have two independent solutions, $\epsilon_1$ and $\epsilon_2$, to the Killing spinor equations. With respect to the $\SU8$ action, the solutions are thus invariant under an $\SU6$ subgroup. In $\Ex{7(7)}\times\bbR^+$ generalised geometry this $\SU8$ action is a local symmetry~\cite{CSW11,CSW14}. From this perspective, as stressed in~\cite{GLSW09,CSW14b}, we can view the $\mathcal{N}=2$ background as defining a generalised $\SU6$ structure
\begin{equation}
   \text{$\mathcal{N}=2$ background $\{\epsilon_1,\epsilon_2\}$} \quad
   \Longleftrightarrow \quad
   \text{generalised $\SU6$ structure} . 
\end{equation}
Understanding how this $\SU6$ structure is defined and its integrability conditions, along with the analogous structures in $D=5$ and $D=6$, will be the central goal of this paper.


\subsection{Generalising the notion of a Calabi--Yau structure}
\label{sec:gen-ing-CY}

The classic example of an $\mathcal{N}=2$ background is, of course, type II string theory with vanishing fluxes, where $M$ is a Calabi--Yau threefold. Generic flux solutions of the $\mathcal{N}=2$ Killing spinor equations can thus be thought of as string-theory generalisations of the conventional notion of a Calabi--Yau manifold to backgrounds including both NS-NS and R-R fluxes. 

Calabi--Yau manifolds admit a single covariantly constant spinor $\chi^+$ which is invariant under the action of an $\SU3$ subgroup of $\Spin6\simeq\SU4$. In this case, the two $\SU8$ Killing spinors of~\eqref{eq:IIspinor} are given by 
\begin{equation}
\label{eq:CY-KS}
   \epsilon_1 = \begin{pmatrix} \chi^+ \\ 0 \end{pmatrix} , \qqq
   \epsilon_2 = \begin{pmatrix} 0 \\ \chi^- \end{pmatrix} ,
\end{equation}
One can build two differential forms as bilinears in $\chi$: a symplectic form $\omega$ and a holomorphic three-form $\Omega$, satisfying the compatibility conditions
\begin{equation}
\label{eq:SU3-consistency}
   \omega\wedge\Omega = 0 , \qqqq
   \tfrac{1}{3!}\omega\wedge\omega\wedge\omega 
      = \tfrac{1}{8}\ii \Omega\wedge\bar{\Omega} . 
\end{equation}
The Killing spinor equations are equivalent to the integrability conditions 
\begin{equation}
   \dd\omega = 0 , \qqq \dd\Omega = 0 ,
\end{equation}
which together define an integrable or torsion-free $\SU3$ structure on $M$. The forms $\omega$ and $\Omega$ are invariant under $\Symp{6;\bbR}$ and $\SL{3;\bbC}$ subgroups of $\GL{6;\bbR}$ respectively. Thus the different structure groups embed as 
\begin{equation}
\label{eq:embedCY}
   \begin{array}{ccccc}
     \GL{6;\bbR} && \supset 
          && \text{$\Symp{6;\bbR}$ for $\omega$}  \\
     \cup &&&& \cup \\
     \text{$\SL{3;\bbC}$ for $\Omega$} && \supset 
          && \text{$\SU3$ for $\{\omega,\Omega\}$} 
   \end{array} 
\end{equation}

The simplest extension is to consider generic NS-NS backgrounds by including  $H=\dd B$ flux and dilaton. These are beautifully described within generalised complex geometry~\cite{Hitchin02,Gualtieri04,GMPT04b,GMPT05}. The two $\SU8$ spinors are taken to have the form\footnote{As we will discuss in some detail, there are also pure NS-NS, $\mathcal{N}=2$ backgrounds where the Killing spinors do not take the form~\eqref{eq:KS-genCY}, and hence are not described by generalised complex structures.}
\begin{equation}
\label{eq:KS-genCY}
   \epsilon_1 = \begin{pmatrix} \chi_1^+ \\ 0 \end{pmatrix} , \qqq
   \epsilon_2 = \begin{pmatrix} 0 \\ \chi_2^- \end{pmatrix} .
\end{equation}
Each spinor $\chi_i^+$ is stabilised by a different $\SU3$ subgroup of $\Spin6\simeq\SU4$. Generically the common subgroup leaving both $\chi_i^+$ invariant is $\SU2$. However, since the norm between the spinors can vary over $M$, there can be points where the spinors are parallel and the stabiliser group enhances to $\SU3$. Backgrounds where this happens are called ``type-changing''~\cite{Hitchin02,Gualtieri04}. The presence of two spinors $\chi^+_i$ means that the differential forms constructed from the spinor bilinears are more intricate than in the Calabi--Yau case. The background can be characterised by two polyforms~\cite{GMPT05} 
\begin{equation}
   \Phi^+ \in \Gamma(\ext^+ T^*M) , \qqq \Phi^- \in \Gamma(\ext^-T^*M) ,
\end{equation}
where $\ext^+T^*M$ and $\ext^-T^*M$ are the bundles of even- and odd-degree forms respectively. They satisfy a pair of consistency conditions~\eqref{eq:Phi-consistency} and the Killing spinor equations are equivalent to the integrability conditions
\begin{equation}
   \dd \Phi^+ = 0 , \qqq \dd \Phi^- = 0 ,
\end{equation}
which define what is known as a generalised Calabi--Yau metric. A conventional Calabi--Yau background is of course a special case, given by taking 
\begin{equation}
\label{eq:IIB_def_pure_spinors}
\Phi^{+}=\ee^{-\phi}\ee^{-B}\ee^{-\ii\omega},\qqq
\Phi^{-}=\ii\ee^{-\phi}\ee^{-B}\Omega,
\end{equation}
with $B$ closed and $\phi$ constant. Thus we see that $\Phi^+$ generalises the symplectic structure and $\Phi^-$ generalises the complex structure. 

These conditions define an integrable structure in $\Orth{6,6}\times\bbR^+$ generalised geometry~\cite{Hitchin02,Gualtieri04,GMPT05}. One considers the generalised tangent bundle $E\simeq TM\oplus T^*M$, which admits a natural $\Orth{6,6}$ metric $\eta$. The two polyforms $\Phi^\pm$ can then be viewed as sections of the positive and negative helicity $\Spin{6,6}$ spinor bundles\footnote{In making this identification there is an arbitrary scaling factor that can be viewed as promoting the $\Orth{6,6}$ action to an $\Orth{6,6}\times\bbR^+$ action, corresponding to the dilaton degree of freedom~\cite{Witt06,GMPW08}.} associated to $E$, each stabilised by a (different) $\SU{3,3}$ subgroup of $\Spin{6,6}$. Therefore, each $\Phi^\pm$ individually defines a generalised $\SU{3,3}$ structure. The compatibility conditions imply that their common stability group is $\SU3\times\SU3$, so 
\begin{equation}
\label{eq:embedGCY}
   \begin{array}{ccccc}
     \Orth{6,6}\times\bbR^+ && \supset 
          && \text{$\SU{3,3}_+$ for $\Phi^+$}  \\
     \cup &&&& \cup \\
     \text{$\SU{3,3}_-$ for $\Phi^-$} && \supset 
          && \text{$\SU3\times\SU3$ for $\{\Phi^+,\Phi^-\}$} 
   \end{array} 
\end{equation}
Note that the two $\SU3$ stabiliser groups are precisely the groups preserving $\chi_1^+$ and $\chi_2^-$ in~\eqref{eq:KS-genCY}. As we discuss in section~\ref{sec:0dd}, the integrability conditions $\dd\Phi^\pm=0$ are equivalent to the existence of a torsion-free generalised connection compatible with the relevant $\SU{3,3}_\pm$ structure. 

It is natural to ask how these structures and their integrability conditions are extended when one considers completely generic backgrounds, for example including R-R fluxes for which the type II Killing spinor equations take the form~\eqref{eq:typeIIspinor}. These are the questions we address in the next two sections. In identifying the relevant objects in the generalised geometry, and how they connect to conventional notions of $G$-structures, it will be useful to have a range of examples of $\mathcal{N}=2$ backgrounds. To this end, a number of simple cases, with and without R-R fluxes and in both type II and M-theory, are summarised in appendix~\ref{sec:examples}, along with more details of the Calabi--Yau and generalised Calabi--Yau metric cases.


\section{\texorpdfstring{$\Ex{7(7)}$}{E7(7)} structures}
\label{sec:Hypermultiplet-structures}

We will now show that a generic $\mathcal{N}=2$, $D=4$ background defines a pair of generalised structures in $\Ex{7(7)}\times\bbR^+$ generalised geometry. For type II backgrounds this pair was first identified in~\cite{GLSW09}. We will turn to the integrability conditions in the next section. 

The idea of a generalised $G$-structure is as follows. In conventional geometry, the generic structure group of the tangent bundle $TM$ of a $d$-dimensional manifold $M$ is $\GL{d;\bbR}$. The existence of a $G$-structure implies that the structure group reduces to $G\subset\GL{d;\bbR}$. It can be defined by a set of tensors $\Xi$ that are stabilised by the action of $G$, or alternatively as a principle $G$-sub-bundle $P_G$ of the $\GL{d;\bbR}$ frame bundle $F$. In generalised geometry, one considers an extended tangent bundle $E$ which admits the action of a group larger than $\GL{d;\bbR}$. For us the relevant generalised geometry will have an action of $\Ex{7(7)}\times\bbR^+$. One can define frames for $E$ and a corresponding principle $\Ex{7(7)}\times\bbR^+$-bundle, called the generalised frame bundle $\tilde{F}$. A generalised $G$-structure is then defined by a set of generalised tensors that are invariant under the action of a subgroup $G\subset\Ex{7(7)}\times\bbR^+$. Equivalently, it is a principle $G$-sub-bundle $\GS{G}$, of the generalised frame bundle $\tilde{F}$. 

The two generalised $G$-structures relevant to $\mathcal{N}=2$, $D=4$ backgrounds are\footnote{In~\cite{GLSW09} these were denoted $K_\alpha$ and $\lambda=2\re L$ respectively.}\textsuperscript{,}\footnote{$\Spinstar{12}$ is the double cover of $\text{SO}^*(12)$, the latter corresponding to a particular real form of the complex $\SO{12;\bbC}$ Lie algebra~\cite{Gilmore08}.} 
\begin{equation}
\begin{aligned}
   &\text{\textit{hypermultiplet structure}, $J_\alpha$} && &
      G &= \Spinstar{12} , \\
   &\text{\textit{vector-multiplet structure}, $K$} && & 
      G &= \Ex{6(2)} . 
\end{aligned}   
\end{equation}
We will often refer to these as \HM{} and \VM{} structures respectively. As we will see, we can impose two compatibility conditions between the structures such that their common stabiliser group is $\Spinstar{12}\cap\Ex{6(2)}=\SU6$, defining  
\begin{equation}
\begin{aligned}
   &\text{\textit{\HV{} structure}, $\{J_\alpha,K\}$} && &
      G &= \SU6 . 
\end{aligned}   
\end{equation}
We see that the generalisation of the embeddings~\eqref{eq:embedCY} and~\eqref{eq:embedGCY} for Calabi--Yau and generalised Calabi--Yau metrics respectively is given by 
\begin{equation}
   \begin{array}{ccccc}
     \Ex{7(7)}\times\bbR^+ && \supset 
          && \text{$\Spinstar{12}$ for $J_\alpha$}  \\
     \cup &&&& \cup \\
     \text{$\Ex{6(2)}$ for $K$} && \supset 
          && \text{$\SU6$ for $\{J_\alpha,K\}$} 
   \end{array}  
\end{equation}
The $\SU6$ group is the same one that stabilises the pair of $\SU8$ Killing spinors $\{\epsilon_1,\epsilon_2\}$. 

These structures are generalisations of the symplectic and complex structures on Calabi--Yau manifolds in type II compactifications. Focussing on type IIB, in table~\ref{table:EGT_EGG_IIB} we list the (generalised) tangent bundles and structures that appear in conventional and generalised $\Orth{d,d}\times\bbR^+$ and $\Ex{7(7)}\times\bbR^+$ geometries. We see that the \HM{} structure generalises the symplectic structure $\omega$ (or the pure spinor $\Phi^{+}$), while the \VM{} structure generalises the complex structure $\Omega$ (or the pure spinor $\Phi^{-}$). For type IIA the situation is reversed, and the \VM{} and \HM{} structures generalise $\omega$ and $\Omega$ respectively. 
\begin{table}
\centering
\begin{tabular}{@{}llllll@{}} 
\toprule
\multicolumn{2}{c}{} & \multicolumn{2}{c}{hyper} & \multicolumn{2}{c}{vector} \\
\cmidrule(lr){3-4} \cmidrule(l){5-6}
$E$ & $G_{\text{frame}}$ & $G$ & $\Xi$ & $G$ & $\Xi$ \\
\midrule
$TM$ & $\GL{6}$ & $\Symp{6;\mathbb{R}}$ & $\omega$
& $\SL{3;\bbC}$ & $\Omega$ \\
$TM\oplus T^{*}M$ & $\Orth{6,6}\times\mathbb{R}^{+}$ & $\SU{3,3}_+$ & $\Phi^{+}$ & $\SU{3,3}_-$ & $\Phi^-$ \\
$TM\oplus T^{*}M\oplus\ext^{-}T^{*}M\oplus\ldots$ & $\Ex{7(7)}\times\mathbb{R}^{+}$ & $\Spinstar{12}$ & $J_{\alpha}$ & $\Ex{6(2)}$ & $K$ \\
\bottomrule 
\end{tabular}
\protect\caption{The (generalised) tangent bundles and $G$-structures in conventional, generalised complex and exceptional generalised geometry for type IIB supergravity. We include the group $G_{\text{frame}}$ that acts on the (generalised) frame bundle, the reduced structure group $G$ of the symplectic, complex, generalised complex, vector- or hypermultiplet structure, and the invariant object $\Xi$ that defines the structure.}
\label{table:EGT_EGG_IIB}
\end{table}

Recall that the moduli spaces of (integrable) symplectic and complex structures of Calabi--Yau manifolds are associated with $\mathcal{N}=2$, $D=4$ hypermultiplets and vector multiplets in type II theories. The same thing happens here: the moduli space of integrable $\Spinstar{12}$ structures defines fields in  hypermultiplets and that of integrable $\Ex{6(2)}$ structures defines fields in vector multiplets, hence the names. In fact, one can also consider the infinite-dimensional space of all such structures, without imposing any integrability conditions, and these too can naturally be associated with hypermultiplets and vector multiplets. As described in~\cite{GLW06,GLW07,GLSW09}, one can view this structure as arising from a rewriting of the full ten- or eleven-dimensional theory, analogous to the construction in~\cite{DN86}, but with only eight supercharges manifest. The local $\SO{9,1}$ Lorentz symmetry is broken and the degrees of freedom can be repackaged into $\mathcal{N}=2$, $D=4$ multiplets. However, since all modes are kept -- there is no Kaluza--Klein truncation -- the hyper- and vector-multiplet spaces become infinite dimensional. 
 
In the rest of this section we first review $\Ex{7(7)}\times\bbR^+$ generalised geometry. We then define \HM{} and \VM{} structures, discuss the infinite-dimensional space of structures, and, in each case, show how the various examples of $\mathcal{N}=2$, $D=4$ backgrounds given in appendix~\ref{sec:examples} define $J_\alpha$ and $K$.


\subsection{\texorpdfstring{$\Ex{d(d)}\times\mathbb{R}^{+}$}{Ed(d) x R+} generalised geometry}\label{gen_geo}

$\Ex{d(d)}\times\mathbb{R}^{+}$ or exceptional generalised geometry is the study of structures on a generalised tangent bundle $E$, where $E$ admits a unique action of the $\Ex{d(d)}$ group~\cite{Hull07,PW08}. The corresponding differential geometry was developed in~\cite{CSW11,CSW14}. Here we summarise the key points, relegating some of the details to appendix~\ref{app:gen-geom}. 

For M-theory on a manifold $M$ of dimension $d\leq7$, the generalised tangent bundle is
\begin{equation}
\label{eq:E-Mtheory}
  E\simeq TM\oplus\ext^{2}T^{*}M\oplus\ext^{5}T^{*}M
     \oplus(T^{*}M\otimes\ext^{7}T^{*}M).
\end{equation}
For a type II theory on a $(d-1)$-dimensional manifold $M$, the generalised tangent bundle is
\begin{equation}
\label{eq:E-typeII}
E\simeq TM\oplus T^{*}M\oplus\ext^{\pm}T^{*}M\oplus\ext^{5}T^{*}M\oplus(T^{*}M\otimes\ext^{6}T^{*}M),
\end{equation}
where $\pm$ refers to even- or odd-degree forms for type IIA or IIB respectively. For type IIA, this is just a dimensional reduction of the M-theory case. For type IIB, this can be rewritten in a way that stresses the $\SL{2;\bbR}$ symmetry as
\begin{equation}
\label{eq:E-IIB}
   E \simeq TM\oplus(S\otimes T^{*}M)    
      \oplus \ext^{3}T^{*}M
      \oplus(S\otimes \ext^{5}T^{*}M)
      \oplus(T^{*}M\otimes\ext^{6}T^{*}M),
\end{equation}
where $S$ is an $\bbR^2$ bundle transforming as a doublet of $\SL{2;\bbR}$. In all cases the generalised tangent bundle is an $\Ex{d(d)}\times\mathbb{R}^{+}$ vector bundle. For example, for $d=7$ it transforms in the $\rep{56}_{\rep{1}}$ representation, where the subscript denotes the $\bbR^+$ weight. By definition, a scalar field of weight $p$, transforming in the representation $\rep{1_p}$, is a section of $(\det T^*M)^{p/(9-d)}$. (Since supersymmetric backgrounds are orientable, we can assume $\det T^*M$ is trivial.) 

The generalised frame bundle $\tilde{F}$ is an $\Ex{d(d)}\times\mathbb{R}^{+}$ principal bundle constructed from frames for $E$. One defines generalised tensors as sections of the vector bundles associated with different $\Ex{d(d)}\times\mathbb{R}^{+}$ representations. Of particular interest is the adjoint bundle $\ad\tilde{F}$, corresponding to the adjoint representation of $\Ex{d(d)}\times\bbR^+$. In M-theory we have
\begin{equation}
\label{eq:ad-M-theory}
   \ad\tilde{F}\simeq\mathbb{R}
      \oplus(TM\otimes T^{*}M)
      \oplus\ext^{3}T^{*}M \oplus\ext^{6}T^{*}M
      \oplus\ext^{3}TM \oplus\ext^{6}TM , 
\end{equation}
while in type II 
\begin{equation}
\label{eq:ad-II}
\begin{aligned}
   \ad\tilde{F} 
   &\simeq \bbR \oplus \left[ \mathbb{R} \oplus \ext^{6}TM 
      \oplus \ext^{6}T^*M \right] \\ & \eqspace
   \oplus\left[(TM\otimes T^{*}M) \oplus \ext^{2}T^*M 
      \oplus \ext^{2}TM \right]
   \oplus \left[ \ext^\mp TM \oplus \ext^\mp T^*M \right] ,
\end{aligned}
\end{equation}
where the upper and lower signs refer to type IIA and type IIB respectively. For IIB this can also be written as 
\begin{equation}
\label{eq:ad-IIB}
\begin{split}
   \ad\tilde{F} 
      &\simeq \mathbb{R}
        \oplus(TM\otimes T^{*}M)
        \oplus(S\otimes S^{*})_0
        \oplus(S\otimes\ext^{2}TM)
        \oplus(S\otimes\ext^{2}T^{*}M) \\  &   
        \eqspace \oplus\ext^{4}TM
        \oplus\ext^{4}T^{*}M
        \oplus(S\otimes\ext^{6}TM)
        \oplus(S\otimes\ext^{6}T^{*}M) , 
\end{split}
\end{equation}
where the subscript on $(S\otimes S^*)_0$ indicates that one takes the traceless part. For $d=7$ these bundles transform in the $\rep{1_0}+\rep{133_0}$ representation, where the singlet is the part generating the $\bbR^+$ action.  

The generalised tangent bundle is actually defined as an extension, so that there is a non-trivial patching between the tensor components. In M-theory, on the overlap of two local patches $U_{i}\cap U_{j}$ of $M$, a generalised vector $V\in\Gamma(E)$ is patched by  
\begin{equation}
V_{(i)}=\ee^{\dd\Lambda_{(ij)}+\dd\tilde{\Lambda}_{(ij)}}V_{(j)},
\end{equation}
where $\Lambda_{(ij)}$ and $\tilde{\Lambda}_{(ij)}$ are locally two- and five-forms respectively, which can be identified as sections of $\ad\tilde{F}$, so that $\ee^{\dd\Lambda_{(ij)}+\dd\tilde{\Lambda}_{(ij)}}$ is the exponentiated adjoint action. The isomorphisms~\eqref{eq:E-Mtheory} and~\eqref{eq:ad-M-theory} depend on a pair of potentials $A\in\Gamma(\ext^3T^*M)$ and $\tilde{A}\in\Gamma(\ext^6T^*M)$ via the exponentiated adjoint action
\begin{equation}
\label{eq:M-twist}
   V = \ee^{A+\tilde{A}}\tilde{V} , \qqq
   R = \ee^{A+\tilde{A}}\tilde{R}\,\ee^{-A-\tilde{A}} , 
\end{equation}
where $V\in\Gamma(E)$ and $R\in\Gamma(\ad\tilde{F})$, the ``untwisted'' objects $\tilde{V}$ and $\tilde{R}$ are sections of $TM\oplus\ext^2T^*M\oplus\cdots$ and $\bbR\oplus(TM\otimes T^*M)\oplus\cdots$ respectively, and $A$ and $\tilde{A}$ are patched by 
\begin{equation}
   A_{(i)} = A_{(j)} + \dd\Lambda_{(ij)} , \qqq
   \tilde{A}_{(i)} = \tilde{A}_{(j)} + \dd\tilde{\Lambda}_{(ij)}
      - \tfrac12 \dd\Lambda_{(ij)} \wedge A_{(j)} . 
\end{equation}
The corresponding gauge-invariant field strengths 
\begin{equation}
   F = \dd A, \qqq
   \tilde{F} = \dd\tilde{A} - \tfrac{1}{2}A\wedge F,
\end{equation}
are precisely the supergravity objects defined in~\eqref{eq:FtF}. The type II theories are similarly patched. For type IIB we have \begin{equation}
V_{(i)}=\ee^{\dd\Lambda_{(ij)}^{i}+\dd\tilde{\Lambda}_{(ij)}}V_{(j)},
\end{equation}
where $\Lambda_{(ij)}^{i}$ and $\tilde{\Lambda}_{(ij)}$ are locally a pair of one-forms and a three-form respectively. The relations between the twisted and untwisted objects are written as 
\begin{equation}
\label{eq:IIB-twist}
   V = \ee^{B^i+C}\tilde{V} , \qqq
   R = \ee^{B^i+C}\tilde{R}\,\ee^{-B^i-C} ,
\end{equation}
with the corresponding three- and five-form field strengths given by 
\begin{equation}
\label{eq:IIB_flux_main}
   F^{i}=\dd B^{i},\qqq 
   F= \dd C+\tfrac{1}{2}\epsilon_{ij}B^i\wedge F^j , 
\end{equation}
where $F^1=H$, $F^2=F_3$ and $F$ are the usual supergravity field strengths. We discuss how to include a non-zero axion-dilaton in appendix \ref{rep_IIB}, following \cite{LSW14}.

The differential structure of the generalised tangent bundle is captured by a generalisation of the Lie derivative that encodes the bosonic symmetries of supergravity, namely diffeomorphisms and form-field gauge transformations. Given a generalised vector field $V\in\Gamma(E)$, one can define the action of the generalised Lie derivative (or Dorfman derivative) $\Dorf_{V}$ on any generalised tensor. For example, its action on generalised vectors is given in~\eqref{eq:M_Dorf_vector} and~\eqref{eq:IIB_Dorf_vector}, and on sections of $\ad\tilde{F}$ in~\eqref{eq:M_Dorf_adjoint} and~\eqref{eq:IIB_Dorf_adjoint}. It endows $E$ with the structure of a Leibniz algebroid~\cite{Baraglia12}. It will play an essential role in defining the integrability conditions on the generalised structures. 


\subsection{Hypermultiplet structures}\label{hyper_structure}

The idea of an hypermultiplet structure (or \HM{} structure) was first introduced in~\cite{GLSW09} in the context of type II theories. Formally we have:
\begin{defn}
An $\Ex{7(7)}$ \emph{hypermultiplet structure} is a
$\Spinstar{12}\subset\Ex{7(7)}\times\bbR^+$ generalised structure, 
\end{defn} 
\noindent
or, in other words, a $\Spinstar{12}$ principle sub-bundle $\GS{\Spinstar{12}}$ of the generalised frame bundle $\tilde{F}$. More concretely, we can define the structure by choosing a set of invariant generalised tensors. The relevant objects are a triplet of sections of a weighted adjoint bundle
\begin{equation}
\label{eq:J-def}
   J_\alpha \in \Gamma(\ad\tilde{F}\otimes(\det T^*M)^{1/2})
   \qquad \alpha=1,2,3,
\end{equation}
such that they transform in the $\rep{133_1}$ representation of $\Ex{7(7)}\times\bbR^+$. We require them to define a \emph{highest weight} $\su{2}$ subalgebra of $\ex{7(7)}$, which is the necessary and sufficient condition for them to be invariant under $\Spinstar{12}$. We can write the algebra as 
\begin{equation}
\label{eq:su2_algebra}
   [J_{\alpha},J_{\beta}] = 2\kappa\epsilon_{\alpha\beta\gamma}J_{\gamma},
\end{equation}
where $\kappa$ is a section of $(\det T^{*}M)^{1/2}$ and the commutator is simply the commutator in the adjoint representation of $\Ex{7(7)}\times\bbR^+$, defined in~\eqref{eq:ad-ad-M} and~\eqref{eq:ad-ad-IIB}. The norms of the $J_{\alpha}$, calculated using the $\ex{7(7)}$ Killing form given in~\eqref{eq:M_Killing} and~\eqref{eq:IIB_Killing}, are then fixed to be
\begin{equation}
\label{eq:J_a_norm}
   \tr ( J_{\alpha}J_{\beta} )=-\kappa^{2}\delta_{\alpha\beta}.
\end{equation}
As described in~\cite{GLSW09}, decomposing under the $\SU8$ subgroup\footnote{The actual subgroup is $\SU8/\bbZ_2$ but throughout this paper, for simplicity, we will ignore discrete group factors.} of $\Ex{7(7)}$, one can view the corresponding ``untwisted'' objects $\tilde{J}_\alpha$ as being constructed from bilinears of the Killing spinors $\epsilon_i$ of the form $\sigma_\alpha^{ij}\epsilon_i\bar{\epsilon}_j$, where $\sigma_\alpha^{ij}$ are the Pauli matrices. 

A key point for us, first noted in~\cite{GLSW09}, is that the infinite-dimensional space of \HM{} structures admits a natural hyper-K\"ahler metric. To define the space of structures, note that, at a particular point $x\in M$, the structure $\left.J_\alpha\right|_x$ is invariant under $\Spinstar{12}$ so it can be viewed as fixing a point in the homogeneous space 
\begin{equation}
   \left.J_\alpha\right|_x \in W = \Ex{7(7)}\times\bbR^+/\Spinstar{12} . 
\end{equation}
One can then consider the fibre bundle of homogeneous spaces 
\begin{equation}
   \begin{tikzcd}
      W \arrow{r} & \ZH \arrow{d} \\
     {} & M 
   \end{tikzcd} 
\end{equation}
constructed by taking a quotient $\ZH=\tilde{F}/G$ of the generalised frame bundle $\tilde{F}$ by the structure group $G=\Spinstar{12}$. Choosing an \HM{} structure is equivalent to choosing a section of $\ZH$. Thus the infinite-dimensional space of all possible \HM{} structures is simply the space of smooth sections, 
\begin{equation}
   \text{space of hypermultiplet structures $\MH=\Gamma(\ZH)$.}
\end{equation}
Crucially, the fibres $W$ of $\ZH$ are themselves pseudo-Riemannian hyper-K\"ahler spaces. In fact $W$ is a hyper-Kähler cone over a pseudo-Riemannian symmetric quaternionic-Kähler space, also known as a Wolf space,
\begin{equation}
\label{eq:E7-wolf}
   W/\bbH^* = \Ex{7(7)}/(\Spinstar{12}\times\SU2) , 
\end{equation}
where the action of the quaternions $\bbH^*$ mods out by $\SU2\times\bbR^+$. The Riemannian symmetric quaternionic-K\"ahler spaces were first considered by Wolf in~\cite{Wolf65} and classified by Alekseevsky in~\cite{Alekseevsky68}, while the pseudo-Riemannian case was analysed by Alekseevsky and Cort\'es~\cite{AC05}, and~\eqref{eq:E7-wolf} is indeed included in their list. Recall that one can always construct a hyper-K\"ahler cone, known as the Swann bundle, over any  quaternionic-K\"ahler space~\cite{Swann91}. In this case the cone directions are simply the $\SU2$ bundle together with the overall $\bbR^+$ scaling. The hyper-K\"ahler geometry on $W$, as first described in~\cite{KS01}, is summarised in appendix~\ref{app:wolf}. 

The hyper-K\"ahler geometry on $\MH$ is inherited directly from the hyper-K\"ahler geometry of the $W$ fibres of $\ZH$. This is in much the same way that the infinite-dimensional space of smooth Riemannian metrics on a compact $d$-dimensional manifold (which can be viewed as the space of sections of a $\GL{d;\bbR}/\Orth{d}$ homogeneous fibre bundle) is itself a Riemannian space~\cite{DeWitt67,Ebin68,GM91}. The construction follows that on $W$. Concretely, consider a point $\sigma\in \MH$, corresponding to a choice of section $\sigma(x)\in\Gamma(\ZH)$. Equivalently, given a point $\sigma\in\MH$ we have a triplet of sections $J_\alpha(x)$. Formally, one can think of $J_\alpha(x)[\sigma]$ as a triplet of functions on $\MH$ taking values in the space of sections $\Gamma(\ad\tilde{F}\otimes(\det T^*M)^{1/2})$
\begin{equation}
\label{eq:J-map}
   J_\alpha : \MH 
     \to \Gamma(\ad\tilde{F}\otimes(\det T^*M)^{1/2}) . 
\end{equation}
The tangent space $T_\sigma \MH$ at $\sigma$ is spanned by vectors $v\in T_\sigma\MH$ that can be viewed as a small deformation of the structure $J_\alpha(x)$. Formally, we can define the change $v_\alpha(x)$ in $J_\alpha(x)$, given by $v$ acting on the section-valued functions $J_\alpha$, that is $v_\alpha=v(J_\alpha)=\imath_v \delta J_\alpha$, where $\delta$ is the exterior derivative on $\MH$. By definition, $v_\alpha(x)$ is a section of $\ad\tilde{F}\otimes(\det T^*M)^{1/2}$. At each point $\sigma$ it can always be written as 
\begin{equation}
v_{\alpha}(x)=[R(x),J_{\alpha}(x)] ,
\end{equation}
where $R(x)$ is a section of the $\ex{7(7)}\oplus\bbR$ adjoint bundle $\ad\tilde{F}$. Note that only elements that are not in $\spinstar{12}$ actually generate non-zero $v_\alpha$. Decomposing $\ad\tilde{F}\simeq\ad\GS{\Spinstar{12}}\oplus\ad\GS{\Spinstar{12}}^\perp$, where $\GS{\Spinstar{12}}$ is the generalised $G$-structure defined by $J_\alpha$, this means formally we can also identify
\begin{equation}
\label{eq:TMH}
   T_\sigma\MH 
      \simeq \Gamma(\ad\GS{\Spinstar{12}}^\perp\otimes(\det T^*M)^{1/2}) .
\end{equation}
Given two tangent vectors $v,w\in T_\sigma \MH$, we then define a triplet of symplectic forms at the point $\sigma\in \MH$, such that the symplectic products between $v$ and $w$ are given by 
\begin{equation}
\label{triplet_symplectic}
   \Omega_{\alpha}(v,w)
      = \epsilon_{\alpha\beta\gamma}\int_{M}\tr (v_{\beta}w_{\gamma}).
\end{equation}
Recall that $v_\alpha(x)$ and $w_\alpha(x)$ are sections of $\ad\tilde{F}\otimes(\det T^*M)^{1/2}$. Thus $\tr (v_{\beta}w_{\gamma})$ is a section of $\det T^*M$ and can indeed be integrated over $M$. These forms define the hyper-K\"ahler structure. 

The geometry on $\MH$ is actually itself a hyper-K\"ahler cone. There is a \emph{global} $\SU2\times\bbR^+$ action that rotates and rescales the structures $J_\alpha$. This means that one can define a hyper-K\"ahler potential~\cite{Swann91}, a real function $\chi$ which is simultaneously a K\"ahler potential for each of the three symplectic structures. On $\MH$ it is given by the functional
\begin{equation}
   \chi = \tfrac{1}{2}\int_M \kappa^2 ,
\end{equation}
where $\kappa^2$ is the density that depends on the choice of structure $\sigma(x)\in\Gamma(\ZH)$ through~\eqref{eq:su2_algebra}. In terms of the Killing spinors $\epsilon_i$, the global $\SU2$ symmetry corresponds to the fact that, under the decompositions~\eqref{eq:mtheoryspinor} and~\eqref{eq:typeIIspinor}, the $\epsilon_i$ are determined only up to global $\mathrm{U}(2)$ rotations of the pair of four-dimensional spinors $\eta_i^+$. Thus the global $\SU2$ action on $J_\alpha$ is simply part of the four-dimensional $\mathcal{N}=2$ R-symmetry. The global $\bbR^+$ rescaling corresponds to shifting the warp factor $\Delta$ in~\eqref{eq:warp-metric} by a constant, and then absorbing this in a constant conformal rescaling of the flat metric $\dd s^2(\bbR^{3,1})$. Modding out by these symmetries, we see that the physical space of structures is actually an infinite-dimensional quaternionic-K\"ahler space. As we have mentioned, this structure on $\MH$ can be viewed, following~\cite{GLW06,GLW07,GLSW09}, as a rewriting of the full ten- or eleven-dimensional supergravity theory as a four-dimensional $\mathcal{N}=2$ theory coupled to an infinite number of hypermultiplets, corresponding to the full tower of Kaluza--Klein modes parametrising $\MH$. Physically, the Swann bundle structure corresponds to coupling hypermultiplets to superconformal gravity~\cite{WKV00,WRV01,WRV01b}. 


\subsection{Vector-multiplet structures}\label{sec:V-structures}

Vector-multiplet structures (or \VM{} structures) were also first introduced in~\cite{GLSW09} in the context of type II theories. Formally we have:
\begin{defn}
An $\Ex{7(7)}$ \emph{vector-multiplet structure} is an
$\Ex{6(2)}\subset\Ex{7(7)}\times\bbR^+$ generalised structure.
\end{defn} 
\noindent
In other words, it is an $\Ex{6(2)}$ principle sub-bundle $\GS{\Ex{6(2)}}$ of the generalised frame bundle $\tilde{F}$. 

The corresponding invariant generalised tensor is a section of the generalised tangent bundle
\begin{equation}
\label{eq:T-def}
   K \in \Gamma(E),
\end{equation}
which we recall transforms in the $\rep{56_1}$ representation of $\Ex{7(7)}\times\bbR^+$. This tensor is almost generic, the only requirement is that it satisfies
\begin{equation}
   q(K) > 0 , 
\end{equation}
where $q$ is the quartic invariant of $\Ex{7(7)}$.\footnote{Recall that $\Ex{7(7)}$ can be defined as the group preserving a symplectic invariant $s$ and a symmetric quartic invariant $q$. Given the $\bbR^+$ weight of $E$, note that $q(K)\in\Gamma((\det T^*M)^2)$.} This ensures that the stabiliser group is $\Ex{6(2)}$~\cite{FG98}. As will see below when we discuss the geometry of the space of \VM{} structures following~\cite{GLSW09}, one can use $q(K)$ to construct a second invariant generalised vector $\hat{K}$, and it is often convenient to consider the complex object 
\begin{equation}
   X = K + \ii\hat{K} . 
\end{equation}
Decomposing under the $\SU8$ subgroup of $\Ex{7(7)}$, one can view the corresponding ``untwisted'' objects $\tilde{X}$ as being constructed from bilinears of the Killing spinors $\epsilon_i$ of the form $\epsilon^{ij}\epsilon_i\epsilon^T_j
=\epsilon_1\epsilon_2^T-\epsilon_2\epsilon_1^T$.

In this case, the infinite-dimensional space of \VM{} structures admits a natural rigid (or affine) special K\"ahler metric~\cite{GLSW09}. The structure $\left.K\right|_x$ at a particular point $x\in M$ fixes a point in the homogeneous space 
\begin{equation}
   \left.K\right|_x \in P = \Ex{7(7)}\times\bbR^+/\Ex{6(2)} . 
\end{equation}
One can then consider the fibre bundle of homogeneous spaces 
\begin{equation}
   \begin{tikzcd}
      P \arrow{r} & \ZV \arrow{d} \\
     {} & M 
   \end{tikzcd} 
\end{equation}
constructed by taking a quotient $\ZV=\tilde{F}/G$ of the generalised frame bundle $\tilde{F}$ by the structure group $G=\Ex{6(2)}$. Choosing a \VM{} structure is equivalent to choosing a section of $\ZV$. Thus the infinite-dimensional space of all possible \VM{} structures is simply the space of smooth sections, 
\begin{equation}
   \text{space of vector-multiplet structures $\MV=\Gamma(\ZV)$.}
\end{equation}
The space of $K$ is an open subset of $\Gamma(E)$, thus we can identify the space of \VM{} structures as 
\begin{equation}
\label{eq:Y-coord}
   \MV = \{ K\in\Gamma(E) : q(K) >0 \} . 
\end{equation}
Note that $\Gamma(E)$ is a vector space, and hence we have a natural set of local flat coordinates on $\MV$, fixed by choosing a frame for $E$. The decomposition into conventional tensors as in~\eqref{eq:V-Mtheory} and~\eqref{eq:V-IIB} is an example of such a choice.

The special K\"ahler metric on $\MV$ is again inherited from the special K\"ahler metric on $P$, the homogeneous space fibres of $\ZV$. (Special K\"ahler geometry is reviewed in~\cite{CRTV97,Freed98} and summarised  in appendix~\ref{sec:SK}.) Recall that one can always define a complex cone over a local special K\"ahler manifold to give the corresponding rigid special K\"ahler manifold. The Riemannian symmetric spaces that admit local special K\"ahler metrics were analysed in~\cite{CFG89,Cecotti89} and include the case $\Ex{7(-25)}/(\Ex6\times\Uni1)$. Here we need a pseudo-Riemannian form based on $\Ex{7(7)}$, so the relevant space is 
\begin{equation}
\label{eq:LSK}
   P/\bbC^* = \Ex{7(7)}/(\Ex{6(2)}\times\Uni1).
\end{equation}
Here the $\bbC^*$ action is generated by the $\Uni1$ bundle together with the overall $\bbR^+$ scaling. The rigid special K\"ahler geometry on $\MV$ can be formulated in analogy to Hitchin's construction of the metric on the space of $\SL{3;\bbC}$ structures~\cite{Hitchin00} and $\SU{3,3}$ structures~\cite{Hitchin02}. The space $P$ is a ``prehomogeneous vector space''~\cite{SK77}, that is, it is an open orbit of $\Ex{7(7)}\times\bbR^+$ in the real $\rep{56}_{\rep{1}}$ representation. The open subset is defined by the condition $q(K)>0$. Consider a point $K\in \MV$. The vectors in the tangent space $T_K \MV$ at $K$ can be viewed as a small deformation of $K$, which are just sections of $E$, hence $T_K \MV\simeq \Gamma(E)$. Given $v,w\in T_K \MV$, the fibre-wise $\Ex{7(7)}$ symplectic invariant $s$ then defines a symplectic form $\Omega$ on $\MV$ by 
\begin{equation}
\label{eq:sympl}
   \Omega(v,w) = \int_M s(v,w) , 
\end{equation}
where, since sections of $E$ are weighted objects, $s(v,w)$ is a section of $\det T^*M$ and hence it can be integrated over $M$. As reviewed in appendix~\ref{sec:SK}, special K\"ahler geometry requires the existence of a flat connection preserving $\Omega$. Here, the vector-space structure of $\Gamma(E)$ provides natural flat coordinates on $\MV$, and hence defines a flat connection with respect to which $\Omega$ is by definition constant. We can then use the quartic invariant to define a function $H$ that determines the complex structure and hence the metric~\eqref{eq:hessian}. We define the real Hitchin functional 
\begin{equation}
\label{eq:HK}
   H = \int_M \sqrt{q(K)} ,
\end{equation}
where again the weight of $K$ means that $\sqrt{q(K)}\in\Gamma(\det T^*M)$. This defines a second invariant tensor $\hat{K}\in\Gamma(E)\simeq T_K \MV$ as the corresponding Hamiltonian vector field
\begin{equation}
\label{eq:hatK-def}
   \imath_{\hat{K}}\Omega = - \delta H , 
\end{equation}
where $\delta$ is the exterior derivative on $\MV$, and hence an invariant complex generalised vector $X=K+\ii\hat{K}$. The two real invariants correspond to the two singlets in the decomposition $\rep{56}=\rep{1}+\rep{1}+\rep{27}+\rep{\overline{27}}$ under $\Ex{6(2)}\subset\Ex{7(7)}$. The metric on $\MV$ is given by the Hessian
\begin{equation}
   H_{MN} = -\frac{\delta H}{\delta K^M\delta K^N} , 
\end{equation}
where $M=1,\dots,56$ denote the components of $K$. The definition of the metric is equivalent to choosing a complex structure given by $\mathcal{I}^M{}_N=-\delta\hat{K}^M/\delta K^N$, and implies that $-H$ is the K\"ahler potential for the special K\"ahler metric on $\MV$.\footnote{Note that our conventions for the $\Ex{7(7)}$ symplectic form mean that the metric here is $\frac{1}{8}$ that in~\cite{GLSW09}. Also our normalisation of the quartic invariant is fixed relative to the symplectic form by the relation~\eqref{eq:LSK-K}.} In these expressions we are using the flat coordinates on $\MV$ defined by the vector space structure on $\Gamma(E)$. To see the more conventional description of special K\"ahler geometry in terms of a holomorphic prepotential $\mathcal{F}$, one needs to switch to a particular class of complex coordinates, as described in~\cite{Freed98}. 

On any rigid special K\"ahler geometry there is a global $\bbC^*$ symmetry, such that the quotient space is, by definition, a local special K\"ahler geometry. On $\MV$, the action of $\bbC^*$ is constant rescaling and phase-rotation of the invariant tensor $X$. The $\Uni1$ part is simply the overall $\Uni1$ factor of the four-dimensional $\mathcal{N}=2$ R-symmetry, while, as for the hypermultiplet structure, the $\bbR^+$ action is a reparametrisation of the warp factor $\Delta$.\footnote{The second compatibility condition in~\eqref{eq:e7_comp_condition} implies that $\bbR^+$ actions on $J_\alpha$ and $X$ are correlated.} Modding out by this symmetry, the physical space of structures $\MV/\bbC^*$ is an infinite-dimensional local special K\"ahler space. This is in line with the discussion of~\cite{GLW06,GLW07,GLSW09}, where we view $\MV/\bbC^*$ as the space of vector-multiplet degrees of freedom, coming from rewriting the full ten- or eleven-dimensional supergravity theory as a four-dimensional $\mathcal{N}=2$ theory. Physically, the cone structure on $\MV$ corresponds to coupling the vector multiplets to superconformal gravity~\cite{WKV00,WRV01,WRV01b}. 

\subsection{Exceptional Calabi--Yau structures}

In the previous sections, we defined two generalised structures that give the extension of complex and symplectic geometry of Calabi--Yau manifolds for generic flux solutions, but alone these are not enough to characterise a supersymmetric background. Recall that $\mathcal{N}=2$ backgrounds define a generalised $\SU6$ structure~\cite{GLSW09,CSW14b}, this $\SU6$ being the same group that stabilises the $\mathcal{N}=2$ Killing spinors. Formally we define:
\begin{defn}
	An $\Ex{7(7)}$ \emph{\HV{} structure} is an
	$\SU{6}\subset\Ex{7(7)}\times\bbR^+$ generalised structure, 
\end{defn} 
\noindent
or, in other words, an $\SU{6}$ principle sub-bundle $\GS{\SU{6}}$ of the generalised frame bundle $\tilde{F}$. Here \HV{} stands for ``exceptional Calabi--Yau''. For type II backgrounds it is the flux generalisation of a Calabi--Yau three-fold, while for M-theory it is the generalisation of the product of a Calabi--Yau three-fold and $\text{S}^1$. Strictly the \HV{} structure will refer to an integrable $\SU6$ structure, but for simplicity we will often not make this distinction. 

As in the simpler Calabi--Yau case, to ensure that the background is indeed $\mathcal{N}=2$ we need to impose a compatibility condition between the \HM{} and \VM{} structures such that together they define a generalised $\SU6$ structure. The common stabiliser group $\Spinstar{12}\cap\Ex{6(2)}$ of the pair $\{J_\alpha,K\}$ is $\SU6$ if and only if $J_\alpha$ and $K$ satisfy two compatibility conditions.
\begin{defn}
The two structures $J_\alpha$ and $K$ are \emph{compatible} if together they define an $\SU{6}\subset\Ex{7(7)}\times\bbR^+$ generalised structure. The necessary and sufficient conditions are~\cite{GLSW09}
\begin{equation}
\label{eq:e7_comp_condition}
\begin{split}
J_{\alpha}\cdot K &= 0, \\
\tr (J_\alpha J_\beta) &= - 2\sqrt{q(K)}\,\delta_{\alpha\beta} ,
\end{split}
\end{equation}
where $\cdot$ is the adjoint action $\rep{133}\times\rep{56}\rightarrow\rep{56}$, given in \eqref{eq:M_adjoint} and \eqref{eq:IIB_adjoint}.
\end{defn}
\noindent These constraints can be thought of as the generalisations of the conditions~\eqref{eq:SU3-consistency} between symplectic and complex structures on a Calabi--Yau manifold. Note that they are equivalent to
\begin{equation}
\label{eq:JT-comp}
J_+ \cdot X = J_- \cdot X = 0 ,
\end{equation}
where $J_\pm=J_1\pm\ii J_2$, and the normalisation condition
\begin{equation}
\label{eq:e7_comp}
\tfrac{1}{2}\ii s(X,\bar{X}) = \kappa^2 , 
\end{equation}
respectively, where $\kappa$ is the factor appearing in~\eqref{eq:su2_algebra} and $s(\cdot,\cdot)$ is the $\Ex{7(7)}$ symplectic invariant, given in~\eqref{eq:M_symplectic} and~\eqref{eq:IIB_symplectic}. 


\subsection{Examples of \texorpdfstring{$\Ex{7(7)}$}{E7(7)} structures}
\label{sec:H-examples}

We now show how the examples of $\mathcal{N}=2$ supersymmetric backgrounds described in appendix~\ref{sec:examples} each define an \HM{} and \VM{} structure. We hope this will give a sense of the variety of geometries that can be described. In the same way that generic generalised complex structures can be thought of as interpolating between complex and symplectic structures, we will see that \HM{} structures can interpolate between these and conventional hyper-Kähler structures. Similarly, \VM{} structures cover a wide range of possibilities, interpolating between complex, symplectic and simple product structures. We will also check that the structures are compatible, and so define an \HV{} or generalised $\SU6$ structure. Although we do not give the details, the structures can be calculated explicitly as Killing spinor bilinears using the decomposition of $\Ex{7(7)}$ under $\SU8$.

Throughout this section we will use the ``musical isomorphism'' to raise indices with the background metric $g$ on $M$. For example, if $\omega$ is a two-form, $\omega^\sharp$ is the corresponding bivector $(\omega^\sharp)^{mn} = g^{mp}g^{nq}\omega_{pq}$. Note that when the flux is non-trivial, since the compatibility and normalisation conditions are $\Ex{7(7)}\times\bbR^+$ covariant, we can always check them using the untwisted structures. For example, the compatibility condition in M-theory is
\begin{equation}
J_\alpha\cdot K = (\ee^{A+\tilde{A}}\tilde{J}_{\alpha}\ee^{-A-\tilde{A}})\cdot(\ee^{A+\tilde{A}}\tilde{K})
=\ee^{A+\tilde{A}}(\tilde{J}_{\alpha}\cdot \tilde{K})=0
\quad\Leftrightarrow\quad 
\tilde{J}_{\alpha}\cdot \tilde{K}=0. 
\end{equation}

For the following examples, one can check the $\su2$ algebra~\eqref{eq:su2_algebra} and normalisation~\eqref{eq:J_a_norm} of the $J_\alpha$ using \eqref{eq:ad-ad-M} and \eqref{eq:M_Killing} for M-theory, and \eqref{eq:ad-ad-IIB} and \eqref{eq:IIB_Killing} for type IIB. The normalisation~\eqref{eq:e7_comp} of $X$ (or $K$) can be checked using the symplectic invariant, given by \eqref{eq:M_symplectic} for M-theory and \eqref{eq:IIB_symplectic} for type IIB. Finally, one can check compatibility of the structures~\eqref{eq:JT-comp} using the adjoint action, given by \eqref{eq:M_adjoint} for M-theory and \eqref{eq:IIB_adjoint} for type IIB.

\subsubsection{Calabi--Yau manifolds in type IIB}\label{CYIIB}

Consider first type IIB on a Calabi--Yau manifold $M$. The \HM{} structure is defined by the symplectic form $\omega$ on $M$. The decomposition of the adjoint bundle $\ad\tilde{F}$ in this case follows~\eqref{eq:ad-IIB}. 
The \HM{} structure is given by 
\begin{equation}
\label{eq:J-CY}
\begin{split}
J_{+} & =\tfrac{1}{2}\kappa n^{i}\omega
- \tfrac{1}{2}\ii\kappa n^{i}\omega^{\sharp}
+ \tfrac{1}{12}\ii\kappa n^{i}\omega\wedge\omega\wedge\omega
+ \tfrac{1}{12}\kappa n^{i} \omega^{\sharp} \wedge \omega^{\sharp} 
\wedge\omega^{\sharp} , \\
J_{3} & =\tfrac{1}{2}\kappa\hat{\tau}^{i}_{\phantom{i}j}
- \tfrac{1}{4}\kappa\omega\wedge\omega
+ \tfrac{1}{4}\kappa\omega^\sharp\wedge\omega^\sharp ,
\end{split}
\end{equation}
where the $\SL{2;\bbR}$ doublet $n^{i}=(-\ii,1)^i$ is a section of $S$, $\hat{\tau}=-\ii\sigma_{2}$ is a section of $(S\otimes S^*)_0$, where $\sigma_2$ is the second Pauli matrix, and the density is simply $\kappa^{2}=\vol_{6}$, where $\vol_6=\frac{1}{3!}\omega\wedge\omega\wedge\omega$ is the volume form on $M$. Note that $J_3$ can be thought of as a combination of two $\Uni{1}$ actions embedded in $\Ex{7(7)}$, the first generated by $\hat{\tau}$ in $\mathfrak{sl}_2$ and the second generated by $\omega\wedge\omega-\omega^\sharp\wedge\omega^\sharp$. Since $\omega^\sharp=\omega^{-1}$, $J_\alpha$ is completely determined by $\omega$ alone. 

Recall that in type IIB the generalised tangent bundle $E$ has a decomposition into tensors, given in~\eqref{eq:E-IIB}. For a Calabi--Yau background, the \VM{} structure is defined by the holomorphic three-form $\Omega$ simply as 
\begin{equation}
\label{T-CY}
X= \Omega .
\end{equation}
We can also check the compatibility condition~\eqref{eq:JT-comp} given the form of $J_\alpha$ in~\eqref{eq:J-CY}. The adjoint action~\eqref{eq:IIB_adjoint} gives
\begin{equation}
J_{+}\cdot X 
\propto -\ii n^{i}\omega^{\sharp}\lrcorner\Omega
+ n^{i}\Omega\wedge\omega, \qqq
J_{-}\cdot X 
\propto-\ii \bar{n}^{i}\omega^{\sharp}\lrcorner\Omega
+ \bar{n}^{i}\Omega\wedge\omega.
\end{equation}
These vanish if and only if $\omega\wedge\Omega=\omega\wedge\bar{\Omega}=0$, from which we recover the standard compatibility condition~\eqref{eq:SU3-consistency} for an $\SU{3}$ structure.

\subsubsection{\texorpdfstring{$\text{CY}_3 \times \text{S}^1$}{CY3 x S1} in M-theory}

For type IIA compactifications on Calabi--Yau three-folds, the complex structure should define the \HM{} structure. If we add the M-theory circle to this case, we expect the holomorphic three-form $\Omega$ and the complex structure $I$ to appear in $J_\alpha$ -- this is indeed the case. Using the decomposition~\eqref{eq:ad-M-theory}, we find 
\begin{equation}
\label{eq:J-MCY}
\begin{aligned}
J_{+} &= \tfrac{1}{2}\kappa\Omega 
- \tfrac{1}{2}\kappa\Omega^{\sharp}, \\
J_{3} &= \tfrac{1}{2}\kappa I
- \tfrac{1}{16}\ii\kappa\Omega\wedge\bar{\Omega}
- \tfrac{1}{16}\ii\kappa\Omega^{\sharp}\wedge\bar{\Omega}^{\sharp} ,
\end{aligned}
\end{equation}
where the density is just the volume form $\kappa^{2}=\vol_{7}=\frac{1}{8}\ii\Omega\wedge\bar{\Omega}\wedge \zeta$. 

The symplectic structure on the Calabi--Yau manifold determines the \VM{} structure. Using the decomposition~\eqref{eq:E-Mtheory}, we find 
\begin{equation}
\label{eq:T-MCY}
X = \zeta^\sharp + \ii\omega 
- \tfrac{1}{2}\zeta\wedge\omega\wedge\omega
- \ii \zeta\otimes\vol_7 .
\end{equation}
Using the adjoint action~\eqref{eq:M_adjoint} and the algebraic conditions $\imath_{\zeta^\sharp}\Omega=0$, $\imath_{\zeta^\sharp}\omega=0$ and $\omega\wedge\Omega=0$, it is straightforward to show that the compatibility conditions~\eqref{eq:JT-comp} are satisfied. 

\subsubsection{Generalised Calabi--Yau metrics in type II}

This is the case first considered in~\cite{GLSW09}. The \HM{} structure is determined by the $\SU{3,3}_\pm$ structure pure spinors $\Phi^-$ and $\Phi^+$ in type IIA and type IIB respectively. To see the embedding it is natural to use the decomposition of $\Ex{7(7)}$ under $\SL{2;\bbR}\times\Orth{6,6}$. The adjoint bundle was given in~\eqref{eq:ad-II}. The three sets of terms in brackets correspond to the decomposition $\rep{133}=(\rep{3},\rep{1}) + (\rep{1},\rep{66}) + (\rep{2},\rep{32}^\mp)$, while the first term is just the singlet $(\rep{1},\rep{1})$ generating the $\bbR^+$ action.

The \HM{} structure is given by\footnote{Note that with our conventions, the $\rep{32}^\mp$ component $C^\mp$ here is equal to $\sqrt{2}$ times the $C^\mp$ used in~\cite{GLSW09}.}
\begin{equation}
\label{eq:J-GCY}
\begin{aligned}
J_{+} &= u^{i}\Phi^{\mp} , \\ 
J_{3} &= \kappa(u^{i}\bar{u}_{j}+\bar{u}^{i}u_{j}) 
- \tfrac{1}{2}\kappa\mathcal{J}^{\mp} , 
\end{aligned}
\end{equation}
where the upper/lower choice of sign in $\Phi^{\mp}$ gives the type IIA/IIB embedding, and we have defined
\begin{equation}
u^i = \frac{1}{2}\begin{pmatrix}
-\ii\kappa \\ \kappa^{-1} \end{pmatrix}^i
\in \Gamma((\det T^*M)^{1/2}\otimes
(\bbR\oplus\ext^6TM )) , 
\end{equation}
with 
\begin{equation}
\kappa^2 = \tfrac{1}{8}\ii\langle\Phi^{\pm},\bar{\Phi}^{\pm}\rangle , 
\end{equation}
where $u_{i}=\epsilon_{ij}u^{j}$, so that $u^{i}\bar{u}_{i}=-\tfrac{\ii}{2}$, and we are using the isomorphism $\ext^\pm TM\simeq\ext^6TM\otimes\ext^\pm T^*M$. The object $\mathcal{J}^\pm$, transforming in the $\Orth{6,6}$ adjoint representation $(\rep{1},\rep{66})$, is the generalised complex structure defined in \eqref{eq:GCS-J-def}. It is important to note that the NS-NS $B$-field is included in the definition of the pure spinors so that the objects $J_\alpha$ are honest sections of the twisted bundle $\ad\tilde{F}$. 

Using the adjoint action and the $\ex{7(7)}$ Killing form in section~3 of~\cite{GLSW09}, one can check that the triplet satisfies the $\su{2}$ algebra~\eqref{eq:su2_algebra} and is correctly normalised~\eqref{eq:J_a_norm}. The embedding reduces to the previous examples in that, for type IIA, the pure spinor $\Phi^{-}$ corresponding to the complex structure embeds in $J_{\alpha}$ and, for type IIB, we find $J_{\alpha}$ contains the symplectic structure. Note that upon taking a conventional symplectic structure, we expect this to reduce to the type IIB case of section \ref{CYIIB}. It is important to note that the $\SL{2;\bbR}$ factor in each case is different: for type IIB it is S-duality, while for the generalised complex structure it is the commutant of the $\Orth{6,6}$ action. Taking this into account, it is straightforward to show the two cases match after including a constant $\SU2$ rotation of the $J_\alpha$.

The \VM{} structure is determined by the generalised complex structure as~\cite{GLSW09}
\begin{equation}
\label{eq:T-GCY}
X= \Phi^{\pm},
\end{equation}
where the upper/lower choice of sign in $\Phi^{\pm}$ gives the type IIA/IIB embedding. Using the symplectic invariant in section~3 of \cite{GLSW09}, rescaled by a factor of $1/4$, one can check this satisfies the normalisation condition (\ref{eq:e7_comp}). Notice that upon taking a conventional complex structure, this does indeed reduce to the case of section \ref{CYIIB}.

For $J_+$ in~\eqref{eq:J-GCY}, the adjoint action in section~3 of~\cite{GLSW09} gives 
\begin{equation}
J_{+}\cdot  X
\propto u^{i}\langle\Phi^{\mp},\Gamma^{A}\Phi^{\pm}\rangle,
\qqq    J_{-}\cdot  X
\propto \bar{u}^{i}\langle\bar{\Phi}^{\mp},\Gamma^{A}\Phi^{\pm}\rangle.
\end{equation}
These vanish if $\langle\Phi^{\pm},\Gamma^{A}\Phi^{\mp}\rangle=\langle\bar{\Phi}^{\pm},\Gamma^{A}\Phi^{\mp}\rangle=0$. We recover the compatibility conditions \eqref{eq:Phi-consistency} for $\{\Phi^{+},\Phi^{-}\}$ to define an $\SU{3}\times\SU{3}$ structure.

\subsubsection{D3-branes on \texorpdfstring{$\text{HK} \times \bbR^2$}{HK x R2} in type IIB}

In this case, the hyper-K\"ahler geometry on $M$ provides a natural candidate for realising the $\su2$ algebra. Using the structures defined in section~\ref{sec:D3-bkgd}, we start by defining the \emph{untwisted} structure
\begin{equation}
\label{eq:J-D3}
\tilde{J}_{\alpha} =
- \tfrac{1}{2}\kappa I_{\alpha} 
- \tfrac{1}{2}\kappa \omega_{\alpha}\wedge
\zeta_{1}\wedge \zeta_{2}
+ \tfrac{1}{2}\kappa \omega^{\sharp}_{\alpha}\wedge
\zeta_{1}^{\sharp}\wedge \zeta_{2}^{\sharp} , 
\end{equation}
where $\kappa^{2}=\ee^{2\Delta}\vol_{6}$ includes the warp factor. The actual structure is a section of the twisted bundle $\ad\tilde{F}$, and includes the four-form potential $C$ and two-form potentials $B^i$ via the adjoint action as in~\eqref{eq:IIB-twist}
\begin{equation}
J_\alpha = \ee^{B^i+C} \tilde{J}_\alpha \ee^{-B^i-C} .
\end{equation}
We see explicitly that \HM{} structures can also encode hyper-K\"ahler geometries. 

$X$ essentially defines the structure of the $\bbR^2$ factor, since the hyper-K\"ahler structure was already encoded in $J_\alpha$. We first define the untwisted object 
\begin{equation}
\label{eq:T-D3}
\tilde{X} = \bar{n}^{i}\ee^\Delta(\zeta_{1}-\ii \zeta_{2})
+ \ii \bar{n}^{i}\ee^\Delta (\zeta_{1}-\ii \zeta_{2})\wedge\vol_{4} , 
\end{equation}
where $n^{i}=(-\ii,1)^i$ and $\frac{1}{2}\omega_{\alpha}\wedge\omega_{\beta}=\delta_{\alpha\beta}\vol_4$. The presence of five- and three-form flux means the actual structure is a section of the twisted bundle $E$
\begin{equation}
X = \ee^{B^i+C} \tilde{X} .
\end{equation}
We can check the compatibility condition with $J_\alpha$ in~\eqref{eq:J-D3}. This can be done using the twisted or untwisted forms, since the twisting is an $\Ex{7(7)}\times\bbR^+$ transformation. We find 
\begin{equation}
\begin{split}
\tilde{J}_{\alpha}\cdot \tilde{X} 
&\propto - \bar{n}^{i} I_{\alpha}\cdot(\zeta_{1}-\ii \zeta_{2})
- \ii \bar{n}^i(\omega^{\sharp}_{\alpha}\wedge \zeta^{\sharp}_{1}
\wedge \zeta^{\sharp}_{2})
\lrcorner\bigl((\zeta_{1}-\ii \zeta_{2})
\wedge\vol_{4}\bigr) \\
& \eqspace
- \ii \bar{n}^{i} I_{\alpha}\cdot\bigl(
(\zeta_{1}-\ii \zeta_{2})\wedge\vol_{4}\bigr) .
\end{split}
\end{equation}
This vanishes as $I_{\alpha}\cdot \zeta_{i}=I_{\alpha}\cdot\vol_{4}=0$ and $\zeta^{\sharp}_{i}\lrcorner\omega_{\alpha}=0$. 

\subsubsection{Wrapped M5-branes on \texorpdfstring{$\text{HK} \times \bbR^3$}{HK x R2} in M-theory}

The final example is that of wrapped M5-branes. As discussed in section~\ref{sec:M5-bkgd}, the geometry admits two different sets of Killing spinors depending on whether the M5-branes wrap $\bbR^2$ or a K\"ahler two-cycle in the hyper-K\"ahler geometry. These lead to two different \HM{} structures. 

Let us consider the K\"ahler two-cycle case first. Using the structure defined in section~\ref{sec:M5-bkgd}, we can define the untwisted \HM{} structure as
\begin{equation}
\label{eq:J-M5K}
\begin{split}
\tilde{J}_\alpha &= -\tfrac{1}{2}\kappa r_\alpha 
+ \tfrac{1}{2} \kappa\omega_3 \wedge \zeta_\alpha 
-\tfrac{1}{2}\kappa\omega_3^\sharp \wedge \zeta_\alpha^\sharp \\ & \eqspace 
- \tfrac{1}{4}\kappa \epsilon_{\alpha\beta\gamma}
\zeta_\beta \wedge \zeta_\gamma \wedge \vol_4
- \tfrac{1}{4}\kappa \epsilon_{\alpha\beta\gamma}
\zeta_\beta^\sharp \wedge \zeta_\gamma^\sharp \wedge \vol_4^\sharp , 
\end{split}
\end{equation}
where $\kappa=\ee^{2\Delta}\vol_7$ and the tensors
\begin{equation}
r_\alpha = \epsilon_{\alpha\beta\gamma}
\zeta_\beta^\sharp\otimes \zeta_\gamma
\in \Gamma(TM\otimes T^*M) , 
\end{equation}
generate the $\SO3$ rotations on $\bbR^3$. The \VM{} structure is defined by the untwisted object
\begin{equation}
\label{eq:T-M5K}
\tilde{X} = \ee^{\Delta} \Omega + \ii \ee^{\Delta} \Omega\wedge \vol_3 ,
\end{equation}
where $\Omega =\omega_2 + \ii \omega_1$. 

For M5-branes wrapped on $\bbR^2$, the untwisted structures are 
\begin{equation}
\label{eq:J-M5R2}
\tilde{J}_\alpha = -\tfrac{1}{2}\kappa I_{\alpha} 
+ k\tfrac{1}{2}\kappa \omega_{\alpha}\wedge \zeta_3 
- k\tfrac{1}{2}\kappa \omega^{\sharp}_{\alpha}\wedge \zeta_3^{\sharp},
\end{equation}
where again $\kappa^2=\ee^{2\Delta}\vol_7$, and 
\begin{equation}
\begin{split}
\label{eq:T-M5R2}
\tilde{X} &= \ee^\Delta(\zeta_1^\sharp + \ii \zeta_2^\sharp) 
+ \ee^\Delta (\zeta_1 + \ii \zeta_2) \wedge \zeta_3
- \ee^\Delta (\zeta_1 + \ii \zeta_2)\wedge\vol_4 \\
& \eqspace - \ii \ee^\Delta (\zeta_1 + \ii \zeta_2) \otimes \vol_7. 
\end{split}
\end{equation}. 

In both cases there is a non-trivial four-form flux, so that the actual twisted structures depend on the three-form potential $A$ and, as in~\eqref{eq:M-twist}, are given by
\begin{equation}
J_\alpha = \ee^A \tilde{J}_\alpha \ee^{-A} , \qquad X = \ee^A \tilde{X} . 
\end{equation}
It is easy to check that in both cases the algebra (\ref{eq:su2_algebra}), normalisation and compatibility conditions are all satisfied. 


\section{Integrability}\label{sec:Integrability}

Having given the algebraic definitions of hyper- and vector-multiplet structures, we now need to find the differential conditions on them that imply the background is supersymmetric. Formulations in terms of specific generalised connections have already appeared in~\cite{GLSW09,GO12}. Here we would like to write conditions that use only the underlying differential geometry, in the same way that $\dd\omega=\dd\Omega=0$ depends only on the exterior derivative. The key ingredient will be the action of the group of generalised diffeomorphisms $\GDiff$. Infinitesimally, this action is generated by the generalised Lie derivative $\Dorf_V$, and we will see that all the conditions are encoded using this operator.

We will show that the hypermultiplet conditions arise as moment maps for the action of $\GDiff$ on the space of structures $\MH$. These maps were already partially identified in~\cite{GLSW09}. As we prove in section~\ref{sec:int-tor}, in the language of $G$-structures, they are equivalent to requiring that the generalised $\Spinstar{12}$ structure is torsion-free. The vector-multiplet condition similarly implies that the generalised $\Ex{6(2)}$ structure is torsion-free. Finally we consider integrability for an \HV{} structure. Given integrable \HM{} and \VM{} structures, there is an additional requirement for the generalised $\SU6$ structure, defined by the pair $\{J_\alpha,K\}$, to be torsion-free. In other words, the existence of compatible torsion-free $\Spinstar{12}$ and $\Ex{6(2)}$ structures is not sufficient to imply that the $\SU6$ structure is torsion-free. While not inconsistent with the general $G$-structure formalism, this is in contrast with the Calabi--Yau case, where the combination of integrable and compatible symplectic and complex structures is enough to imply the manifold is Calabi--Yau.


\subsection{Integrability of the hypermultiplet structure}\label{int_hyp}

We now introduce moment maps for the action of generalised diffeomorphisms on the infinite-dimensional space of \HM{} structures. An \HM{} structure is then integrable if the corresponding moment maps vanish.

We denote the group of generalised diffeomorphisms -- diffeomorphisms and form-field gauge transformations -- by $\GDiff$. Infinitesimally it is generated by the generalised Lie derivative $\Dorf_V$, where $V$ is a generalised vector, that is, a section of $E$. Thus roughly we can identify the Lie algebra $\gdiff$ with the space of sections $\Gamma(E)$. Actually this is not quite correct since there is a kernel in the map $\Gamma(E)\to\gdiff$. For example, in M-theory, on a local patch $U_i$ of $M$, we see from~\eqref{eq:M_Dorf_vector} that the component $\tau\in\Gamma(T^*U_i\otimes\ext^7T^*U_i)$ in $V$ does not contribute to $\Dorf_V$. Similarly, if the components $\omega\in\Gamma(\ext^2T^*U_i)$ and $\sigma\in\Gamma(\ext^5T^*U_i)$ are closed they do not contribute. In what follows, it is nonetheless convenient to parametrise elements of $\gdiff$ by $V\in\Gamma(E)$ remembering that this map is not an isomorphism. 

Suppose that $\sigma(x)\in \MH$ is a particular choice of \HM{} structure parametrised by the triplet $J_\alpha$. The change in structure generated by $\gdiff$ is $\delta J_\alpha=\Dorf_V J_\alpha$, which can be viewed as an element of the tangent space $T_\sigma \MH$. Thus we have a map
\begin{equation}
   \rho\colon\mathfrak{gdiff}\rightarrow \Gamma(T\MH) ,
\end{equation}
such that, acting on the triplet of section-valued functions $J_\alpha$ defined in~\eqref{eq:J-map}, the vector $\rho_V$ generates a change in $J_\alpha$  
\begin{equation}
   \rho_V(J_\alpha) = \Dorf_{V}J_{\alpha} . 
\end{equation}
Given an arbitrary vector field $w\in\Gamma(T\MH)$, we have, from~\eqref{triplet_symplectic}, that
\begin{equation}
   \imath_{\rho_V}\Omega_\alpha(w) = \Omega_\alpha(\rho_V,w)
      = \epsilon_{\alpha\beta\gamma} \int_M \tr \bigl((
          \Dorf_VJ_\beta) w_\gamma\bigr) . 
\end{equation}
If $\pi\in\Gamma(\ext^7T^*M)$ is a top-form, so that it transforms in the $\rep{1_2}$ representation of $\Ex{7(7)}\times\bbR^+$, then by definition
\begin{equation}
\label{eq:int-dorf}
   \int_M \Dorf_V \pi = \int_M \mathcal{L}_v \pi = 0 , 
\end{equation}
where $\mathcal{L}_v$ is the conventional Lie derivative and $v\in\Gamma(TM)$ is the vector component of the generalised vector $V\in\Gamma(E)$. Using the Leibniz property of $\Dorf_V$, we then have 
\begin{equation}
\begin{split}\label{exact}
   \imath_{\rho_V}\Omega_\alpha(w) 
      &= \tfrac{1}{2}\epsilon_{\alpha\beta\gamma} \int_M 
           \tr \bigl[(\Dorf_VJ_\beta) w_\gamma 
           - J_\beta (\Dorf_V w_\gamma) \bigr] \\
      &= - \tfrac{1}{2}\epsilon_{\alpha\beta\gamma} \int_M 
           \tr \bigl[ w_\beta (\Dorf_VJ_\gamma) 
           + J_\beta (\Dorf_V w_\gamma) \bigr] \\
      &= \imath_{w} \delta \mu_\alpha(V),
\end{split} 
\end{equation}
where $\delta$ is the exterior derivative on $\MH$, that is, a functional derivative such that by definition $\imath_w\delta J_\alpha=w_\alpha$, and 
\begin{equation}
\label{eq:mm}
  \mu_{\alpha}(V)
  \coloneqq -\tfrac{1}{2}\epsilon_{\alpha\beta\gamma}
      \int_{M}\tr \bigl(J_{\beta}(\Dorf_{V}J_{\gamma})\bigr) , 
\end{equation}
is a triplet of moment maps.

With this result we can define what we mean by an integrable structure:
\begin{defn}
An \emph{integrable} or \emph{torsion-free} hypermultiplet structure $J_\alpha$ is one satisfying  
\begin{equation}
\label{eq:e7_mm}
   \mu_{\alpha}(V) = 0
   \qquad\text{for all $V\in\Gamma(E)$} , 
\end{equation}
where $\mu_\alpha(V)$ is given by~\eqref{eq:mm}. 
\end{defn}
\noindent As we will show in section~\ref{sec:int-tor}, these conditions are equivalent to $J_\alpha$ admitting a torsion-free, compatible generalised connection. They are also the differential conditions on $J_\alpha$ implied by the requirement that the background admits Killing spinors preserving $\mathcal{N}=2$ supersymmetry in four dimensions. 


\subsection{Integrability of the vector-multiplet structure}

The integrability condition for the vector-multiplet structure $K$ also depends on the generalised Lie derivative, but in a very direct way. Recall that $K\in\Gamma(E)$, thus we can consider the generalised Lie derivative along $K$, namely $\Dorf_K$.
\begin{defn}
An \emph{integrable} or \emph{torsion-free} vector-multiplet structure $K$ is one satisfying
\begin{equation}
\label{eq:e7_comp_int1}
\Dorf_KK=0 , 
\end{equation}
or, in other words, $K$ is invariant under the generalised diffeomorphism generated by itself. 
\end{defn}
\noindent As we will show in section~\ref{sec:int-tor}, these conditions are equivalent to there being a torsion-free generalised connection compatible with the generalised $\Ex{6(2)}$ structure defined by $K$. Furthermore, it is easy to see that it implies $\Dorf_K\hat{K}=0$. In addition, using the results of appendix~\ref{sec:int-tor-su6}, we see that the generalised Lie derivative $\Dorf_X X$, where $X=K+\ii\hat{K}$, is identically zero for any vector-multiplet structure $K$. Hence the integrability condition~\eqref{eq:e7_comp_int1} is equivalent to 
\begin{equation}
\label{eq:e7_comp_int}
   \Dorf_X \bar{X} = 0 . 
\end{equation}
Again, \eqref{eq:e7_comp_int1} is implied by the existence of $\mathcal{N}=2$ Killing spinors. 

In section~\ref{sec:m-vector}, we will show that~\eqref{eq:e7_comp_int1} is actually equivalent to the vanishing of a moment map for the action of $\GDiff$ on $\MV$. 


\subsection{Integrability of the \HV{} structure}

Finally, we can consider the integrability conditions for the \HV{} structure, defined by a compatible pair $\{J_\alpha,K\}$. 
\begin{defn}
An \emph{integrable} or \emph{torsion-free} \HV{} structure $\{J_\alpha,K\}$ is one such that $J_\alpha$ and $K$ are separately integrable and in addition
\begin{equation}
\label{eq:e7_intersection_int}
   \Dorf_X J_\alpha = 0 , 
\end{equation}
or, in other words, the $J_\alpha$ are also invariant under the generalised diffeomorphisms generated by $K$ and $\hat{K}$.
\end{defn}
\noindent 
As we will show in section~\ref{sec:int-tor}, these conditions are equivalent to there being a torsion-free generalised connection compatible with the generalised $\SU6$ structure, defined by $\{J_\alpha,K\}$. Using the results of~\cite{CSW14b}, this implies that these conditions are equivalent to the existence of $\mathcal{N}=2$ Killing spinors. 

It is important to note that the pair of compatible and integrable \HM{} and \VM{} structures is not enough to imply that the \HV{} structure is integrable. This is because there can be a kernel in the torsion map, as can happen for conventional $G$-structures.\footnote{See appendix C of \cite{GMW03} for an explicit example of a non-integrable product structure defined by the product of two compatible, integrable complex structures.}


\subsection{Examples of integrable structures}\label{sec:int_examples}

We now return to our examples of supersymmetric $\mathcal{N}=2$ backgrounds and show in each case that the relevant integrability conditions~\eqref{eq:e7_mm},~\eqref{eq:e7_comp_int} and~\eqref{eq:e7_intersection_int} are satisfied. For the examples of Calabi--Yau in type IIB and $\text{CY}_3\times\text{S}^1$ in M-theory, we show that the conditions are necessary and sufficient using a decomposition into $\SU3$ torsion classes. The torsion classes are more complicated for the other examples, and so we show only that the supersymmetric backgrounds give examples of integrable structures. Instead, the equivalence of integrability and $\mathcal{N}=2$ supersymmetry is shown using generalised intrinsic torsion in section~\ref{sec:int-tor}.

There are a number of convenient calculational tools we will use. First note that in the $(J_+,J_-,J_3)$ basis, the moment map conditions are naturally written as the combinations 
\begin{equation}
   \mu_{3}\coloneqq\tfrac{\ii}{2}\int_{M}\tr \bigl(J_{-}(\Dorf_{V}J_{+})\bigr)=0,
   \qqq
   \mu_{+}\coloneqq-\ii\int_{M}\tr \bigl(J_{3}(\Dorf_{V}J_{+})\bigr)=0, 
\end{equation}
and $\Dorf_X J_\alpha$ is equivalent to $\Dorf_X J_+=\Dorf_X J_-=0$. We also note that, from the form of the generalised Lie derivative~\eqref{eq:Dorf-def-M} and the adjoint projection~\eqref{eq:del-ad-M} (and the corresponding expressions~\eqref{eq:Dorf-def-IIB} and~\eqref{eq:del-ad-IIB} in type IIB), acting on any generalised tensor $\alpha$
\begin{equation}
   \Dorf_V \alpha = \mathcal{L}_v \alpha - R \cdot \alpha , 
\end{equation}
where $R\in\Gamma(\ad\tilde{F})$, $R\cdot\alpha$ is the adjoint action, $v$ is the vector component of $V$, $\mathcal{L}_v$ is the conventional Lie derivative and 
\begin{equation}
   R = \begin{cases}
           \dd\omega + \dd\sigma & \text{for M-theory} \\
            \dd\lambda^i + \dd\rho + \dd\sigma^i & \text{for type IIB}
        \end{cases} ,
\end{equation}
where we are using the standard decompositions of $V$ given in~\eqref{eq:V-Mtheory} and~\eqref{eq:V-IIB}. Using the identity $\tr (A[B,C])=\tr (B[C,A])$ and the algebra~\eqref{eq:su2_algebra}, this allows us to rewrite the moment maps~\eqref{eq:mm} as
\begin{equation}
\label{eq:Dorf-split}
\begin{split}
  \mu_{\alpha}(V) 
     &= - \tfrac{1}{2}\epsilon_{\alpha\beta\gamma} \int_{M}
        \tr \bigl(J_{\beta}(\mathcal{L}_vJ_{\gamma}-[R,J_\gamma])\bigr) \\
     &= - \tfrac{1}{2}\epsilon_{\alpha\beta\gamma} \int_{M}
        \tr (J_{\beta}\mathcal{L}_v J_{\gamma})
       -2 \int_M \kappa \tr (R J_\alpha) .
\end{split}
\end{equation}

The final tool is that, when the background has flux, it is often useful to write the conditions using the untwisted structures $\tilde{J}_\alpha$ and $\tilde{X}$. For this we need the twisted generalised Lie derivative $\hat{\Dorf}_{\tilde{V}}$.\footnote{The nomenclature here is confusing: the twisted generalised Lie derivative acts on untwisted fields.} This is just the induced action of $\Dorf_V$ on untwisted fields, and is given in~\eqref{eq:M_twisted_dorf} for M-theory and~\eqref{eq:IIB_twisted_dorf} for type IIB. It has the same form as $\Dorf_{V}$ but includes correction terms involving the fluxes due to the $p$-form potentials. This can be written as a modified $R$ in~\eqref{eq:Dorf-split}, given by 
\begin{equation}
\label{eq:tildeR}
   \tilde{R} = \begin{cases}
           \dd\tilde{\omega} - \imath_{\tilde{v}}F
           + \dd\tilde{\sigma} -\imath_{\tilde{v}}\tilde{F} 
           + \tilde{\omega}\wedge F & \text{for M-theory} \\
            \dd\tilde{\lambda}^i - \imath_{\tilde{v}}F^{i}
           + \dd\tilde{\rho} - \imath_{\tilde{v}}F 
           - \epsilon_{ij}\tilde{\lambda}^{i}\wedge F^{j}
           + \dd\tilde{\sigma}^i 
           + \tilde{\lambda}^{i}\wedge F
           - \tilde{\rho}\wedge F^{i} & \text{for type IIB}
        \end{cases} .
\end{equation}
The conditions for integrability on the untwisted structures are simply
\begin{equation}
   \mu_a(\tilde{V}) = -\tfrac{1}{2}\epsilon_{\alpha\beta\gamma}
        \int_M \tr\bigl(\tilde{J}_{\beta}(\hat{\Dorf}_{\tilde{V}}\tilde{J}_{\gamma})\bigr)
        = 0  \quad \forall \tilde{V} , \qqq
   \hat{\Dorf}_{\tilde{X}}\bar{\tilde{X}} = 0 , \qqq
   \hat{\Dorf}_{\tilde{X}}\tilde{J}_\alpha = 0 .
\end{equation}

\subsubsection{Calabi--Yau in type IIB}

Consider first the hypermultiplet structure~\eqref{eq:J-CY}. Parametrising $\tilde{V}$ as in~\eqref{eq:V-IIB}, we get conditions for each component $\tilde{v}$, $\tilde{\lambda}^i$, $\tilde{\rho}$ and $\tilde{\sigma}^i$. From the second term in~\eqref{eq:Dorf-split}, taking each of the form-field components in turn, we find the non-zero moment maps are 
\begin{equation}
\begin{aligned}
   \mu_+(\tilde{\lambda}^i) & \propto \int_M \epsilon_{ij}u^j 
       \kappa^2 \omega^\sharp\lrcorner \dd\tilde{\lambda}^i 
       \propto \int_M \epsilon_{ij}u^j 
       \omega \wedge \omega \wedge \dd\tilde{\lambda}^i 
       \propto \int_M \epsilon_{ij}u^j 
       \dd \omega \wedge \omega \wedge \tilde{\lambda}^i = 0, \\
   \mu_+(\tilde{\sigma}^i) & \propto \int_M \epsilon_{ij}u^j 
       \kappa^2 \vol_6^\sharp\lrcorner \dd\tilde{\sigma}^i 
       \propto \int_M \epsilon_{ij}u^j 
       \dd\tilde{\sigma}^i = 0 ,
\end{aligned}
\end{equation}
where we use $\kappa^2=\vol_6$ so $\kappa^2\omega^\sharp\propto \omega\wedge\omega$, and for $\tilde{\rho}$ 
\begin{equation}
   \mu_3(\tilde{\rho}) \propto \int_M \kappa^2
          (\omega^\sharp\wedge\omega^\sharp)\lrcorner\dd\tilde{\rho} 
       \propto \int_M \omega\wedge \dd\tilde{\rho} 
       \propto \int_M \dd\omega\wedge \tilde{\rho} = 0 ,
\end{equation}
where we use $\kappa^2\omega^\sharp\wedge\omega^\sharp\propto \omega$. From this we recover $\dd\omega=0$. For the vector component $\tilde{v}$ the only non-zero contribution is 
\begin{equation}
\begin{split}
   \mu_{3}(\tilde{v}) & \propto \int_{M} 
       \kappa\omega^\sharp\lrcorner\mathcal{L}_{\tilde{v}}(\kappa\omega)
       - \mathcal{L}_{\tilde{v}}(\kappa\omega^\sharp)\lrcorner\kappa\omega
       + \kappa\vol_6^\sharp\lrcorner\mathcal{L}_{\tilde{v}}(\kappa\vol_6)
       - \mathcal{L}_{\tilde{v}}(\kappa\vol_6^\sharp)\lrcorner\kappa\vol_6 \\
       & \propto \int_{M} 
       \tfrac{1}{2}\omega\wedge\omega\wedge\mathcal{L}_{\tilde{v}}\omega
       + \mathcal{L}_{\tilde{v}}\vol_6
       = 0 ,
\end{split}
\end{equation}
which can be seen to vanish using $\dd\omega=0$, $\mathcal{L}_{\tilde{v}}\omega=\imath_{\tilde{v}}\dd\omega+\dd\imath_{\tilde{v}}\omega$ and integrating by parts.

Turning to the conditions on $X$ given by~\eqref{T-CY}, from~\eqref{eq:IIB_Dorf_vector} only the $\tau$ component of the Dorfman derivative is non-trivial 
\begin{equation}
\Dorf_{X}\bar{X}  =j\bar{\Omega}\wedge\dd\Omega = 0 .
\end{equation}
Notice that the integrability condition is considerable weaker than requiring an integrable $\SL{3;\bbC}$ structure -- it only requires that the type-$(3,1)$ part of $\dd\Omega$ vanishes. In the intrinsic torsion language of~\cite{CS02}, only the $\mathcal{W}_5$ component is set to zero, so that the underlying almost complex structure is unconstrained. 

The pair $\{J_\alpha,K\}$ define an integrable generalised $\SU6$ structure if they are individually integrable and also satisfy~\eqref{eq:e7_intersection_int}. From \eqref{eq:IIB_Dorf_adjoint}, we have
\begin{equation}
\begin{aligned}
   \Dorf_{X}J_{+} &\propto
      \ii u^{i}\omega^{\sharp}\lrcorner\dd\Omega
      - u^{i}\omega\wedge\dd\Omega = 0 , \\
   \Dorf_{X}J_3 &\propto
      - \tfrac{1}{2}(\omega^\sharp\wedge\omega^\sharp)
          \lrcorner \dd\Omega
      - j(\omega^\sharp\wedge\omega^\sharp)
          \lrcorner j\dd\Omega
       + \tfrac{1}{2}\id (\omega^\sharp\wedge\omega^\sharp)
          \lrcorner \dd\Omega = 0 , 
\end{aligned}
\end{equation}
which sets the remaining type-$(2,2)$ components of $\dd\Omega$ to zero. Taken together, we have $\dd\omega=\dd\Omega=0$, as expected.

\subsubsection{\texorpdfstring{$\text{CY}_3 \times \text{S}^1$}{CY3 x S1} in M-theory}

Consider first the hypermultiplet structure~\eqref{eq:J-MCY}. Parametrising $\tilde{V}$ as in~\eqref{eq:V-Mtheory}, the form field components $\tilde{\omega}$ and $\tilde{\sigma}$ in the second term in~\eqref{eq:Dorf-split} give the non-zero moment maps
\begin{equation}
\begin{aligned}
   \mu_+(\tilde{\omega}) & \propto \int_M \kappa^2 \Omega^\sharp\lrcorner \dd\tilde{\omega} 
       \propto \int_M \zeta \wedge \Omega \wedge \dd \tilde{\omega} 
       \propto \int_M \dd(\zeta\wedge\Omega) \wedge \tilde{\omega} = 0 , \\
   \mu_3(\tilde{\sigma}) & \propto \int_M \kappa^2
        (\Omega^\sharp\wedge\bar{\Omega}^\sharp) \lrcorner \dd\tilde{\sigma}
       \propto \int_M \zeta \wedge \dd \tilde{\sigma}
       \propto \int_M \dd \zeta \wedge \tilde{\sigma} = 0 ,
\end{aligned}
\end{equation}
which give $\dd \zeta=0$ and $\zeta\wedge\dd\Omega=0$ as conditions, and where we have used $\kappa^2\Omega^\sharp\propto \zeta\wedge\Omega$ and $\kappa^2\Omega^\sharp\wedge\bar{\Omega}^\sharp\propto \zeta$. In the intrinsic torsion language of \cite{BCL06}, this fixes the components $\{\mathcal{W}_1,\mathcal{W}_2,\mathcal{W}_5\}$ and $\{R,V_1,T_1,\mathcal{W}_0\}$ to zero. The vector contribution is 
\begin{equation}
\begin{split}
   \mu_{3}(\tilde{v}) & \propto \int_{M} 
       \kappa\bar{\Omega}^\sharp\lrcorner\mathcal{L}_{\tilde{v}}(\kappa\Omega)
       + \mathcal{L}_{\tilde{v}}(\kappa\Omega^\sharp)\lrcorner\kappa\bar{\Omega} \\
       & \propto \int_{M} 
       \zeta\wedge\bar{\Omega} \wedge \mathcal{L}_{\tilde{v}}\Omega
       + \zeta\wedge\Omega \wedge \mathcal{L}_{\tilde{v}}\bar{\Omega} \\
       & \propto \int_{M} 
       \imath_{\tilde{v}}\zeta \, \dd\Omega\wedge\bar{\Omega}
       = 0 ,
\end{split}
\end{equation}
where we have used $\int\mathcal{L}_{\tilde{v}}\kappa^2=0$ and the previous conditions to reach the final line. This fixes the torsion class $E$ to zero.

Turning to the conditions on $X$ given by~\eqref{eq:T-MCY}, upon using the algebraic relations we find~\eqref{eq:e7_comp_int} simplifies to $\dd\omega\wedge\omega=0$, which requires the torsion classes $\{\mathcal{W}_4,E+\bar{E},V_2,T_2\}$ to vanish. Notice that this is weaker than requiring an integrable $\Symp{6;\bbR}$ structure. One can also explicitly check that \eqref{eq:e7_comp_int1} and \eqref{eq:e7_comp_int} constrain the same torsion classes, and that $\Dorf_X X=0$ vanishes identically.

Finally, we have the additional condition that ensures the \HV{} structure is integrable~\eqref{eq:e7_intersection_int}. Upon imposing the previous conditions, this forces the remaining torsion classes to vanish. Taken together, we find $\zeta$, $\omega$ and $\Omega$ are closed, and that $\zeta^\sharp$ is a Killing vector
\begin{equation}
   \mathcal{L}_{\zeta^\sharp}\omega=0 , \qqqq
   \mathcal{L}_{\zeta^\sharp}\Omega= 0 .
\end{equation}

\subsubsection{Generalised Calabi--Yau metrics in type II}

Throughout we will use the expressions given in appendix~B of~\cite{GLSW09}, generalised to describe both type IIA and IIB. The generalised vector decomposes as $V=v+\Lambda+\tilde{\Lambda}+\tau+\Lambda^\pm$ where $v$ is a vector, $\Lambda$ a one-form, $\tilde{\Lambda}$ a five-form, $\tau$ is a one-form density and $\Lambda^\pm$ are sums of even or odd forms. From the $\ee^{B+\tilde{B}+C^\pm}$ action we conclude that  in the splitting~\eqref{eq:Dorf-split} we have
\begin{equation}
   R = \dd\Lambda + (\dd\tilde{\Lambda})_{1\dots 6}v^iv_j 
        + v^i\dd\Lambda^\pm ,  
\end{equation}
where $v^i=(1,0)$, $\dd\Lambda$ acts as a ``$B$-transform'', and the upper sign refers to type IIA and the lower to type IIB. Thus in the moment maps for $J_\alpha$ given in~\eqref{eq:J-GCY}, we have the non-zero contributions, using the trace formula given in section~3.1 of~\cite{GLSW09} and $u^iv_i=\kappa^{-1}$,
\begin{equation}
   \mu_+(\Lambda^\pm) 
      \propto \int_M \langle\dd\Lambda^{\pm},\Phi^{\mp}\rangle
      \propto \int_M \langle\Lambda^{\pm},\dd\Phi^{\mp}\rangle
      = 0 ,
\end{equation}
and 
\begin{equation}
\begin{aligned}
   \mu_3(\Lambda) &\propto \int_M
       \langle\Phi^\mp,\dd\Lambda\wedge\Phi^{\mp}\rangle
       \propto \int_M \langle\dd\Phi^\mp,\Lambda\wedge\Phi^{\mp}\rangle 
           + \langle\Phi^\mp,\Lambda\wedge\dd\Phi^{\mp}\rangle = 0 , \\ 
   \mu_3(\tilde{\Lambda}) &\propto \int_M \dd\tilde{\Lambda} = 0 ,
\end{aligned}
\end{equation}
where in the first line we have used the expression~\eqref{eq:GCS-J-def} for $\mathcal{J}^{\pm A}_{\phantom{\mp A}B}$. From these we recover $\dd\Phi^\mp=0$.  For the vector component we have 
\begin{equation}
   \mu_3(v) \propto \int_M \epsilon_{ij} 
        \langle\bar{u}^i\bar{\Phi}^\mp,\mathcal{L}_v(u^i\Phi^{\mp})\rangle
        \propto \int_M \langle\bar{\Phi}^\mp,\mathcal{L}_v\Phi^{\mp}\rangle
        = 0 ,
\end{equation}
where we have used the identity $\epsilon_{ij}\bar{u}^i\mathcal{L}_vu^j=0$. Using $\mathcal{L}_v\Phi^\mp=\imath_v\dd\Phi^\mp+\dd\imath_v\Phi^\mp$ and integration by parts, we see that this indeed vanishes. 

For the conditions involving $X$ given by~\eqref{eq:T-GCY}, using~\eqref{eq:Dorf-split} we have
\begin{equation}
   \Dorf_{X} \alpha
        \propto (v^i \dd\Phi^\pm) \cdot \alpha = 0 , 
\end{equation}
where $\alpha$ is any generalised tensor, $\cdot$ is the relevant adjoint action and we have imposed $\dd\Phi^\pm=0$. Hence~\eqref{eq:e7_comp_int} and~\eqref{eq:e7_intersection_int} are both satisfied. 

\subsubsection{D3-branes on \texorpdfstring{$\text{HK} \times \bbR^2$}{HK x R2} in type IIB} 

We have a non-trivial five-form flux $F$ in this case, so it is convenient to use the untwisted structures and twisted generalised Lie derivative. Focussing on $\tilde{J}_\alpha$ given in~\eqref{eq:J-D3}, from~\eqref{eq:Dorf-split} and~\eqref{eq:tildeR}, the only non-zero form-field contribution to the moment maps is 
\begin{equation}
   \mu_\alpha(\tilde{\rho}) \propto \int_M \kappa^2 
       (\omega_\alpha^\sharp\wedge \zeta_1^\sharp\wedge \zeta_2^\sharp)
            \lrcorner\dd\tilde{\rho}
       \propto \int_M \ee^{2\Delta} \omega_\alpha \wedge \dd\tilde{\rho}
       \propto \int_M \dd \big(\ee^{2\Delta} \omega_\alpha\bigr)
           \wedge \tilde{\rho} = 0 . 
\end{equation}
We recover $\dd (\ee^{2\Delta}\omega_\alpha)=0$. The $\tilde{v}$ condition is considerably more complicated and involves the five-form flux $F$ through the term $\imath_{\tilde{v}}F$ in~\eqref{eq:tildeR}. After some manipulation, using in particular that $\epsilon_{\alpha\beta\gamma}\tr(I_\beta\mathcal{L}_vI_\gamma)= -\epsilon_{\alpha\beta\gamma}(\omega_\beta^\sharp\lrcorner
\mathcal{L}_v\omega_\gamma)$,  one finds 
\begin{equation}
\mu_{\alpha}(\tilde{v}) \propto \int_M \ee^{2\Delta}\omega_{\alpha}\wedge\imath_{\tilde{v}}F+2\ee^{2\Delta}\epsilon_{\alpha\beta\gamma}\dd\Delta\wedge\omega_{\beta}\wedge\imath_{\tilde{v}}\omega_{\gamma}\wedge \zeta_1 \wedge \zeta_2.
\end{equation}
This vanishes for $\dd\Delta=-\frac{1}{4}\star F$, or more precisely it fixes the components of $\dd\Delta$ that are transverse to $\zeta_{1,2}$.

For $\tilde{X}$ given in~\eqref{eq:T-D3} and using~\eqref{eq:tildeR}, acting on any untwisted generalised tensor $\tilde{\alpha}$ we have
\begin{equation}
   \hat{\Dorf}_{\tilde{X}} \tilde{\alpha} 
        = - \tilde{R}\cdot\tilde{\alpha} = 0 , 
\end{equation}
since we have
\begin{equation}
\begin{aligned}
   \tilde{R} = \bar{n}^i \dd\bigl(\ee^{\Delta}(\zeta_{1} - \ii \zeta_{2})\bigr)
       + \ii \bar{n}^i \dd \bigl(\ee^{\Delta}(\zeta_{1} - \ii \zeta_{2})\wedge\vol_{4}\bigr)
       + \bar{n}^i \ee^{\Delta}(\zeta_{1} - \ii \zeta_{2})\wedge F = 0 .
\end{aligned}
\end{equation}
We have used $\dd(\ee^\Delta \zeta_i)=0$ and $\dd(\ee^{4\Delta}\vol_4)=0$ so that the last two terms simplify to 
\begin{equation}
4 \ii \, \dd\Delta\wedge(\zeta_1 - \ii \zeta_2)\wedge \vol_4 = (\zeta_1 - \ii \zeta_2) \wedge F ,
\end{equation}
which vanishes for $\dd\Delta=-\frac{1}{4}\star F$, or more precisely it fixes the components of $\dd\Delta$ that are in the direction of $\zeta_{1,2}$. Hence the conditions~\eqref{eq:e7_comp_int} and~\eqref{eq:e7_intersection_int} are both satisfied. 

We also note that it is simple to extend our description to include imaginary self-dual three-form flux, as first considered in~\cite{GP01,Gubser00,GP02} and analysed in detail in the case of hyper-K\"ahler manifolds times $\mathbb{R}^2$ in~\cite{BCFFST01}. The metric, five-form flux and axion-dilaton are of the same form as for our example, but the warp factor is no longer harmonic and there is a non-zero three-form flux on $M$ 
\begin{equation}
F^1+\ii F^2 = \dd \gamma_I(z)\wedge\tau_I,
\end{equation}
where $\gamma_I(z)$ are analytic functions of $z=x+\ii y$, and $\tau_I$ are harmonic anti-self-dual two-forms on the hyper-K\"ahler space. The moment maps are altered only in the $\tilde\rho$ component, thanks to the $-\epsilon_{ij}\tilde{\lambda}^i \wedge F^j$ contribution to $\tilde{R}$ in the presence of three-form flux, giving a term 
\begin{equation}
\int_M(\tilde{\lambda}^{1}\wedge\omega_\alpha\wedge F^2-\tilde{\lambda}^{2}\wedge\omega_\alpha\wedge F^1),
\end{equation}
which vanishes as the wedge product of a self-dual two-form $\omega_\alpha$ with an anti-self-dual two-form $\tau_I$ is zero. The $\hat{\Dorf}_{\tilde{X}}$ expression is also altered thanks to the same correction, giving an extra term
\begin{equation}
\epsilon_{ij}\bar{n}^i (\zeta_1 - \ii \zeta_2) \wedge F^j = - (\zeta_1 - \ii \zeta_2) \wedge (F^1 - \ii F^2).
\end{equation}
But this also vanishes as $F^1+\ii F^2 = \gamma'_I(z) \dd z\wedge\tau_I$, and $\dd z = \ee^{-\Delta}(\zeta_1+\ii \zeta_2)$. Hence we still have $\hat{\Dorf}_{\tilde{X}}\tilde{\alpha}=0$ for any tensor $\tilde{\alpha}$.

\subsubsection{Wrapped M5-branes on \texorpdfstring{$\text{HK} \times \bbR^3$}{HK x R2} in M-theory}

In both cases we have a non-trivial four-form flux $F$, and so it is convenient to use untwisted structures and the twisted generalised Lie derivative.

We first consider M5-branes wrapping a K\"{a}hler two-cycle in the hyper-K\"ahler. Using the form of $\tilde{J}_{\alpha}$ given in (\ref{eq:J-M5K}), together with (\ref{eq:Dorf-split}) and (\ref{eq:tildeR}), the contribution to the moment maps from $\tilde{\sigma}$ is
\begin{equation}
\mu_{\alpha}(\tilde{\sigma})\propto\int_{M}\kappa^{2}\epsilon_{\alpha\beta\gamma}(\vol_{4}^{\sharp}\wedge \zeta_{\beta}^{\sharp}\wedge \zeta_{\gamma}^{\sharp})\lrcorner\dd\tilde{\sigma}\propto\int_{M}\ee^{2\Delta}\zeta_{\alpha}\wedge\dd\tilde{\sigma}\propto\int_{M}\dd(\ee^{2\Delta}\zeta_{\alpha})\wedge\tilde{\sigma}.
\end{equation}
We recover $\dd(\ee^{2\Delta}\zeta_{i})=0$ for $i=1,2,3$. The terms in the moment maps due to $\tilde{\omega}$ are
\begin{equation}
\begin{split}\mu_{\alpha}(\tilde{\omega}) & \propto\int_{M}\tfrac{1}{2}\epsilon_{\alpha\beta\gamma}\kappa^{2}(\vol_{4}^{\sharp}\wedge \zeta_{\beta}^{\sharp}\wedge \zeta_{\gamma}^{\sharp})\lrcorner(\tilde{\omega}\wedge F)-\kappa^{2}(\omega_{3}^{\sharp}\wedge \zeta_{\alpha}^{\sharp})\lrcorner\dd\tilde{\omega}\\
 & \propto\int_{M}\ee^{2\Delta}\zeta_{\alpha}\wedge F\wedge\tilde{\omega}+\tfrac{1}{2}\epsilon_{\alpha\beta\gamma}\dd(\ee^{2\Delta}\omega_{3}\wedge \zeta_{\beta}\wedge \zeta_{\gamma})\wedge\tilde{\omega}.
\end{split}
\end{equation}
This vanishes upon using the expressions for the flux $F=\ee^{-4\Delta}\star\dd(\ee^{4\Delta}\omega_{3})$ and the exterior derivatives of the $\zeta_{i}$. Again, the $\tilde{v}$ condition is more complicated and involves the four-form flux $F$ through the term $\imath_{\tilde{v}}F$ in (\ref{eq:tildeR}). After some manipulation, one finds
\begin{equation}
\begin{split}
\mu_{\alpha}(\tilde{v})&\propto\int_M 12\dd\Delta\wedge\vol_{4}\wedge \zeta_{\alpha}\wedge(\zeta_{1}\wedge\imath_{\tilde{v}}\zeta_{1}+\zeta_{2}\wedge\imath_{\tilde{v}}\zeta_{2}+\zeta_{3}\wedge\imath_{\tilde{v}}\zeta_{3}) \\
&\phantom{{}\propto{}\int_M}+\epsilon_{\alpha\beta\gamma}\omega_{3}\wedge \zeta_{\beta}\wedge \zeta_{\gamma}\wedge\imath_{\tilde{v}}F
\end{split}
\end{equation}
Again, this vanishes after imposing the conditions from (\ref{K2_torsion}).

Now consider the conditions that depend on $\tilde{X}$. For $\tilde{X}$ given in (\ref{eq:T-M5K}), acting on any untwisted generalised tensor $\tilde{\alpha}$ we have
\begin{equation}
\Dorf_{\tilde{X}}\tilde{\alpha}=-\tilde{R}\cdot\tilde{\alpha},
\end{equation}
where $\tilde{R}$ is given by
\begin{equation}
\tilde{R}=\dd(\ee^{\Delta}\Omega)+\dd(\ii\ee^{\Delta}\Omega\wedge\vol_{3})+\ee^{\Delta}\Omega\wedge F.
\end{equation}
But $\tilde{R}$ itself vanishes as
\begin{equation}
\begin{aligned}\dd(\ee^{\Delta}\Omega) & =0,\\
\dd(\ii\ee^{\Delta}\Omega\wedge\vol_{3})+\ee^{\Delta}\Omega\wedge F & =0,
\end{aligned}
\end{equation}
where we have used the expressions for the flux $F$ and the torsion conditions on $\omega_{\alpha}$ and $\zeta_{i}$ from (\ref{K2_torsion}). Hence, both (\ref{eq:e7_comp_int}) and (\ref{eq:e7_intersection_int}) are satisfied.

Next we consider M5-branes wrapping an $\mathbb{R}^{2}$ plane in $\mathbb{R}^{3}$. Using the form of $\tilde{J}_{\alpha}$ given in (\ref{eq:J-M5R2}), together with (\ref{eq:Dorf-split}) and (\ref{eq:tildeR}), the non-zero form-field contribution to the moment maps is
\begin{equation}
\mu_{\alpha}(\tilde{\omega})\propto\int_{M}\kappa^{2}(\omega_{\alpha}^{\sharp}\wedge \zeta_{3}^{\sharp})\lrcorner\dd\tilde{\omega}\propto\int_{M}\ee^{2\Delta}\omega_{\alpha}\wedge \zeta_{1}\wedge \zeta_{2}\wedge\dd\tilde{\omega}\propto\int_{M}\dd(\ee^{2\Delta}\omega_{\alpha}\wedge \zeta_{1}\wedge \zeta_{2})\wedge\tilde{\omega}.
\end{equation}
This vanishes after using the expressions in \eqref{R2_torsion}. Again, the $\tilde{v}$ condition is more complicated and involves the four-form flux $F$ through the term $\imath_{\tilde{v}}F$ in (\ref{eq:tildeR}). After some manipulation, one finds
\begin{equation}
\mu_{\alpha}(V)=\int12\epsilon_{\alpha\beta\gamma}\ee^{2\Delta}\dd\Delta\wedge\omega_{\beta}\wedge\vol_{3}\wedge\imath_{\tilde{v}}\omega_{\gamma}-4\ee^{2\Delta}\omega_{\alpha}\wedge \zeta_{1}\wedge \zeta_{2}\wedge\imath_{\tilde{v}}F.
\end{equation}
This vanishes for $\star F=\ee^{-4\Delta}\dd(\ee^{4\Delta}\zeta_{1}\wedge \zeta_{2}),$ or more precisely it fixes the components of $\dd\Delta$ that are transverse to $\zeta_{1,2,3}$.

For $\tilde{X}$ given in (\ref{eq:T-M5R2}), acting on any untwisted generalised tensor $\tilde{\alpha}$ we have
\begin{equation}
\Dorf_{\tilde{X}}\tilde{\alpha}=\mathcal{L}_{\ee^{\Delta}(\zeta_{1}^{\sharp}+\ii \zeta_{2}^{\sharp})}\tilde{\alpha}-\tilde{R}\cdot\tilde{\alpha},
\end{equation}
where $\tilde{R}$ is given by
\begin{equation}
\begin{split}
\tilde{R}&=\dd\bigl[\ee^{\Delta}(\zeta_{1}+\ii \zeta_{2})\wedge \zeta_{3}\bigr]-\ee^{\Delta}(\zeta_{1}^{\sharp}+\ii \zeta_{2}^{\sharp})\lrcorner F-\dd\bigr[\ee^{\Delta}(\zeta_{1}+\ii \zeta_{2})\wedge\vol_{4}\bigr]\\
&\eqspace+\ee^{\Delta}(\zeta_{1}+\ii \zeta_{2})\wedge \zeta_{3}\wedge F.
\end{split}
\end{equation}
But $\tilde{R}$ vanishes as
\begin{equation}
\begin{aligned}\dd(\ee^{\Delta}\zeta_{1}\wedge \zeta_{3}) & =0,\\
\zeta_{1}^{\sharp}\lrcorner F & =0,\\
\dd(\ee^{\Delta}\zeta_{1}\wedge\vol_{4})-\ee^{\Delta}\zeta_{1}\wedge \zeta_{3}\wedge F & =0,
\end{aligned}
\end{equation}
with similar expressions for $\zeta_{2}$. The generalised Lie derivative along $\tilde{X}$ then reduces to the Lie derivative along $\ee^{\Delta}(\zeta_{1}^{\sharp}+\ii \zeta_{2}^{\sharp})$, and we note that $\Delta$ does not depend on the coordinates $x$ or $y$, so that $\zeta_{1,2}^{\sharp}\lrcorner\dd\Delta=0$. It is then simple to check that the Lie derivative along $\ee^{\Delta}(\zeta_{1}^{\sharp}+\ii \zeta_{2}^{\sharp})$ preserves both $\bar{\tilde{X}}$ and $J_{\alpha}$, and so both (\ref{eq:e7_comp_int}) and (\ref{eq:e7_intersection_int}) are satisfied.


\section{Generalised intrinsic torsion, supersymmetry and moduli spaces}
\label{sec:int-tor}

In this section, we analyse the integrability conditions for the hyper- and vector-multiplet structures using the notion of generalised intrinsic torsion, first introduced in generality in~\cite{CSW14b} and for a specific heterotic extension of $\Orth{d,d}\times\bbR^+$ generalised geometry in~\cite{CMTW14}. This will allow us to do two things: first to show that each condition is equivalent to the existence of a torsion-free generalised connection compatible with the relevant structure, and second to prove, using the results of~\cite{CSW14b}, that the full set of conditions defining an integrable \HV{} structure are equivalent to solving the $\mathcal{N}=2$ Killing spinor equations. 

We then show that the integrability conditions have a simple interpretation in terms of rewriting the full ten- or eleven-dimensional supergravity theory in terms of an $\mathcal{N}=2$, $D=4$ gauged supergravity coupled to an infinite number of hyper- and vector-multiplets, as considered in~\cite{GLW06,GLW07,GLSW09}. Finally we discuss some general aspects of the moduli spaces of structures. 


\subsection{Generalised intrinsic torsion and integrability}

We start by recalling the definition of generalised intrinsic torsion given in~\cite{CSW14b}. Let $\GS{G}\subset\tilde{F}$ be a principal sub-bundle of the generalised frame bundle $\tilde{F}$ defining a generalised $G$-structure. It is always possible to find a generalised connection $\hat{D}$ that is compatible with $\GS{G}$, however in general it will not be torsion-free.  Recall that the generalised torsion $T$ of $\hat{D}$ is defined, given any generalised tensor $\alpha$ and generalised vector $V\in\Gamma(E)$, by~\cite{CSW11}
\begin{equation}
\label{eq:torsion-def}
   T(V)\cdot\alpha = \Dorf^{\hat{D}}_V \alpha - \Dorf_V \alpha,
\end{equation}
where the torsion is viewed as a map $T:E\to\ad\tilde{F}$ and $T(V)$ acts in the adjoint representation on $\alpha$. Here $\Dorf^{\hat{D}}_V$ is the generalised Lie derivative with the partial derivative replaced with the covariant derivative $\hat{D}$, that is, acting on any generalised tensor $\alpha$,
\begin{equation}
\label{eq:dorfD-def}
   \Dorf^{\hat{D}}_V \alpha 
      = (V\cdot\hat{D}) \alpha - (\hat{D}\oadj V)\cdot \alpha .
\end{equation}
Let $W\subset E^*\otimes\ad\tilde{F}$ be the space of generalised torsions. For $\Ex{7(7)}\times\bbR^+$ generalised geometry, we have~\cite{CSW11}
\begin{equation}
   W \simeq E^* \oplus K,
\end{equation}
where the dual generalised tangent bundle $E^*$ transforms as $\rep{56}_{-\rep{1}}$ and $K$ is the generalised tensor bundle corresponding to the $\rep{912}_{-\rep{1}}$ representation. For other $\Ex{d(d)}\times\bbR^+$ groups the representations appearing in $W$ are listed in~\cite{CSW11}. 

By definition, any other generalised connection $\hat{D}'$ compatible with $\tilde{P}_G$ can be written as $\hat{D}'=\hat{D}+\Sigma$, where 
\begin{equation}
   \Sigma = \hat{D} - \hat{D}' \in \Gamma(K_G) , \qquad
   \text{with } K_G=E^*\otimes \ad{\tilde{P}_G} .
\end{equation}
We then define a map $\tau:K_G\to W$ as the difference of the torsions of $\hat{D}$ and $\hat{D}'$, 
\begin{equation}
   \tau(\Sigma) = T - T' \in \Gamma(W) .
\end{equation}
In general, the map $\tau$ will not fill out the whole of $W$. Defining the image 
\begin{equation}
   W_G = \image \tau \subseteq W ,
\end{equation}
we can then define the space of the generalised intrinsic torsion, in exact analogy to ordinary geometry, as the part of $W$ not spanned by $W_G$, that is 
\begin{equation}
   \Tint^G = W / W_G.
\end{equation}
Given any $G$-compatible connection $\hat{D}$, we say that the generalised intrinsic torsion $T_{\text{int}}^G$, of the generalised $G$-structure $\tilde{P}_G$, is the projection of the torsion $T$ onto $\Tint^G$. By definition this is independent of the choice of $\hat{D}$. It is the part of the torsion that cannot be changed by varying our choice of compatible connection. 

The intrinsic torsion $T_\text{int}^G$ is the obstruction to finding a connection which is simultaneously torsion-free and compatible with the generalised $G$-structure. Hence, if it vanishes we say that $\tilde{P}_G$ is an \emph{integrable} or \emph{torsion-free} generalised $G$-structure. 


\subsubsection{Intrinsic torsion for hypermultiplet structures}

Let us calculate the intrinsic torsion for a $\Spinstar{12}$ structure. Decomposing $W$ under $\SU2\times\Spinstar{12}$ we have\footnote{Since calculating intrinsic torsion reduces to linear algebra at a point in the manifold, in what follows we do not distinguish between vector bundles and their representations.} 
\begin{equation}
\label{eq:W-Hstruc}
   W = \rep{56} + \rep{912}
      =  2(\rep{2},\rep{12}) + (\rep{1},\rep{32}) 
         + (\rep{3},\rep{32}) + (\rep{1},\rep{352}) + (\rep{2},\rep{220}),
\end{equation}
while for the space of $\Spinstar{12}$ connections we have
\begin{equation}
   K_{\Spinstar{12}} 
      = \bigl( (\rep{2},\rep{12}) + (\rep{1},\rep{32}) \bigr) 
         \times (\rep{1},\rep{66})
      = (\rep{2},\rep{12}) + (\rep{2},\rep{220}) 
           + (\rep{1},\rep{32}) + (\rep{1},\rep{352}). 
\end{equation}
This implies $W_{\Spinstar{12}}\subseteq(\rep{2},\rep{12})+(\rep{1},\rep{32})+(\rep{1},\rep{352})+ (\rep{2},\rep{220})$. Using the explicit form of the map $\tau$, we can show that this is actually an equality, hence
\begin{equation}
\label{eq:Tint-Hstruc}
   \Tint^{\Spinstar{12}} = (\rep{2},\rep{12}) + (\rep{3},\rep{32}).
\end{equation}

We will now show that the triplet of moment maps constrain the same representations. Since $\hat{D}$ is compatible with the $\Spinstar{12}$ structure, by definition $\hat{D}J_\alpha=0$. Using~\eqref{eq:torsion-def} and~\eqref{eq:dorfD-def}, and integrating by parts to move $\hat{D}$ from $V$ to $J_\alpha$, we have 
\begin{equation}
\label{eq:mu-ex}
\begin{split}
   \mu_\alpha(V) 
     &\propto \epsilon_{\alpha\beta\gamma}\int_M 
        \tr \Bigl(J_\beta\bigl( (V\cdot\hat{D})J_\gamma 
           - [(\hat{D}\oadj V),J_\gamma] 
           - [ T(V), J_\gamma]\bigr)\Bigr) \\
     &\propto \int_M \tr \kappa\bigl(J_\alpha T(V) \bigr)
         + \tr \kappa\bigl(J_\alpha (\hat{D}\oadj V) \bigr) \\
     &\propto \int_M \tr \kappa\bigl(
            J_\alpha \, \TTint^{\Spinstar{12}}(V) \bigr) 
         + \int_M \tfrac{1}{2} \TTint^{\Spinstar{12}}(J_\alpha\cdot V)\cdot \kappa^2, 
\end{split}
\end{equation}
where the second term in the last line comes from the torsion of $\hat{D}$ when evaluating the total derivative in the integration by parts. We have also used the fact that the expression is independent of the choice of compatible connection $\hat{D}$ and so only depends on the intrinsic torsion $\TTint^{\Spinstar{12}}$. We see that the moment maps vanish if and only if the $(\rep{3},\rep{1})$ component of $\TTint^{\Spinstar{12}}(V)$ vanish for all $V$. Recall that $V$ transforms in the $\rep{56}=(\rep{2},\rep{12})+(\rep{1},\rep{32})$ representation. Given the decomposition~\eqref{eq:Tint-Hstruc}, we see that the $(\rep{3},\rep{1})$ component of $\TTint^{\Spinstar{12}}(V)$ vanishes if and only if both the $(\rep{2},\rep{12})$ and $(\rep{3},\rep{32})$ components of the intrinsic torsion vanish. Thus the vanishing of the moment maps is equivalent to the existence of a torsion-free $\Spinstar{12}$ structure. 


\subsubsection{Intrinsic torsion for vector-multiplet structures}
 
Repeating the analysis for vector-multiplet structures by decomposing under $\Ex{6(2)}$, we have 
\begin{equation}
\label{eq:W-Vstruc}
   W = \rep{56} + \rep{912}
      =  \rep{1} + 2\cdot\rep{27} + \rep{78} + \rep{351} + \CC ,
\end{equation}
while for the space of $\Ex{6(2)}$ connections we have
\begin{equation}
   K_{\Ex{6(2)}} 
      = ( \rep{1} + \rep{27} + \CC )
           \times \rep{78} 
      =  \rep{27} + \rep{78} + \rep{351} + \rep{1728} + \CC
\end{equation}
This implies $W_{\Ex{6(2)}}\subseteq \rep{27}+\rep{78}+\rep{351}+\text{c.c}$. Using the explicit form of the map $\tau$, we can show again that this is actually an equality, hence
\begin{equation}
\label{eq:Tint-Vstruc}
   \Tint^{\Ex{6(2)}} = \rep{1} + \rep{27} + \CC
\end{equation}

We will now show that the $\Dorf_K K=0$ condition is equivalent to vanishing generalised intrinsic torsion. Using \eqref{eq:torsion-def}, \eqref{eq:dorfD-def} and $\hat{D}K=0$, we have 
\begin{equation}
\label{eq:LTT-ex}
\Dorf_K K = \Dorf^{\hat{D}}_K K - T(K)\cdot K 
   = - T_\text{int}^{\Ex{6(2)}}(K)\cdot K . 
\end{equation}
Since $K$ is a singlet under $\Ex{6(2)}$ and $\Dorf_KK$ is a generalised vector transforming in the $\rep{56}=\rep{1}+\rep{27}+\CC$ representation, this condition implies that the $\rep{1}+\rep{27}+\CC$ components of $T_\text{int}^{\Ex{6(2)}}$ vanish. However, these are precisely the components in the intrinsic torsion~\eqref{eq:Tint-Vstruc}. Thus the vanishing of $\Dorf_KK$ is equivalent to the existence of a torsion-free $\Ex{6(2)}$ structure.  


\subsubsection{Intrinsic torsion for \HV{} structures}

It was shown in~\cite{CSW14b} that solutions of the $\mathcal{N}=2$ Killing spinor equations are in one-to-one correspondence with torsion-free $\SU6$ structures. We now show that the full set of integrability conditions on compatible pairs of structures $\{J_\alpha,K\}$ are equivalent to vanishing $\SU6$ intrinsic torsion and hence to solutions of the $\mathcal{N}=2$ Killing spinor equations. 

The analysis in~\cite{CSW14b} primarily considered minimally supersymmetric backgrounds. However, it was also noted that the same simple group theory analysis can be used for $\mathcal{N}=2$ backgrounds in $\Ex{7(7)}\times\bbR^+$ generalised geometry. Explicitly we have, decomposing under $\SU2\times\SU6$, 
\begin{equation}
\begin{split}
   W = \rep{56} + \rep{912}
      &=  (\rep{1},\rep{1}) + 2(\rep{1},\rep{15}) + (\rep{1},\rep{21})
         + (\rep{1},\rep{35}) + (\rep{1},\rep{105})
         \\ & \quad
         + 3(\rep{2},\rep{6}) + (\rep{2},\rep{20}) + (\rep{2},\rep{84})
         + (\rep{3},\rep{1}) + (\rep{3},\rep{15}) + \CC
\end{split}
\end{equation}
From the analysis in~\cite{CSW14b} we have 
\begin{equation}
\label{eq:Tint-HVstruc}
\begin{split}
   \Tint^{\SU6} 
      &= (\rep{2},\rep{1}) \times (S + J) + \CC \\
      &= (\rep{1},\rep{1}) + (\rep{3},\rep{1}) + 2(\rep{2},\rep{6}) 
        + (\rep{1},\rep{15}) + (\rep{3},\rep{15}) 
        + (\rep{2},\rep{20}) + \CC ,
\end{split}
\end{equation}
where $S+J=\rep{8}+\rep{56}=(\rep{2},\rep{1}) +2(\rep{1},\rep{6}) +(\rep{2},\rep{15}) +(\rep{1},\rep{20})$ are the representations in which the Killing spinor equations transform. Note that we can also decompose the hyper- and vector-multiplet intrinsic torsions as 
\begin{equation}
\begin{aligned}
   \Tint^{\Spinstar{12}}
      &= (\rep{2},\rep{6}) + (\rep{3},\rep{1}) 
         + (\rep{3},\rep{15}) + \CC , \\
   \Tint^{\Ex{6(2)}}
      &= (\rep{1},\rep{1}) + (\rep{2},\rep{6}) 
         + (\rep{1},\rep{15}) + \CC   
\end{aligned}   
\end{equation}
Since the $(\rep{2},\rep{20})$ is missing from these decompositions, it is immediately clear that having an integrable hypermultiplet structure $J_\alpha$ and a compatible integrable vector-multiplet structure $K$ is not sufficient to imply we have an integrable $\SU6$ structure. 

As we will now see, the missing components are set to zero by the extra conditions $\Dorf_X J_\alpha=0$. As before, given an $\SU6$-compatible generalised connection, from \eqref{eq:torsion-def}, \eqref{eq:dorfD-def} and $\hat{D}K=\hat{D}J_\alpha=0$ we have 
\begin{equation}
\label{eq:LTJ-ex}
\Dorf_X J_\alpha = \Dorf^{\hat{D}}_X J_\alpha - [T(X), J_\alpha]
   = - [ \TTint^{\SU6}(X), J_\alpha ] . 
\end{equation}
Since $X$ is a singlet under $\SU6$ and $\Dorf_X J_\alpha$ transforms in the $\rep{133}$ representation, we see that $\Dorf_X J_\alpha$ indeed includes the missing $\repp{2}{20}$ component.  In appendix~\ref{sec:int-tor-su6}, we calculate which components of the intrinsic torsion appear in which of the three supersymmetry conditions~\eqref{eq:mu-ex},~\eqref{eq:LTT-ex} and~\eqref{eq:LTJ-ex}. The results are summarised in table~\ref{tab:SU6-int-tor}. 

\begin{table}
\centering 
\begin{tabular}{@{}lccccccc@{}} 
\toprule
\multicolumn{1}{l}{} & \multicolumn{7}{c}{$\Tint^{\SU6}$ component} \\
\cmidrule(l){2-8}
Integrability condition & $\repp{1}{1}$ & $\repp{3}{1}$ 
   & $\repp{2}{6}$ & $\repp{2}{6}'$ 
   & $\repp{1}{15}$ & $\repp{3}{15}$ & $\repp{2}{20}$ \tabularnewline
\midrule
$\mu_\alpha=0$ & & $\times$ & $\times$ & & & $\times$ & \tabularnewline 
$\Dorf_K K=0$ & $\times$ & & & $\times$ & $\times$ & & \tabularnewline 
$\Dorf_X J_\alpha=0$ & $\times$ & $\times$ & & $\times$ & 
   & $\times$ & $\times$ \tabularnewline
\bottomrule
\end{tabular}
\protect\caption{The components of the generalised intrinsic torsion $\Tint^{\SU6}$ appearing in each of the $\mathcal{N}=2$ supersymmetry conditions.}
\label{tab:SU6-int-tor}
\end{table}

We see that collectively the three integrability conditions on $\{J_\alpha,K\}$ are equivalent to solving the $\mathcal{N}=2$ Killing spinor equations. Since an $\SU6$-compatible connection is a special case of both a $\Spinstar{12}$- and an $\Ex{6(2)}$-compatible connection, this decomposition also provides a direct proof that there are indeed no unexpected kernels in the $\tau$ map in these two cases, and that $\mu_\alpha=0$ and $\Dorf_K K$ are equivalent to the existence of a torsion-free $\Spinstar{12}$ and $\Ex{6(2)}$ generalised structure respectively. 

We also see that certain components of $\Tint^{\SU6}$ appear in multiple conditions. The $\mu_\alpha(V)$ and $\Dorf_K K$ conditions are complementary. However the $\Dorf_X J_\alpha$ condition shares components with each of the other conditions. The relation between $\repp{1}{1}$ components comes from taking $\Dorf_X$ of the second compatibility condition in~\eqref{eq:e7_comp_condition} and using $\Dorf_X X=0$
\begin{equation}
   \tr(J_\alpha\Dorf_X J_\beta+J_\beta\Dorf_X J_\alpha)
      = - \tfrac{1}{2} \ii s(X, \Dorf_X \bar{X}) \, \delta_{\alpha\beta}  ,
\end{equation}
while the relation between $\repp{2}{6}'$ components comes from taking $\Dorf_X$ of the first condition in~\eqref{eq:e7_comp_condition}
\begin{equation}
   (\Dorf_X J_\alpha) \cdot K  + J_\alpha\cdot \Dorf_X K = 0 . 
\end{equation}
The relation between the $\repp{3}{1}$ and $\repp{3}{15}$ components arises from evaluating the moment maps on $X$
\begin{equation}
   \mu_\alpha(X) = - \tfrac{1}{2}\epsilon_{\alpha\beta\gamma}
      \int_M \tr \bigl(J_\beta(\Dorf_X J_\gamma)\bigr) . 
\end{equation}

Let us end this section by briefly noting how the conventional $\SU3$ intrinsic torsion, which vanishes for type II Calabi--Yau backgrounds, embeds into the generalised case. The combined $\SU8$ spinor~\eqref{eq:IIspinor} defines two different embeddings of $\Spin6\simeq\SU4_\pm\subset\SU8$: one for type IIA and one for type IIB, corresponding to the decompositions $\rep{8}=\rep{4}+\overline{\rep{4}}$ and $\rep{8}=\rep{4}+\rep{4}$ respectively. There are hence two different embeddings of $\SU3_\pm\subset\SU6$, giving the embeddings of the torsion classes defined in~\cite{CS02} for type IIA 
\begin{equation}
\begin{aligned}
   \mathcal{W}_1 &: \rep{1}_\bbC \subset \repp{3}{1} , & 
   \mathcal{W}_2 &: \rep{8}_\bbC \subset \repp{3}{15} , & 
   \mathcal{W}_3 &: \rep{6} \subset \repp{2}{20} , \\
   \mathcal{W}_4 &: \rep{3} \subset \repp{2}{6}' , & 
   \mathcal{W}_5 &: \rep{3} \subset \repp{2}{6} ,
\end{aligned}
\end{equation}
and for type IIB 
\begin{equation}
\begin{aligned}
   \mathcal{W}_1 &: \rep{1}_\bbC \subset \repp{3}{1} , & 
   \mathcal{W}_2 &: \rep{8}_\bbC \subset \repp{2}{20} , & 
   \mathcal{W}_3 &: \rep{6} \subset \repp{3}{15} , \\
   \mathcal{W}_4 &: \rep{3} \subset \repp{2}{6} , & 
   \mathcal{W}_5 &: \rep{3} \subset \repp{2}{6}' ,
\end{aligned}
\end{equation}
which in each case is consistent with the analysis of section~\ref{sec:int_examples}. 


\subsection{Supersymmetry conditions from gauged supergravity}

As we have already noted, there is a natural physical interpretation of the spaces of hypermultiplet and vector-multiplet structures. We can view them as arising from a rewriting of the full ten- or eleven-dimensional theory as in~\cite{DN86} but with only eight supercharges manifest~\cite{GLW06,GLW07,GLSW09}. The local $\SO{9,1}$ Lorentz symmetry is broken and the degrees of freedom can be repackaged into $\mathcal{N}=2$, $D=4$ multiplets. However, since all modes are kept -- there is no Kaluza--Klein truncation -- the vector- and hypermultiplet spaces $\MV$ and $\MH$ become infinite dimensional.  As previously argued for  $\mathcal{N}=1$ backgrounds in $\Orth{6,6}$ generalised geometry in~\cite{KM07} and in $\Ex{7(7)}$ generalised geometry in~\cite{PW08,GLSW09,GO11}, the integrability conditions can be similarly interpreted in a four-dimensional language. The interactions of the four-dimensional theory are encoded in the gauging of isometries on $\MH$ and $\MV$, together with the concomitant moment maps, as summarised in~\cite{ABCDFFM97}. From the form~\eqref{eq:mm} of the hyper-K\"ahler moment maps, we see that we are gauging generalised diffeomorphisms. The general conditions, coming from the vanishing of the gaugino, hyperino and gravitino variations, for the four-dimensional theory to admit a supersymmetric $\mathcal{N}=2$ Minkowski vacuum have been analysed in~\cite{HLV09,LST12}. As we now show, these translate directly into the three integrability conditions for $J_\alpha$ and $K$.

Recall that the scalar components of the hypermultiplets describe a quaternionic-K\"ahler space. Let $\MH$ be the associated hyper-K\"ahler cone. Similarly, the scalar components of the vector multiplets describe a local special K\"ahler space. Let $\MV$ be the associated rigid special K\"ahler cone. The gauging is a product of an action of a group $\mathcal{G}_\text{H}$ on the quaternionic-Kähler space and of a group $\mathcal{G}_\text{V}$ on the local special K\"ahler space, that each preserve the corresponding structures. These can always be lifted to an action on $\MH$ that preserves the triplet of symplectic forms and commutes with the $\SU2$ action on the cone, and an action on $\MV$ that preserves the K\"ahler form and complex structure and commutes with the $\Uni1$ action on the cone. Following~\cite{LST12}, the conditions for a Minkowski vacuum in a generic gauged $\mathcal{N}=2$ theory, lifted to $\MH$ and $\MV$, can be written as 
\begin{equation}
\label{eq:sugra-conds}
   \Theta_\Lambda^\lambda \mu_{\alpha,\lambda} = 0 , \qqq
   X^\Lambda\Theta_\Lambda^{\lambda} k_\lambda^u = 0 , \qqq
   \bar{X}^\Lambda\hat{\Theta}_\Lambda^{\hat{\lambda}} \hat{k}_{\hat{\lambda}}^i = 0 . 
\end{equation}
Here $\lambda$ parametrises the Lie algebra $\mathfrak{g}_\text{H}$ of $\mathcal{G}_\text{H}$ while $\hat{\lambda}$ parametrises the Lie algebra $\mathfrak{g}_\text{V}$ of $\mathcal{G}_\text{V}$, and $k_\lambda$ and $\hat{k}_{\hat{\lambda}}$ are the corresponding sets of vector fields generating the actions on $\MH$ and $\MV$ (see also appendix~\ref{sec:SK}). The label $u$ is a coordinate index on $\MH$ and $i$ is a holomorphic coordinate index on $\MV$, so that $\hat{k}_{\hat{\lambda}}$ is actually the holomorphic part of the real vector generating the action. The $\mu_{\alpha,\lambda}$ are a triplet of moment maps $\mu_\alpha:\MH\to\mathfrak{g}_\text{H}^*$. As discussed in appendix~\ref{sec:SK}, the complex vector $X^\Lambda$ is a particular non-zero holomorphic vector on $\MV$, written in flat coordinates, that defines the special K\"ahler geometry and also generates the $\bbC^*$ action on the cone. The indices $\Lambda$ denote components in the natural flat coordinates on $\MV$. The matrices $\Theta_\Lambda^\lambda$ and $\hat{\Theta}_\Lambda^{\hat{\lambda}}$ are the corresponding embedding tensors~\cite{WST05b,WM11}. 

Let us now translate this formalism into the geometrical objects defined previously when $\MH$ and $\MV$ are the infinite-dimensional spaces of hyper- and vector-multiplet structures. In this case, the gauging is by generalised diffeomorphisms $\mathcal{G}_\text{H}=\mathcal{G}_\text{V}=\GDiff$. Recall that we parametrised the Lie algebra $\gdiff$ by sections $V\in\Gamma(E)$ even though there was actually a kernel in this map. Furthermore, from~\eqref{eq:Y-coord}, we saw that generalised vectors defined flat coordinates on $\MV$. Thus we can identify the embedding tensors with the map
\begin{equation}
\label{eq:Theta-maps}
   \Theta = \hat{\Theta} : \Gamma(E) \to \gdiff .
\end{equation}
The vectors $k_\lambda$ and $\hat{k}_{\hat{\lambda}}$ generate the action of $\GDiff$ on $\MH$ and $\MV$, so we can view them as maps 
\begin{equation}
\label{eq:k-maps}
   k : \gdiff \to \Gamma(T\MH) , \qqqq 
   \hat{k} : \gdiff \to \Gamma(T\MV) .
\end{equation}
Hence we can identify the composite maps $k\circ \Theta$ and $\hat{k}\circ \hat{\Theta}$, acting on an arbitrary generalised vector $V$, with \begin{equation}
\begin{aligned}
   V^\Sigma \Theta_\Sigma^\lambda k_\lambda &= \Dorf_V J_\alpha , \\
   V^\Sigma \hat{\Theta}_\Sigma^{\hat{\lambda}} 
       \hat{k}_{\hat{\lambda}} &= \Dorf_V X . 
\end{aligned}
\end{equation}
From appendix~\ref{sec:SK}, note that $\hat{k}\circ\hat{\Theta}$ is just the set of generators $\XX_{\Lambda\Xi}{}^\Sigma $ acting on $X$. Thus, as first noted in~\cite{CMTW14}, in the infinite-dimensional gauging, we can identify a generic combination of generators $V^\Lambda \XX_{\Lambda\Xi}{}^\Sigma$ with the generalised Lie derivative $\Dorf_V$. Similarly we have 
\begin{equation}
   V^\Sigma \Theta_\Sigma^\lambda \mu_{\alpha,\lambda} = \mu_\alpha(V) .
\end{equation}
Finally, recall from the discussion in section~\ref{sec:V-structures} that our notation is consistent and the holomorphic vector field $X^\Lambda$ is indeed the complexified vector-multiplet structure $X=K+\ii\hat{K}$. Thus the three conditions~\eqref{eq:sugra-conds} are precisely 
\begin{equation}
   \mu_\alpha(V) = 0 \quad \text{for all $V$}, \qqqq
   \Dorf_X J_\alpha = 0 , \qqqq
   \Dorf_{\bar{X}} X = 0 .
\end{equation}

We see that the integrability conditions on the structures have a very simple interpretation in terms of the gauged supergravity. This analysis is useful when looking for integrability conditions in other situations, in particular the backgrounds in $D=5$ and $D=6$ with eight supercharges which we discuss in later sections.


\subsection{Moduli spaces}
\label{sec:moduli-spaces}

In this section, we will discuss some simple aspects of the moduli spaces of \HM{}, \VM{} and \HV{} structures. In the Calabi--Yau case, these come from deformations of the complex and symplectic structures. For example in type IIA, the \HM{}-structure moduli space describes the complex moduli together with harmonic three-form potentials $C$, while the \VM{}-structure moduli space describes the K\"ahler moduli. The main point here is that the \HM{} and \VM{} moduli spaces appear as hyper-K\"ahler and symplectic quotients respectively, and so by construction describe quaternionic and special K\"ahler geometries as required by supersymmetry. 


\subsubsection{Moduli space of hypermultiplet structures}
\label{sec:m-hyper}

We have already seen that the differential conditions~\eqref{eq:e7_mm} that define integrable \HM{} structures can be viewed as the vanishing of a triplet of moment maps for the action of the generalised diffeomorphism group $\GDiff$ on the space $\MH$. Acting on the moment maps with the vector field $\rho_W\in\Gamma(T\MH)$, corresponding to an element of $\gdiff$ labelled by $W$, we have, using integration by parts and Leibniz,
\begin{equation}
\begin{split}
   \imath_{\rho_W} \delta \mu_\alpha(V) 
      &= -\tfrac{1}{2}\epsilon_{\alpha\beta\gamma}
      \int_{M}\tr \bigl[
      (\Dorf_WJ_{\beta})(\Dorf_{V}J_{\gamma}) 
      + J_{\beta}\Dorf_V(\Dorf_WJ_{\gamma}) \bigr] \\
   &= -\tfrac{1}{2}\epsilon_{\alpha\beta\gamma}
      \int_{M}\tr \bigl(J_{\beta} (\Dorf_{\Dorf_VW}J_{\gamma})\bigr) \\
   &= \mu_\alpha(\Dorf_V W) ,
\end{split}
\end{equation}
where we have used~\eqref{eq:int-dorf} and the Leibniz property. However the Lie bracket on $\gdiff$ is 
\begin{equation}
	\label{eq:dorf-lie}
	[ \Dorf_V, \Dorf_W ] = \Dorf_{\Dorf_V W} = \Dorf_{\Bgen{V}{W}},
\end{equation}
where $\Bgen{V}{W}$ is the antisymmetric Courant bracket for $\Ex{7(7)}\times\bbR^+$~\cite{PW08,CSW11}. Thus we see that the moment maps~\eqref{eq:mm} are equivariant. Since any two structures that are related by a generalised diffeomorphism -- a combination of diffeomorphism and gauge transformation --- are physically  equivalent, the moduli space of integrable structures is naturally a hyper-K\"ahler quotient, defined as
\begin{equation}
\ModH = \MH /\!\!/\!\!/ \GDiff = 
\mu_1^{-1}(0)\cap\mu_2^{-1}(0)\cap\mu_3^{-1}(0) / \GDiff.
\end{equation}
By construction $\ModH$ is also hyper-K\"ahler.

The space of structures $\MH$ is actually a hyper-K\"ahler cone, and for the quotient space to also be a hyper-K\"ahler cone one needs to take the zero level set\footnote{More generally, one requires that the level set is invariant under the $\SU2$ action.} of the moment maps, as we do, and ensure that the $\GDiff$ action commutes with the $\SU2$ action on the cone. We can check that this is indeed that case. Under the $\SU2$ action we have $\delta J_\alpha=\epsilon_{\alpha\beta\gamma}\theta_\beta J_\gamma$, or, in other words, the action is generated by a triplet of vectors $\xi^\alpha\in \Gamma(T\MH)$ such that 
\begin{equation}
   \xi^\alpha(J_\beta) = \epsilon_{\alpha\beta\gamma}J_\gamma . 
\end{equation}
Acting on the section-valued functions $J_\alpha$, we see the Lie bracket is given by 
\begin{equation}
   \bigl[\rho_V,\xi^\alpha\bigr](J_\beta)
      = \Dorf_V(\epsilon_{\alpha\beta\gamma}J_\gamma)
           - \epsilon_{\alpha\beta\gamma}\Dorf_V J_\gamma
      = 0 . 
\end{equation}
Hence the action of $\GDiff$ does indeed commute with the $\SU2$ action. This means that $\ModH$ is also a hyper-K\"ahler cone~\cite{Swann91}, and we identify the physical moduli space with the base of the cone $\ModH/\bbH^*$. By construction, as required by supersymmetry, this space is quaternionic-K\"ahler. 

It is worth noting that the action of $\GDiff$ on $\mu_1^{-1}(0)\cap\mu_2^{-1}(0)\cap\mu_3^{-1}(0)$ is not generally free. For example, in a type IIB Calabi--Yau background, the integrable structure $J_\alpha$, given in~\eqref{eq:J-CY}, is invariant under symplectomorphisms. Thus we expect that the moduli space $\ModH$ is not generically a manifold, but has a complicated structure as a union of hyper-K\"ahler spaces~\cite{DS97}. We could still try to calculate the dimension of $\ModH$ in a neighbourhood by considering the linear deformation away from a point $\sigma\in\MH$ corresponding to an integrable structure $J_\alpha$. The variation of the moment maps is just $\delta\mu_\alpha$, where $\delta$ is the exterior derivative on $\MH$, while the infinitesimal generalised diffeomorphisms are generated by $\Dorf_V$. We can identify $\gdiff$ with $\Gamma(E)$ and $T_\sigma\MH$ with sections of a bundle $\ad\GS{\Spinstar{12}}^\perp$ as in~\eqref{eq:TMH}. We then have the exact sequence of maps 
\begin{equation}
   \begin{tikzcd}
      \Gamma(E) \arrow{r}{\Dorf_{\bullet}J_\alpha}   
      & \Gamma\bigl(\ad\GS{\Spinstar{12}}^\perp\bigr) \arrow{r}{\delta\mu_\alpha} 
      & \Gamma(E^*)\otimes\bbR^3 .
   \end{tikzcd} 
\end{equation}
Again, this is complicated by the existence of fixed points. From our examples, it appears that generically the sequence is not elliptic, and hence the moduli space is not finite-dimensional.


\subsubsection{Moduli space of vector-multiplet structures}
\label{sec:m-vector}

For the vector-multiplet structures we need to understand the constraint $\Dorf_KK=0$ on the space of structures $\MV$ and again mod out by generalised diffeomorphisms. It turns out that the integrability condition can again be interpreted as the vanishing of a moment map as we now describe. In fact, this reformulation is not specific to this infinite-dimensional set up, but applies to any flat, supersymmetric vacuum of gauged $\mathcal{N}=2$, $D=4$ supergravity, giving a new interpretation of the conditions derived in~\cite{HLV09,LST12}.

We have argued that from a gauged supergravity perspective the condition $\Dorf_KK$ arises from a gauging of the generalised diffeomorphism group on $\MV$. As discussed in appendix~\ref{sec:SK}, there are a number of requirements of the action of the gauge group on $\MV$ for it to preserve the special K\"ahler structure. First it must leave the symplectic form invariant. Let $\rho_V\in \Gamma(T\MV)$ be the vector field on $\MV$ generating the action of a generalised diffeomorphism parametrised by $V\in\Gamma(E)$. Recall that the structure $K$ can be viewed as a coordinate on $\MV$, as given in~\eqref{eq:Y-coord}, thus associating $T_K\MV\simeq\Gamma(E)$ we have 
\begin{equation}
   \rho_V = \Dorf_V K \in \Gamma(T\MV) . 
\end{equation}
Given an arbitrary vector field $W\in\Gamma(T\MV)$ we have, from~\eqref{eq:sympl}, that
\begin{equation}
   \imath_{\rho_V}\Omega(W) = \Omega(\rho_V,W)
      = \int_M s(\Dorf_VK,W) . 
\end{equation}
Using~\eqref{eq:int-dorf} and the Leibniz property of $\Dorf_V$ we have 
\begin{equation}
\begin{split}
   \imath_{\rho_V}\Omega(W) 
      &= \tfrac{1}{2}\int_M s(\Dorf_VK,W) - s(K,\Dorf_VW) \\
      &= - \tfrac{1}{2}\int_M s(W,\Dorf_VK) + s(K,\Dorf_VW) \\
      &= \imath_{w} \delta \mu(V),
\end{split} 
\end{equation}
where $\delta$ is the exterior derivative on $\MV$ and $\mu(V)$ is a the moment map 
\begin{equation}
\label{eq:mm-V}
  \mu(V)\coloneqq 
     -\tfrac{1}{2}\int_{M} s(K,\Dorf_VK) . 
\end{equation}
Thus the action of $\GDiff$ preserves the symplectic structure on $\MV$. 

Acting on the moment map by the vector field $\rho_W$, corresponding to an element of $\gdiff$ labelled by $W$, we have
\begin{equation}
\begin{split}
   \imath_{\rho_W} \delta \mu(V) 
       &= - \tfrac{1}{2}\int_M s(\Dorf_WK,\Dorf_VK) 
           + s(K,\Dorf_V(\Dorf_WK) \\
       &= - \tfrac{1}{2}\int_M s(K,\Dorf_{\Dorf_VW}K) \\
       &= \mu(\Dorf_V W) ,
\end{split}
\end{equation}
where we have used~\eqref{eq:int-dorf} and the Leibniz property. Thus from the Lie bracket~\eqref{eq:dorf-lie} on $\gdiff$ we see that the moment map~\eqref{eq:mm-V} is equivariant. We also see, using~\eqref{eq:HK} and~\eqref{eq:int-dorf}, that the Hitchin functional $H$ is invariant under the action of $\rho_V$, since 
\begin{equation}
   \mathcal{L}_{\rho_V}H = \rho_V(H) 
       = \int_M (\Dorf_VK)^M \frac{\partial}{\partial K^M}\sqrt{q(K)}
       = \int_M \Dorf_V \sqrt{q(K)} 
       = 0 . 
\end{equation}
In addition, $\rho_V=\Dorf_VK$ is clearly linear in $K$ and so maps flat coordinates to flat coordinates. This is enough to show that the $\GDiff$ action also preserves the complex structure. Finally recall that the coordinate $K^M(x)$ can also be regarded as the components of a vector field and that the $\bbC^*$ action on $\MV$ is generated by $X=K+\ii\hat{K}=K-\ii \mathcal{I}\cdot K\in\Gamma(T\MV)$, where $\mathcal{I}$ is the complex structure on $\MV$. As a vector field we have $\mathcal{L}_{\rho_V}K=[\rho_V,K]=0$ and so, since $\mathcal{L}_{\rho_V}\mathcal{I}=0$, we have $[\rho_V,X]=0$ and hence the action of $\GDiff$ on $\MV$ commutes with the $\bbC^*$ action. These means that this gauging satisfies all the conditions necessary to preserve the special K\"ahler structure. 

We now show that the condition $\Dorf_KK=0$ is actually equivalent to the vanishing of the moment map $\mu(V)$ for all $V$. To do this we first note some identities using~\eqref{eq:M_Dorf_vector},~\eqref{eq:M_symplectic} and~\eqref{eq:int-dorf}
\begin{equation}
\begin{gathered}
   \Omega(V,\Dorf_VV) = \tfrac{1}{4}\int_M 
       \mathcal{L}_v(\imath_v\tau) + \tfrac{1}{3}\dd (\omega^3)
       + \dd (\imath_v\omega\wedge\sigma) 
       - \dd (\imath_v\sigma\wedge\omega) 
       \equiv 0 , \\
   \Omega(U,\Dorf_VW) - \Omega(W,\Dorf_VU)
       = \int_M \Dorf_V s(U,W) \equiv 0 . 
\end{gathered}
\end{equation}
Under the identification of $V^\Lambda \XX_{\Lambda\Xi}{}^\Sigma$ with $\Dorf_V$, these are just the representation constraint~\eqref{eq:X-ident-rep} and the first constraint of~\eqref{eq:X-idents}. These imply
\begin{equation}
   \mu(V) = -\tfrac{1}{2}\Omega(K,\Dorf_VK)
      = \Omega(V,\Dorf_KK) = \int_M s(V,\Dorf_KK) , 
\end{equation}
and hence we have the alternative definition for the integrability of $K$: 
\begin{defn}
An \emph{integrable} or \emph{torsion-free} $\Ex{7(7)}$ vector-multiplet structure $K$ is one satisfying
\begin{equation}
\label{eq:e7_Kmm}
   \mu(V)= 0 \quad \text{for all $V\in\Gamma(E)$} , 
\end{equation}
where $\mu(V)$ is given by \eqref{eq:mm-V}. 
\end{defn}

\noindent 
This reformulation is actually generic for any gauged $\mathcal{N}=2$, $D=4$ theory as we now show. Using~\eqref{eq:k-flat} and~\eqref{eq:X-k}, we see that the third condition of~\eqref{eq:sugra-conds} can be rewritten as 
\begin{equation}
   \bar{X}^\Lambda \hat{\Theta}_\Lambda^{\hat{\lambda}}
       \hat{k}_{\hat{\lambda}}^i (\partial_i X^\Gamma) \Omega_{\Sigma\Gamma}
       = \bar{X}^\Lambda \XX_{\Lambda\Xi\Sigma} X^\Xi
       = \tfrac{1}{2}\XX_{\Lambda\Xi\Sigma} X^\Xi\bar{X}^\Sigma
       = 2 \hat{\Theta}_\Lambda^{\hat{\lambda}} \mu_{\hat{\lambda}} ,
\end{equation}
where we have used the identities~\eqref{eq:X-idents},~\eqref{eq:X-ident-rep} and~\eqref{eq:SK-mm-1}. We see that the condition $\bar{X}^\Lambda \hat{\Theta}_\Lambda^{\hat{\lambda}} \hat{k}_{\hat{\lambda}}^i=0$ is generically equivalent to the vanishing of the moment map $\mu_{\hat{\lambda}}=0$. 

This reformulation gives a simple realisation of the moduli space of vector-multiplet structures. Since any two structures related by a generalised diffeomorphism are equivalent, it is naturally given by the symplectic quotient 
\begin{equation}
   \ModV = \MV \qquotient \GDiff = 
       \mu^{-1}(0) \quotient \GDiff.
\end{equation}
By construction $\ModV$ is also special K\"ahler. In fact it is a cone over a local special K\"ahler space, as required by supersymmetry. As usual for symplectic quotients of K\"ahler spaces, we can also view $\ModV$ as a quotient by the complexified group $\ModV=\MV/\GDiff_\bbC$. As for the case of hypermultiplet structures, generically $\GDiff$ does not act freely on $\mu^{-1}(0)$ and hence $\ModV$ is not necessarily a manifold, but rather has a stratified structure~\cite{SL91}.


\subsubsection{Moduli space of \HV{} structures}

Finally we consider the moduli space of \HV{} structures. We first define the space of compatible \HV{} structures, though without the restriction on the norms. Formally, this is
\begin{equation}
   \MHV= \{ ( J_\alpha , K )\in\MH\times\MV : 
       J_\alpha\cdot K = 0 \} . 
\end{equation}
The moduli space is then given by 
\begin{equation}
   \ModHV=\{ ( J_\alpha , K )\in\MHV : \mu_\alpha = 0, \,
      \mu=0 , \, \Dorf_X J_\alpha=0 , \, \kappa^2=-2\sqrt{q(K)} \} \quotient \GDiff .
\end{equation}
The reason for dropping the norm compatibility condition from the definition of $\MHV$ is that it then has a fibred structure as we now discuss. One can imagine first choosing $K$ and then $J_\alpha$ subject to the condition $J_\alpha\cdot K=0$, or vice versa. At each point $x\in M$, we can then view the coset $\Ex{7(7)}/\SU6$ as a fibration in two different ways:
\begin{equation}
   \begin{tikzcd}
      \Spinstar{12}/\SU6\arrow{r} & \Ex{7(7)}/\SU6 \arrow{d} \\
      & \Ex{7(7)}/\Spinstar{12}
   \end{tikzcd}  \quad
   \begin{tikzcd}
      \Ex{6(2)}/\SU6 \arrow{r} & \Ex{7(7)}/\SU6 \arrow{d} \\
      & \Ex{7(7)}/\Ex{6(2)} 
   \end{tikzcd} 
\end{equation}
In both cases the fibres admit the appropriate geometry
\begin{equation}
\begin{aligned}
   P &= \bbR^+ \times \Spinstar{12}/\SU6 & && && 
      &  \text{$P$ is a special K\"ahler space} , \\
   W &= \bbR^+ \times \Ex{6(2)}/\SU6 & && && 
      & \text{$W/\bbH^*$ is a Wolf space} .
\end{aligned}
\end{equation}
Thus we can use exactly the same construction as in sections~\ref{hyper_structure} and~\ref{sec:V-structures} to define the corresponding infinite-dimensional spaces of structures as hyper-K\"ahler and special K\"ahler manifolds. If we label these $\MV^J$ for the space of \VM{} structures given a fixed \HM{} structure $J_\alpha$, and $\MH^K$ for the space of \HM{} structures given a fixed \VM{} structure $K$, the space $\MHV$ then has two different fibrations:
\begin{equation}
   \begin{tikzcd}
      \MV^J \arrow{r} & \MHV \arrow{d} \\
      & \MH
   \end{tikzcd}  \qqqq
   \begin{tikzcd}
      \MH^K \arrow{r} & \MHV \arrow{d} \\
      & \MV 
   \end{tikzcd}  
\end{equation}

Even with this fibred structure on $\MHV$, the structure of the moduli space $\ModHV$ appears to be very complicated. Nonetheless, $\mathcal{N}=2$ supergravity implies that it should become a product of the hyper- and vector-multiplet moduli spaces. Let us now comment on how this might translate into conditions on our structures. The product structure suggests that, at least locally, the moduli space of hypermultiplet structures is independent of the choice of vector-multiplet structure, and vice versa. One is tempted to conjecture that 
\begin{equation}
\label{eq:mod-conj}
   \ModHV = \ModH^K \times \ModV^J , 
\end{equation}
with $\ModH^K$ and $\ModV^J$ given by the quotients 
\begin{equation}
   \ModH^K = \MH^K \qqquotient \GDiff_K , \qqqq  
   \ModV^J = \MV^J \qquotient \GDiff_J ,
\end{equation}
where $\GDiff_K\subset\GDiff$ is the subset of generalised diffeomorphisms preserving $K$ and $\GDiff_J\subset\GDiff$ is the subset preserving $J_\alpha$. The point here is that $\MH^K$ and $\MV^J$ admit moment maps for $\GDiff_K$ and $\GDiff_J$ respectively, in complete analogy to section~\ref{sec:Integrability}. For this to work the spaces $\ModH^K$ and $\ModV^J$ must (locally) be independent of the choice of $K$ and $J_\alpha$ respectively. 

We end with a few further comments. First, using the results of~\cite{CSW14b}, the integrability conditions on $K$ and $J_\alpha$ are equivalent to the Killing spinor equations, and we identically satisfy the Bianchi identities by defining the structure in terms of the gauge potentials. We then recall that for warped backgrounds of the form~\eqref{eq:warp-metric}, the Killing spinor equations together with the Bianchi identities imply the equations of motion~\cite{GP03,GMW03,KT07,CSW14b}. Consequently, since the equations of motion on $M$ are elliptic, the moduli space $\ModHV$ must always be finite dimensional. 

The second point relates to the generalised metric $G$. Recall that this defines an $\SU8\subset\Ex{7(7)}\times\bbR^+$ structure and encodes the bosonic fields of the supergravity theory, restricted to $M$, along with the warp factor $\Delta$~\cite{Hull07,CSW11}. Since $\SU6\subset\SU8$, the \HV{} structure $\{J_\alpha,K\}$ determines the generalised metric $G$. Given a Lie subalgebra $\mathfrak{g}$ we can decompose $\ex{7(7)}\oplus\bbR=\mathfrak{g}\oplus\mathfrak{g}^\perp$. Decomposing into $\SU2\times\SU6$ representations we find 
\begin{equation}
\begin{split}
   {\spinstar{12}}^\perp &= \repp{1}{1} + \repp{2}{6} 
       + (\rep{2},\overline{\rep{6}}) + \repp{2}{20} + \repp{3}{1} ,\\
   {\ex{6(2)}}^\perp &= 2\repp{1}{1} + \repp{1}{15} 
       + (\rep{1},\overline{\rep{15}}) + \repp{2}{6} 
       + (\rep{2},\overline{\rep{6}}), \\
   {\su8}^\perp &= \repp{1}{1} + \repp{1}{15} + (\rep{1},\overline{\rep{15}})
       +  \repp{2}{20} , \\
   {\su6}^\perp &= 3\repp{1}{1} + \repp{1}{15} 
       + (\rep{1},\overline{\rep{15}}) + \repp{2}{6} 
       + (\rep{2},\overline{\rep{6}}) + \repp{2}{20} + \repp{3}{1} . 
\end{split}
\end{equation}
Thus the deformations of $\{J_\alpha,K\}$ that do not change the generalised metric $G$ are those in the $\repp{1}{1}+\repp{2}{6}+(\rep{2},\overline{\rep{6}})+\repp{3}{1}$ representations. The first and last are the $\Uni1$ and $\SU2$ symmetries acting on $K$ and $J_\alpha$ respectively. It is easy to see that the moment maps vanish only for constant rotations. The remaining $\repp{2}{6}+(\rep{2},\overline{\rep{6}})$ deformations correspond to deforming the Killing spinors for a fixed background. If such solutions exist, they imply that the background actually admits more supersymmetries than the $\mathcal{N}=2$ our formalism picks out. We also note that these deformations appear in both the deformations of $K$ and $J_\alpha$, and are related through the constraint $J_\alpha\cdot K=0$. Thus we conclude that if the background is honestly $\mathcal{N}=2$, then, up to a global $\SU2\times\Uni1$ rotation, there is a unique structure $\{J_\alpha,K\}$ for each generalised metric $G$ and, infinitesimally, we can consider independent $J_\alpha$ and $K$ deformations. This gives some credence to the conjecture that the moduli space takes the form~\eqref{eq:mod-conj}.

Finally we note that the conditions $\Dorf_X\bar{X}=\Dorf_X J_\alpha=0$ imply that 
\begin{equation}
   \Dorf_X G = 0 ,
\end{equation}
and so $K$ and $\hat{K}$ are generalised Killing vectors. This means there is a combination of diffeomorphism and gauge transformations under which all the supergravity fields are invariant. Hence, locally, one can always choose a gauge in which $\Dorf_X=\mathcal{L}_v$, where $v$ is the vector component of $X$. If the metric $g$ has no conventional Killing vectors, then $v=0$ and the integrability conditions involving $X$ are equivalent to 
\begin{equation}
\label{eq:T-Killing-cond}
   \Dorf_X (\text{anything}) = 0 ,
\end{equation}
independent of the choice of $J_\alpha$, as we saw happen explicitly in a number of our examples. In this case, an alternative approach to calculating $\ModHV$ is to solve~\eqref{eq:T-Killing-cond} and the moment map conditions on $J_\alpha$ independently, impose the compatibility conditions, $J_\alpha\cdot K=0$ and $\kappa^2=-2\sqrt{q(K)}$, and then quotient by $\GDiff$. 


\section{ \texorpdfstring{$\Ex{d(d)}$}{Ed(d)} structures for \texorpdfstring{$D=5,6$}{D=5,6} supersymmetric flux backgrounds}\label{sec:Ed_structures}

In this section we consider generic $D=5,6$ type II and M-theory flux backgrounds preserving eight supercharges. In complete analogy with the $D=4$ case, they define a pair of integrable generalised structures, though now in $\Ex{6(6)}$ for $\mathcal{N}=1$, $D=5$ backgrounds and $\Ex{5(5)}\simeq\Spin{5,5}$ for $\mathcal{N}=(1,0)$, $D=6$ backgrounds. In both cases there is a \HM{} structure naturally associated to hypermultiplet degrees of freedom. In $D=5$ there is also a \VM{} structure, though now the space of structures admits a very special real geometry rather than a special K\"ahler geometry, in line with the requirements of $\mathcal{N}=1$, $D=5$ gauged supergravity. In $D=6$ we find the second structure is naturally associated to $\mathcal{N}=(1,0)$ tensor multiplets. 

Since much of the analysis follows mutatis mutandis the $D=4$ case, we will be relatively terse in summarising the constructions. 


\subsection{\texorpdfstring{$\Ex{6(6)}$}{E6(6)} hyper- and vector-multiplet structures}
\label{sec:E6_structures}

For compactifications to $D=5$, the relevant generalised geometry~\cite{CSW11,CSW14} has an action of $\Ex{6(6)}\times\bbR^+$. The generalised tangent bundle transforms in the $\rep{27'_1}$ representation and decomposes under the relevant $\GL{d}$ group as \eqref{eq:E-Mtheory} or \eqref{eq:E-IIB}, where the one-form density terms are not present. The adjoint bundle transforms in the $\rep{1_0}+\rep{78_0}$ representation and decomposes as in \eqref{eq:ad-M-theory} or \eqref{eq:ad-IIB}, where the doublet of six-forms and six-vectors are not present for type IIB. In both type II and M-theory, the spinors transform under the $\USp8$ subgroup of $\Ex{6(6)}\times\bbR^+$. For $\mathcal{N}=1$ backgrounds in $D=5$, the single Killing spinor is stabilised by a $\USp6$ subgroup. 


\subsubsection{Structures and invariant tensors}

The $\Ex{6(6)}$ generalised $G$-structures are defined as follows. 
\begin{defn}
Let $G$ be a subgroup of $\Ex{6(6)}$. We define 
\begin{itemize}
\item a \emph{hypermultiplet structure} is a generalised structure with  $G=\SUstar{6}$
\item a \emph{vector-multiplet structure} is a generalised structure with $G=\Fx{4(4)}$
\item an \emph{\HV{} structure} is a generalised structure with $G=\USp6$
\end{itemize}
\end{defn}
\noindent
As before, an \HM{} structure is defined by a triplet of sections $J_\alpha$ of a weighted adjoint bundle, as in~\eqref{eq:J-def}, such that they transform in the $\rep{78}_{\rep{3}/\rep{2}}$ representation of $\Ex{6(6)}\times\bbR^+$ and define a highest weight $\su{2}$ subalgebra of $\ex{6(6)}$. The algebra and norms of the $J_\alpha$ are the same as for the $D=4$ case, given in \eqref{eq:su2_algebra} and \eqref{eq:J_a_norm}. A \VM{} structure is defined by a generalised vector\footnote{There are two distinct $\Ex{6(6)}$ orbits preserving $\Fx{4(4)}$, distinguished by the sign of $c(K,K,K)$~\cite{FG98}.} 
\begin{equation}
\label{eq:K-def}
   K \in \Gamma(E), \quad \text{such that} \quad c(K,K,K) \neq 0 ,
\end{equation}
where $c$ is the $\Ex{6(6)}$ cubic invariant given in~\eqref{eq:M_cubic} and~\eqref{eq:IIB_cubic}. Compatibility between the vector- and hypermultiplet structures implies that the common stabiliser group $\SUstar{6}\cap\Fx{4(4)}$ of the pair $\{J_\alpha,K\}$ is $\USp6$. The necessary and sufficient conditions are 
\begin{equation}
\label{eq:e6_comp_condition}
\begin{aligned}
   J_{\alpha}\cdot K &= 0, \\
   \tr (J_\alpha J_\beta) &= - c(K,K,K)\,\delta_{\alpha\beta} ,
\end{aligned}
\end{equation}
where $\cdot$ is the adjoint action. Note that the second condition implies $K$ is in the orbit where $c(K,K,K)>0$. They are equivalent to
\begin{equation}
\label{eq:JK-comp}
   J_+ \cdot K = 0 ,  \qquad c(K,K,K) = \kappa^2 , 
\end{equation}
respectively, where $\kappa$ is the factor appearing in~\eqref{eq:su2_algebra}. We again refer to this as an \HV{} structure since in M-theory it is the flux generalisation of a compactification on a Calabi--Yau three-fold. 

As with the $D=4$ case, the infinite-dimensional space of \HM{} structures $\MH$ is the space of smooth sections of a bundle over $M$ with fibre $W=\Ex{6(6)}\times\bbR^+/\SUstar{6}$. This fibre is a hyper-K\"ahler cone over a pseudo-Riemannian Wolf space~\cite{AC05} 
\begin{equation}
\label{eq:E6-wolf}
   W/\bbH^* = \Ex{6(6)}/(\SUstar{6}\times\SU2) . 
\end{equation}
The hyper-K\"ahler geometry on $\MH$ is again inherited directly from the hyper-K\"ahler geometry on $W$. The details follow exactly the $D=4$ case in section~\ref{hyper_structure} upon exchanging the relevant groups.

The infinite-dimensional space of \VM{} structures 
\begin{equation}
   \MV = \{ K \in\Gamma(E) : c(K,K,K) >0 \}
\end{equation}
can also be viewed as the space of smooth sections of a bundle over $M$ with fibre $P=\Ex{6(6)}\times\bbR^+/\Fx{4(4)}$. It admits a natural (rigid) very special real metric, which again is inherited from the very special real metric on the homogeneous-space fibres.\footnote{For reviews of very special real geometry see for example~\cite{WV91,WV96}.} The Riemannian symmetric spaces that admit (local) very special real metrics were analysed in~\cite{GST84} and include the case $\Ex{6(-26)}/\Fx4$. Here we need a pseudo-Riemannian form based on $\Ex{6(6)}$, where the relevant space is again a prehomogeneous vector space~\cite{SK77}
\begin{equation}
\label{eq:LVSR}
   P/\bbR^+ = \Ex{6(6)}/\Fx{4(4)}.
\end{equation}
The geometry on $\MV$ can be constructed as follows. Consider a point $K\in \MV$. Given $u,v,w\in T_K \MV\simeq\Gamma(E)$, the fibre-wise cubic invariant $c$ defines a cubic form on $\MV$ by
\begin{equation}
\label{eq:Cdef}
   C(u,v,w) = \int_M c(u,v,w) , 
\end{equation}
where, since sections of $E$ are weighted objects, we have $c(u,v,w)\in\Gamma(\det T^*M)$ and hence it can be integrated over $M$. The metric on $\MV$ is then defined as the Hessian of $C(K,K,K)$
\begin{equation}
   C_{MN} = \frac{\delta C}{\delta K^M\delta K^N} .
\end{equation}
In these expressions we are using the flat coordinates on $\MV$ defined by the vector-space structure of $\Gamma(E)$. On any rigid very special real geometry there is a global $\bbR^+$ symmetry, such that the quotient space is, by definition, a local very special real geometry. On $\MV$, the action of $\bbR^+$ is constant rescaling of the invariant tensor $K$. As for the hypermultiplet structure, the $\bbR^+$ action is simply a physically irrelevant constant shift in the warp factor $\Delta$. Modding out by the symmetry, the physical space of structures $\MV/\bbR^+$ is an infinite-dimensional local very special real space. 

In analogy with the $D=4$ discussion of~\cite{GLW06,GLW07,GLSW09}, we can view $\MV/\bbR^+$ and the quaternionic-K\"ahler base of $\MH$ as the  infinite-dimensional spaces of vector- and hypermultiplet degrees of freedom, coming from rewriting the full ten- or eleven-dimensional supergravity theory as a five-dimensional $\mathcal{N}=1$ theory. 


\subsubsection{Integrability}

The integrability conditions for the $\Ex{6(6)}$ generalised $G$-structures again arise from gauging the generalised diffeomorphism group, and are almost identical to those in $D=4$ given in~\eqref{eq:e7_mm},~\eqref{eq:e7_comp_int1} and~\eqref{eq:e7_intersection_int}, namely
\begin{align}
   \mu_{\alpha}(V) &= 0 
      \qquad\text{for all $V\in\Gamma(E)$} , \label{eq:e6_mm} \\
   \Dorf_K K &=0 , \label{eq:e6_comp_int} \\
   \Dorf_K J_\alpha &= 0 . \label{eq:e6_intersection_int}
\end{align}
In each case they are equivalent to the structure admitting a torsion-free, compatible generalised connection: if the first condition holds, $J_\alpha$ defines a torsion-free $\SUstar6$ structure; if the second condition holds, $K$ defines a torsion-free $\Fx{4(4)}$ structure; if all three conditions are satisfied, $\{J_\alpha,K\}$ define a torsion-free $\USp6$ structure. In the latter case, using the results of~\cite{CSW14b}, this implies that these conditions are equivalent to the existence of an $\mathcal{N}=1$ Killing spinor. Note that the pair of compatible and integrable \HM{} and \VM{} structures is again not enough to imply that the \HV{} structure is integrable.

To see that these differential conditions constrain the generalised intrinsic torsion for the different generalised structures, we start by noting that for $\Ex{6(6)}\times\mathbb{R}^+$ generalised geometry the space of generalised torsions is~\cite{CSW11}
\begin{equation}
W = \rep{27} + \rep{351}'.
\end{equation}
Repeating the analysis of section~\ref{sec:int-tor}, we find, decomposing under $\SU{2}\times\SUstar{6}$, 
\begin{equation}
\Tint^{\SUstar6}=(\rep{2},\rep{6})+(\rep{3},\rep{15}) , 
\end{equation}
while decomposing under $\Fx{4(4)}$ 
\begin{equation}
\Tint^{\Fx{4(4)}}=\rep{1}+\rep{26}.
\end{equation}
The intrinsic torsion components of an \HV{} or $\USp{6}$ structure, decomposed under $\SU{2}\times\USp{6}$, along with which integrability conditions they constrain, are summarised in table~\ref{tab:USp6-int-tor}. We note that it is equal to $\repp{2}{1}\times(S+ J)$, where $S+ J=\rep{8}+\rep{48}$ is the $\USp{8}$ representation in which the Killing spinor equations transform. From the results of~\cite{CSW11}, we see that, again, the Killing spinor equations are equivalent to the differential conditions for an integrable $\USp{6}$ structure. 

\begin{table}
\centering
\begin{tabular}{@{}lccccccc@{}} 
\toprule
\multicolumn{1}{l}{} & \multicolumn{7}{c}{$\Tint^{\USp6}$ component} \\
\cmidrule(l){2-8}
Integrability condition & $\repp{1}{1}$ & $\repp{3}{1}$ 
   & $\repp{2}{6}$ & $\repp{2}{6}'$ 
   & $\repp{1}{14}$ & $\repp{3}{14}$ & $(\rep{2},\rep{14}')$ 
     \tabularnewline
\midrule
$\mu_\alpha=0$ & & $\times$ & $\times$ & & & $\times$ & \tabularnewline 
$\Dorf_K K=0$ & $\times$ & & & $\times$ & $\times$ & & \tabularnewline 
$\Dorf_KJ_\alpha=0$ & $\times$ & $\times$ & & $\times$ & 
   & $\times$ & $\times$ \tabularnewline
\bottomrule
\end{tabular}
\protect\caption{The components of the generalised intrinsic torsion $\Tint^{\USp6}$ appearing in each of the $\mathcal{N}=1$, $D=5$ supersymmetry conditions.}
\label{tab:USp6-int-tor}
\end{table}

As in the $D=4$ case, the integrability conditions have a direct interpretation in terms of $D=5$ gauged supergravity. Following~\cite{LST12}, the conditions for a Minkowski vacuum in a generic gauged $\mathcal{N}=1$ theory can be written as\footnote{Note that the third condition comes from the term in $W^{xAB}$ proportional to $\epsilon^{AB}$~\cite{BCWGVP06b}, which was assumed to vanish in~\cite{LST12}.}
\begin{equation}
\label{eq:sugra-conds-5d}
   \Theta_I^\lambda \mu_{\alpha,\lambda} = 0 , \qqq
   h^\Lambda\Theta_\Lambda^{\lambda} k_\lambda^u = 0 , \qqq
   h^\Lambda\hat{\Theta}_\Lambda^{\hat{\lambda}} \hat{k}_{\hat{\lambda}}^i = 0 . 
\end{equation}
The only difference compared with the $D=4$ case is that the vector $h^\Lambda$ is now the coordinate vector in the real special geometry on $\MV$, written in flat coordinates, which here we identify with $K$. The three conditions~\eqref{eq:sugra-conds-5d} then translate directly into the three integrability conditions~\eqref{eq:e6_mm}--\eqref{eq:e6_intersection_int}. 

We can again consider the moduli spaces of structures. The integrability conditions for the \HM{} structure are identical to those in $D=4$, and again the moduli space is a hyper-K\"ahler quotient, exactly as discussed in section~\ref{sec:m-hyper}. The arguments leading to the identification of the moduli space of \VM{} structures are also similar to those of $D=4$, and so we simply summarise the relevant observations and results.

As discussed in~\cite{AC09}, rigid very special real geometry requires the existence of a flat torsion-free connection $\hat{\nabla}$ preserving a metric tensor $C_{mn}$ that, with respect to the flat coordinates, can be written as the Hessian of a cubic function $C$. For us, the vector-space structure of $\Gamma(E)$ defines natural flat coordinates on $\MV$ and the cubic function is given by~\eqref{eq:Cdef}. The function is invariant under the action of generalised diffeomorphisms, and since $\rho_V=\Dorf_V K$ is linear in $K$, it maps flat coordinates to flat coordinates. Thus $\GDiff$ preserves the very special real structure. Furthermore, we observe that given an integrable structure $K$ such that $\Dorf_K K=0$, any other choice of structure related to $K$ by the action of $\GDiff$ is automatically integrable too. This means that integrability of the structure is well defined under equivalence by $\GDiff$, so that both the very special real structure and the integrability condition descend to the quotient space. Thus the moduli space of integrable vector-multiplet structures is
\begin{equation}
\ModV=\{ K \in \MV : \Dorf_K K = 0 \} \quotient \GDiff ,
\end{equation}
which, as the $\bbR^+$ action generated by $K$ commutes with $\GDiff$, is a rigid very special real space. As required by supersymmetry, it is a cone over a local very special real space. The moduli space of \HV{} structures is again more complicated, though all the comments made in the $D=4$ case also apply here. 


\subsubsection{Example: Calabi--Yau manifold in M-theory}

Just for orientation, we consider the simplest example of a generalised $\USp6$ structure, namely M-theory on a six-dimensional Calabi--Yau manifold $M$. In fact, assuming $M$ has only an $\SU3$ structure, supersymmetry implies that the metric is Calabi--Yau and that the warp factor $\Delta$ and four-form flux $F$ vanish~\cite{BG00,GMSW04c}. The goal here is to see how these conditions arise from the integrability conditions on the \HM{} and \VM{} structures. 

The untwisted \HM{} and \VM{} structures are encoded by $\Omega$ and $\omega$ respectively. We have  
\begin{equation}
\begin{split}
   \tilde{J}_{+} &= - \tfrac{1}{2}\kappa\Omega
              + \tfrac{1}{2}\kappa\Omega^{\sharp},\\
   \tilde{J}_{3} &= \tfrac{1}{2}\kappa I
              - \tfrac{1}{16}\ii\kappa\Omega\wedge\bar{\Omega}
              - \tfrac{1}{16}\ii\kappa\Omega^{\sharp}
                   \wedge \bar{\Omega}^{\sharp} ,
\end{split}
\end{equation}
where $I$ is the almost complex structure~\eqref{eq:acs} and the $\Ex{6(6)}$-invariant volume is $\kappa^{2}=\ee^{3\Delta}\vol_{6}$, while 
\begin{equation}
   \tilde{K} = -\ee^{\Delta}\omega.
\end{equation}
It is easy to check, using the expressions in appendix~\ref{app:gen-geom}, that $J_\alpha$ generate an $\su2$ algebra and that the structures satisfy the correct normalisation and compatibility conditions, given~\eqref{eq:SU3-consistency}. As previously, the actual structure will include the three-form potential $A$ via the adjoint action: $J_\alpha=\ee^A\tilde{J}_\alpha\ee^{-A}$ and $K=\ee^A\tilde{K}$. In what follows it will be easiest to use the untwisted forms with the twisted Dorfman derivative in the differential conditions. 

The hypermultiplet structure is integrable if the triplet of moment maps vanish. We start with $\mu_{3}$. The moment map is a sum of terms that depend on arbitrary $\tilde{v}$, $\tilde{\omega}$ and $\tilde{\sigma}$. Considering each component in turn we find
\begin{equation}
\mu_3(\tilde{\sigma}) 
   \propto \int_M\dd(\ee^{3\Delta})\wedge\tilde{\sigma} , \qquad 
\mu_3(\tilde{\omega})
   \propto \int_{M}\ee^{3\Delta}\tilde{\omega}\wedge F.
\end{equation}
implying $\dd\Delta=F=0$. Using the fact that $\Delta$ is constant, the $\tilde{v}$ component of $\mu_3$ and the $\tilde{\omega}$ component of $\mu_+$ simplify to 
\begin{equation}
\mu_3(\tilde{v})  
   \propto \int_M\ee^{3\Delta}\left(
       \imath_{\tilde{v}}\bar{\Omega}\wedge\dd\Omega
       - \imath_{\tilde{v}}\Omega\wedge\dd\bar{\Omega} \right), \qquad
   \mu_+(\tilde{\omega}) 
   \propto \int_M\ee^{3\Delta}\dd\Omega\wedge\tilde{\omega}.
\end{equation}
The first requires the $(3,1)$-component of $\dd\Omega$ to vanish or, in the language of~\cite{CS02}, the $\mathcal{W}_5$ component of the $\SU{3}$ torsion is set to zero, while the second vanishes if and only if the $(2,2)$-component of $\dd\Omega$ vanishes, that is, $\mathcal{W}_1=\mathcal{W}_2=0$. Finally the $\tilde{\sigma}$ component of $\mu_+$ vanishes identically, while the $\tilde{v}$ term vanishes if $F$ vanishes. Together, we see that the integrability of the hypermultiplet structure requires a constant warp factor, a vanishing four-form flux and that $\Omega$ is closed.

For the \VM{} structure we have 
\begin{equation}
\hat{\Dorf}_{\tilde{K}}\tilde{K} =
   -\omega\wedge\dd\omega=0 ,
\end{equation}
which requires the $\mathcal{W}_4$ component of the $\SU3$ torsion to vanish. Note that requiring $K$ to define a integrable $\Fx{4(4)}$ structure is considerably weaker than the condition for $\omega$ to define an integrable symplectic structure. Finally, the $\Dorf_K J_\alpha=0$ condition required for an integrable $\USp6$ structure is equivalent to 
\begin{equation}
\hat{\Dorf}_K J_+ \propto j\Omega^\sharp\lrcorner j\dd\omega - \tfrac{1}{3}\id  \Omega^\sharp\lrcorner\dd\omega-\dd\omega\wedge\Omega=0.
\end{equation}
One can show this vanishes if and only if $\dd\omega$ vanishes, that is $\mathcal{W}_1=\mathcal{W}_3=\mathcal{W}_4=0$. 

We have shown that for this restricted $\SU3$ ansatz, integrability of the generalised $\USp6$ structure requires $M$ to be Calabi--Yau, that is $\dd\omega=\dd\Omega=0$, with a constant warp factor and a vanishing four-form flux.


\subsection{\texorpdfstring{$\Ex{5(5)}$}{E5(5)} hyper- and tensor-multiplet structures}
\label{sec:E5_structures}

For compactifications to $D=6$ the relevant generalised geometry~\cite{CSW11,CSW14} has an action of $\Ex{5(5)}\times\mathbb{R}^+\simeq\Spin{5,5}\times\bbR^+$. The generalised tangent bundle transforms in the $\rep{16}_{\rep{1}}$ representation and decomposes under the relevant $\GL{d}$ group as \eqref{eq:E-Mtheory} or \eqref{eq:E-IIB}, where the doublet of five-forms are not present for type IIB and the one-form density terms are not present for type IIB or M-theory. The adjoint bundle transforms in the $\rep{1_0}+\rep{45_0}$ representation and decomposes as in \eqref{eq:ad-M-theory} or \eqref{eq:ad-IIB}, where the six-forms and six-vectors are not present for type IIB or M-theory. In both type II and M-theory, spinors transform under a $\USp4\times\USp4\simeq\Spin5\times\Spin5$ subgroup of $\Ex{5(5)}\times\mathbb{R}^+$. For $\mathcal{N}=(1,0)$ backgrounds, the Killing spinor is stabilised by an $\SU2\times\USp4$ subgroup.


\subsubsection{Structures and invariant tensors}

The $\Ex{6(6)}$ generalised $G$-structures are defined as follows. 
\begin{defn}
Let $G$ be a subgroup of $\Ex{5(5)}$. We define 
\begin{itemize}
\item a \emph{hypermultiplet structure} is a generalised structure with  $G=\SU2\times\Spin{1,5} $
\item a \emph{tensor-multiplet structure} is a generalised structure with $G=\Spin{4,5}$
\item an \emph{\HV{} structure} is a generalised structure with $G=\SU2\times\USp4$
\end{itemize}
\end{defn}
\noindent
As before, the \HM{} structure is defined by a triplet of sections $J_\alpha$ of a weighted adjoint bundle, as in~\eqref{eq:J-def}, such that they transform in the $\rep{45_{2}}$ representation of $\Ex{5(5)}\times\bbR^+$ and define a highest weight $\su{2}$ subalgebra. The algebra and norms of the $J_\alpha$ are the same as for the $D=4$ case, given in \eqref{eq:su2_algebra} and \eqref{eq:J_a_norm}. 

The \TM{} or tensor-multiplet structure is new. It is defined by choosing a section of the bundle $N$ transforming in the $\rep{10_2}$ representation of $\Ex{5(5)}\times\bbR^+$. For M-theory on a five-dimensional manifold $M$, 
\begin{equation}
N\simeq T^{*}M\oplus\ext^{4}T^{*}M,
\end{equation}
while for type IIB on a four-dimensional manifold $M$ it is
\begin{equation}
N\simeq S\oplus\ext^{2}T^{*}M\oplus S\otimes\ext^{4}T^{*}M.
\end{equation}
The invariant generalised tensor for a $\Spin{4,5}$ structure is a section of $N$:
\begin{equation}
\label{eq:Q-def}
   Q \in \Gamma(N), \quad \text{such that} \quad \eta(Q,Q) > 0   
\end{equation}
where $\eta$ is the $\SO{5,5}$ metric given in~\eqref{eq:M_quadratic} and~\eqref{eq:IIB_quadratic}.

A pair of compatible structures define an $\SU2\times\USp4$ structure and satisfy 
\begin{equation}
\label{eq:e5_comp_condition}
\begin{aligned}
   J_{\alpha}\cdot Q &= 0, \\
   \tr (J_\alpha J_\beta) &= - \eta(Q,Q)\,\delta_{\alpha\beta} ,
\end{aligned}
\end{equation}
where $\cdot$ is the adjoint action. They are equivalent to
\begin{equation}
\label{eq:JQ-comp}
   J_+ \cdot Q = 0 ,
\end{equation}
and the normalisation condition
\begin{equation}
\label{eq:e5_comp}
   \eta(Q,Q) = \kappa^2 , 
\end{equation}
respectively, where $\kappa$ is the factor appearing in~\eqref{eq:su2_algebra}. We again still refer to this as an \HV{} structure since it preserves eight supercharges and is that analogue of the corresponding structures in $D=4$ and $D=5$. In this case, there is no example without flux that is a Calabi--Yau space so the nomenclature is somewhat misleading, although the simplest flux example discussed in section~\ref{sec:NS5-IIB} does have an underlying Calabi--Yau two-fold. 

As before, the infinite-dimensional space of \HM{} structures $\MH$ is the space of smooth sections of a bundle over $M$ with fibre $W=\Ex{5(5)}\times\bbR^+/(\SU2\times\Spin{1,5})$. This fibre is a hyper-Kähler cone over a pseudo-Riemannian Wolf space~\cite{AC05}\footnote{Recall $\Ex{5(5)}\simeq\Spin{5,5}$,  $\Spin{4}\simeq\SU2\times\SU2$ and $\USp2\simeq\SU2$, and note we have not been careful here to keep track of any discrete group factors.}
\begin{equation}
\label{eq:E5-wolf}
   W/\bbH^* = \SO{5,5}/(\SO{4}\times\SO{1,5}) . 
\end{equation}
The hyper-K\"ahler geometry on $\MH$ is again inherited directly from the hyper-K\"ahler geometry of $W$. The details of this exactly follow the $D=4$ case in section~\ref{hyper_structure} upon exchanging the relevant groups.

The infinite-dimensional space of \TM{} structures 
\begin{equation}
   \MT = \{ Q\in\Gamma(N) : \eta(Q,Q) >0 \} 
\end{equation}
can also be viewed as the space of smooth sections of a homogeneous-space bundle $\ZT$ over $M$ with fibre $P = \Ex{5(5)}\times\bbR^+/\Spin{4,5}     \simeq\SO{5,5}\times\bbR^+/\SO{4,5} \simeq\bbR^{5,5}$. 
It admits a natural flat metric, which again is inherited from the flat metric on the fibres $P$. In $\mathcal{N}=(1,0)$ gauged supergravity, the scalar fields in the tensor multiplets describe Riemannian geometries of the form $\SO{n,1}/\SO{n}$, where the cone over this space is just flat $\bbR^{n,1}$~\cite{Romans86}. Here our fibres $P$ are isomorphic to $\bbR^{5,5}$ with a flat pseudo-Riemannian metric, with the base of the cone given by the hyperboloid 
\begin{equation}
\label{eq:LF}
   P/\bbR^+ = \SO{5,5}/\SO{4,5},
\end{equation}
where the $\mathbb{R}^+$ action is just the overall scaling. The flat metric is given by the quadratic form on $\MT$ 
\begin{equation}
   \Sigma(v,w) = \int_M \eta(v,w) , 
\end{equation}
where $v,w\in\Gamma(T_Q\MT)\simeq\Gamma(N)$, and since sections of $N$ are weighted objects, we have $\eta(v,w)\in\Gamma(\det T^*M)$ and hence it can be integrated over $M$. The flat metric on $\MT$ is simply $\Sigma$. On $\MT$, the action of $\bbR^+$ is constant rescaling of the invariant tensor $Q$. As for the hypermultiplet structure, the $\bbR^+$ action is simply a reparametrisation of the warp factor $\Delta$. Modding out by the symmetry, the physical space of structures $\MT/\bbR^+$ is an infinite-dimensional hyperbolic space.

As in the discussion of~\cite{GLW06,GLW07,GLSW09}, we view $\MT/\bbR^+$  and the quaternionic-K\"ahler base of $\MH$ as an infinite-dimensional spaces of tensor-  and hypermultiplet degrees of freedom, coming from rewriting the full ten- or eleven-dimensional supergravity theory as a six-dimensional $\mathcal{N}=(1,0)$ theory.

\subsubsection{Integrability}

The integrability conditions again arise from gauging the generalised diffeomorphism group and, for the \HM{} structures, are identical to those in $D=4,5$ given in~\eqref{eq:e7_mm}, namely
\begin{equation}
\label{eq:mm_e5} 
   \mu_{\alpha}(V) = 0 
      \qquad\text{for all $V\in\Gamma(E)$} , 
\end{equation}
which is equivalent to the structure admitting a torsion-free, compatible generalised connection.

The integrability condition for the \TM{} structure $Q$ is new and does \emph{not} require the generalised Lie derivative. Instead, it appears in much the same way as the integrability of the pure spinors $\Phi^\pm$ describing generalised complex structures in $\Orth{d,d}\times\bbR^+$ generalised geometry. Recall that the usual derivative operator $\partial$ embeds in $E^*$ which transforms in the $\rep{16}^c_{-\rep{1}}$ representation of $\Spin{5,5}$. We can use the $\rep{16}^c_{-\rep{1}}\times\rep{10_2}\to\rep{16}_{\rep{1}}$ action to form the projection $E^*\otimes N\to E$, given in \eqref{e5_proj_M} and \eqref{e5_proj_IIB}. This means there is a natural action of $\dd$ on $Q$ which results in a generalised vector, and furthermore, in this case, it is covariant. We then have
\begin{defn}
An \emph{integrable} or \emph{torsion-free} tensor-multiplet structure $Q$ is one satisfying
\begin{equation}
\label{eq:e5_comp_int}
\dd Q=0 , 
\end{equation}
or in other words $Q$ is closed under the exterior derivative. 
\end{defn}
\noindent These conditions are equivalent to there being a torsion-free generalised connection compatible with the generalised $\Spin{4,5}$ structure defined by $Q$. 

We can also consider the integrability conditions for the \HV{}  generalised structure defined by the compatible pair $\{J_\alpha,Q\}$. We find 
\begin{defn}
An \emph{integrable} or \emph{torsion-free} \HV{} structure $\{J_\alpha,Q\}$ is one such that $J_\alpha$ and $Q$ are separately integrable. There are no further conditions.
\end{defn}
\noindent In contrast to the case of compatible \VM{} and \HM{} structures, the existence of a pair of compatible and integrable \HM{} and \TM{} structures is enough to imply that the \HV{} structure is integrable. These conditions are equivalent to there being a torsion-free generalised connection compatible with the generalised $\SU2\times\USp4$ structure defined by $\{J_\alpha,Q\}$. Using the results of~\cite{CSW14b}, this implies that the conditions are equivalent to the existence of an $\mathcal{N}=(1,0)$ Killing spinor. 

To see that these differential conditions constrain the appropriate generalised intrinsic torsion for the different generalised structures, we start by noting that for $\Ex{5(5)}\times\mathbb{R}^+$ generalised geometry the space of generalised torsions is~\cite{CSW11}
\begin{equation}
W = \rep{16}^{c}+\rep{144}^{c}.
\end{equation}
Repeating the analysis of section~\ref{sec:int-tor}, we find, decomposing under $\SU{2}\times\SU2\times\Spin{1,5}$ where the first factor is the $\SU2$ generated by $J_\alpha$, 
\begin{equation}
   \Tint^{\SU{2}\times\Spin{1,5}}
      =(\rep{2},\rep{1},\rep{4}^{c})+(\rep{3},\rep{2},\rep{4}),
\end{equation}
while decomposing under $\Spin{4,5}$ 
\begin{equation}
   \Tint^{\Spin{4,5}}=\rep{16}.
\end{equation}
The intrinsic torsion components of an \HV{} or $\SU2\times\USp4$ structure, decomposed under $\SU{2}\times\SU2\times\USp{4}$, along with which integrability conditions they constrain, are summarised in table~\ref{tab:USp2USp4-int-tor}. We note that the intrinsic torsion is equal to $(\rep{2},\rep{1},\rep{1})\times(S^- + J^-)$, where $S^-+ J^-=(\rep{1},\rep{4})+(\rep{5},\rep{4})$ are the $\USp{4}\times\USp{4}$ representations in which the Killing spinor equations transform for $\mathcal{N}=(1,0)$ supersymmetry~\cite{CSW14b}. Again, from the results of~\cite{CSW11}, the Killing spinor equations are equivalent to the differential conditions for an integrable $\SU{2}\times\USp{4}$ structure.
\begin{table}
\centering
\begin{tabular}{@{}lcccc@{}} 
\toprule
\multicolumn{1}{l}{} & \multicolumn{4}{c}{$\Tint^{\SU2\times\USp4}$ component} \\
\cmidrule(l){2-5}
Integrability condition & $(\rep{1},\rep{2},\rep{4})$ 
   & $(\rep{2},\rep{1},\rep{4})$ 
   & $(\rep{2},\rep{1},\rep{4})'$ 
   & $(\rep{3},\rep{2},\rep{4})$ 
     \tabularnewline
\midrule
$\mu_\alpha=0$ & & $\times$ & & $\times$  \tabularnewline 
$\dd Q=0$ & $\times$ & & $\times$ & \tabularnewline 
\bottomrule
\end{tabular}
\protect\caption{The components of the generalised intrinsic torsion $\Tint^{\SU2\times\USp4}$ appearing in each of the $\mathcal{N}=(1,0)$, $D=6$ supersymmetry conditions.}
\label{tab:USp2USp4-int-tor}
\end{table}

As in the $D=4$ and $D=5$ cases, the integrability conditions have a interpretation in terms of $D=6$ gauged supergravity as we now sketch. The gauging of $D=6$ supergravity coupled to tensor-, vector- and hypermultiplets using the embedding tensor formalism is discussed in the context of ``magical supergravities'' in~\cite{GSS10}. The conditions for a supersymmetric Minkowski background coming from the vanishing of the gaugino variations read\footnote{Note we use a different index notation from~\cite{GSS10} to match the notation used in $D=4$ and $D=5$. Also, the first term in~\eqref{eq:sugra-conds-6d} comes contracted with a matrix $m^{\Sigma\Lambda}$ in the gaugino variation, but using the fact that $m^2\propto (L^IL_I) \id$ we see that this term can only vanish if the first term in~\eqref{eq:sugra-conds-6d} vanishes.}
\begin{equation}
\label{eq:sugra-conds-6d}
   \Theta_{\Lambda}^\lambda \mu_{\alpha,\lambda} = 0 , \qqq
   L^I\theta_I^\Lambda = 0 .
\end{equation}
The key difference compared with the $D=4$ case is that the vector $L^I$ is now the coordinate vector on the flat tensor-multiplet space $\MT$. Note that the first condition was previously discussed in~\cite{LST12}. In making the translation to the integrability conditions we note that $L^I$ corresponds to $Q$, while the matrix $\theta_I^\Lambda$ is a map
\begin{equation}
   \theta : \Gamma(N) \to \Gamma(E),
\end{equation}
which we can identify with the action of the exterior derivative $\dd$ discussed above, so that
\begin{equation}
   L^I\theta_I = \dd Q . 
\end{equation}
Hence the conditions~\eqref{eq:sugra-conds-6d} are precisely~\eqref{eq:mm_e5} and~\eqref{eq:e5_comp_int}. Note that there are number of conditions on $\theta_I^\Lambda$, as well as on the intertwiner between $\Gamma(N)$, $\Gamma(E)$ and $\Gamma(E^*)$, related to the tensor hierarchy and necessary for the supersymmetry algebra to close. It would be interesting to see how these are satisfied by the exterior derivative $\dd$ in the infinite-dimensional case. The fact that the geometry on each fibre $\SO{5,5}\times\bbR^+/\SO{4,5}$ of the homogeneous-space bundle $\ZT$ is a pseudo-Riemannian variant of that appearing in one of the magical supergravity theories suggests that the structure will essentially be inherited fibre-wise. 

The moduli spaces in this case are much the same as the previous examples we have seen. The moduli space of \HM{} structures is again as discussed in section~\ref{sec:m-hyper}. The space of \TM{} structures $\MT$ admits a flat geometry, defined by the metric $\Sigma$. Again, the vector-space structure of $\Gamma(N)$ defines natural flat coordinates on $\MT$ and hence a flat connection that, by definition, preserves $\Sigma$. $\GDiff$ preserves the flat structure, and furthermore an integrable structure $Q$ remains integrable under the action of $\GDiff$. Thus the moduli space of integrable tensor-multiplet structures is
\begin{equation}
\ModT=\{ Q \in \MT : \dd Q = 0 \} \quotient \GDiff ,
\end{equation}
which is again flat. As required by supersymmetry, it is a cone over a hyperbolic space. As for the previous cases, generically $\GDiff$ does not act freely on $\MT$ and hence $\ModT$ is not necessarily a manifold.  The moduli space of \HV{} structures is again more complicated, though all the comments made in the $D=4$ case also apply here.


\subsubsection{Example: NS5-branes on a hyper-Kähler space in type IIB}
\label{sec:NS5-IIB}

The standard NS5-brane solution is a warped product of six-dimensional Minkowski space with a flat four-dimensional transverse space and preserves sixteen supercharges~\cite{CHS92}. Exchanging the flat transverse space for a four-dimensional hyper-Kähler space breaks supersymmetry further, leaving eight supercharges~\cite{Strominger86}. Thus, we expect it can be formulated as an integrable $\SU{2}\times\USp{4}$ structure within $\Ex{5(5)}\times\mathbb{R}^{+}$ generalised geometry. The metric takes the standard form~\eqref{eq:warp-metric} with $D=6$, and the four-dimensional space $M$ admits an $\SU{2}$ structure, with a triplet of two-forms $\omega_\alpha$ as in~\eqref{eq:SU2-structure}, and a canonical volume form $\frac{1}{2}\omega_\alpha\wedge\omega_\beta=\delta_{\alpha\beta} \vol_4$. The solution also has non-trivial NS-NS three-form flux $H$ and dilaton $\phi$, but the warp factor $\Delta$ is zero. The solution is supersymmetric if the $\SU{2}$ structure and three-form flux satisfy~\cite{Strominger86,GMW03}
\begin{equation}
   \dd(\ee^{-2\phi}\omega_{\alpha})=0, 
   \qqq \star H=-\ee^{2\phi}\dd(\ee^{-2\phi}).
\end{equation}
Recall that in type II theories there are two types of ten-dimensional spinors. The NS5-brane solutions are an example of a pure NS-NS geometry preserving eight supercharges where the preserved Killing spinors are all of one type: they have $\nabla^+$ special holonomy in the language of~\cite{GMW03}. As such they cannot be described by generalised complex structures~\cite{GMPT05}. For this reason it is interesting to see how they do appear in the $\Ex{5(5)}\times\bbR^+$ generalised geometry. (Note that we described the same solution wrapped on $\bbR^2$ in $\Ex{7(7)}\times\bbR^+$ generalised geometry in terms of the wrapped M5-brane background of section~\ref{sec:M5-bkgd}.) In the next section we will also see how they can be embedded in $\Orth{4,4}$ generalised geometry. 

Embedding in type IIB, the \HM{} structure is determined by the $\omega_\alpha$, such that the untwisted objects are
\begin{equation}
\label{NS5_J}
   \tilde{J}_{\alpha}
     = - \tfrac{1}{2}\kappa I_{\alpha}
        + \tfrac{1}{2}\kappa u^{i}\omega_{\alpha}
        + \tfrac{1}{2}\kappa v^{i}\omega_{\alpha}^{\sharp},
\end{equation}
where $(I_\alpha)^m{}_n=-(\omega_\alpha)^m{}_n$ is the triplet of almost complex structures, $u^i=(-1,0)$ and $v^i=(0,-1)$, and $\kappa^{2}=\ee^{-2\phi}\vol_{4}$ is the $\Ex{5(5)}$-invariant volume. The untwisted \TM{} structure depends only on the volume form and dilaton through
\begin{equation}\label{NS5_Q}
   \tilde{Q}= u^{i}+\ee^{-2\phi} v^{i}\vol_{4},
\end{equation}
where $u$ and $v$ are the same as above. It is easy to check from the results of appendix~\ref{app:gen-geom} that the $J_\alpha$ generate an $\su{2}$ algebra, and that the normalisation and compatibility conditions are satisfied. The NS-NS three-form flux $H$ embeds in the first component of $F^{i}_{3}$ (see~\eqref{eq:IIB_flux_main}). Thus, as previously, the actual structures will include the NS-NS two-form potential via the adjoint action: $J_\alpha=\ee^{B^1}\tilde{J}_\alpha\ee^{-B^1}$ and $Q=\ee^{B^1}\tilde{Q}$. In what follows it will be easiest to use the untwisted forms with the twisted Dorfman derivative in the differential conditions. 

For the moment maps the $\tilde{\lambda}^{i}$ terms give
\begin{equation}
\mu_{\alpha}(\tilde{\lambda}^{i})\propto\int_{M_{4}}\ee^{-2\phi}\omega_{\alpha}\wedge\dd\tilde{\lambda}^{1},
\end{equation}
which vanishes for $\dd(\ee^{-2\phi}\omega_{\alpha})=0$, completely fixing the intrinsic torsion of the underlying $\SU2$ structure. Using this condition, the $\tilde{v}$ terms simplify to
\begin{equation}
\mu_{\alpha}(\tilde{v})\propto\int_{M_{4}}\epsilon_{\alpha\beta\gamma}\ee^{-2\phi}\dd\phi\wedge\omega_{\beta}\wedge\imath_{\tilde{v}}\omega_{\gamma}-\ee^{-2\phi}\omega_{\alpha}\wedge\imath_{\tilde{v}}H.
\end{equation}
This vanishes for $\star H=-\ee^{2\phi}\dd(\ee^{-2\phi})$. In terms of the untwisted objects, the integrability of the \TM{} structure is given by
\begin{equation}
\dd_{F^i}\tilde{Q} = 0,
\end{equation}
where the action of $\dd_{F^i}$ on $\tilde{Q}\in\Gamma(N)$ is defined in \eqref{dQ_twisted}. Using the explicit form of $\tilde{Q}$, we have
\begin{equation}
\dd_{F^i}\tilde{Q}=\dd u^i+\epsilon_{ij}u^i F^j
\end{equation}
The one-form term vanishes as $u^i$ has constant entries. The three-form term also vanishes as the contraction of $u^i$ with $F^j$ picks out $F^2=F_3$, which is zero for the NS5-brane background.

Finally, note that we can embed the D5-brane solution in a similar way. The dilaton now appears as a warp factor $\Delta$, so the $\Ex{5(5)}$-invariant volume is $\kappa^{2}=\vol_{4}$. We also take $u^{i}=(0,1)$ and $v^{i}=(-1,0)$, and drop the factor of $\ee^{-2\phi}$ in $\tilde{Q}$. The moment maps then vanish if
\begin{equation}
\dd(\ee^{\phi}\omega_{\alpha})=0,\qqq F_{3}=-2\star\dd(\ee^{-\phi}).
\end{equation}
The first of these is the correct differential condition for the $\SU{2}$ structure. The second is the correct three-form flux, coming from the dual of the seven-form flux due to the D5-brane~\cite{GMW03}. The integrability of the \TM{} structure takes the same form as for the NS5-brane, but now  the contraction of $u^i$ with $F^j$ picks out $F^1=H$, which is zero for the D5-brane background.


\section{\texorpdfstring{$\Orth{d,d}$}{O(d,d)} hyper- and vector-multiplet structures}
\label{sec:0dd}

Thus far we have focussed on vector- and hypermultiplet structures in the exceptional $\Ex{d(d)}\times\bbR^+$ generalised geometries. However, as we now describe, they also appear in the $\Orth{d,d}\times\bbR^+$ generalised geometry of Hitchin and Gualtieri~\cite{Hitchin02,Gualtieri04}, describing the NS-NS sector of type II supergravity. The main point of physical interest is that it will give us an alternative description of the NS5-brane backgrounds of~\cite{CHS92,Strominger86,GMW03}. 

Let us step back for a moment and recall that there are three basic classes of (generalised) geometries based on the split real forms $\SL{d;\bbR}$, $\Orth{d,d}$ and $\Ex{d(d)}$ of the simply laced groups as summarised in table~\ref{tab:gen-geom}. In each case, the (generalised) Lie derivative takes a standard form~\cite{CSW11b,CSW11} as in~\eqref{eq:Dorf-def-M}, such that, given $V\in\Gamma(E)$ and any generalised tensor $\alpha$, we have
\begin{equation}
   \Dorf_V \alpha = (V\cdot\partial) \alpha 
       - (\partial\oadj V) \cdot \alpha ,
\end{equation}
where $\oadj$ projects on the adjoint representation of $G_\text{frame}$. One can introduce generalised connections and define torsion (see~\cite{Siegel93a,HK11,Gualtieri10,JLP10,JLP11,CSW11} for $\Orth{d,d}$ and~\cite{CSW11b} for $\Ex{d(d)}$), where the torsion tensor is a section of 
\begin{equation}
   W \simeq E^* \oplus K , 
\end{equation}
with
\begin{equation}
   K = \begin{cases}
         (TM\otimes\ext^2T^*M)_0 
            & \text{for $\SL{d;\bbR}\times\bbR^+$,} \\
         \ext^3E 
            & \text{for $\Orth{d,d}\times\bbR^+$,} \\
         \rep{144}^c, \rep{351}', \rep{912}  
            & \text{for $\Ex{d(d)}\times\bbR^+$ with $d=5,6,7$,} 
         \end{cases}
\end{equation}
and where the subscript on $(TM\otimes\ext^2T^*M)_0$ denotes the traceless sub-bundle. Note that other generalised geometries are possible~\cite{Baraglia12,S-C13}, both as a result of choosing different structure groups and depending on under which representation the generalised tangent space transforms. 

\begin{table}
\centering
\begin{tabular}{@{}lll@{}} 
\toprule
& Structure group $G_\text{frame}$ & Gen.~tangent space $E$ \tabularnewline 
\midrule
Conventional geom. & $\SL{d;\bbR}\times\bbR^+$ & $TM$ 
\tabularnewline  
Hitchin generalised geom. & $\Orth{d,d}\times\bbR^+$ & $TM\oplus T^*M$ 
\tabularnewline  
Exceptional generalised geom. & $\Ex{d(d)}\times\bbR^+$ & $TM\oplus\ext^2T^*M\oplus\ldots$ 
\tabularnewline 
\bottomrule
\end{tabular}
\protect\caption{The three basic classes of (generalised) geometries based on the split real forms of the simply laced Lie groups.}
\label{tab:gen-geom}
\end{table}

The basic point, stressed both in this paper and in~\cite{CSW14b}, is that one can use the generalised analogue of intrinsic torsion to define integrable generalised structures in each case. The original generalised complex structures of~\cite{Hitchin02,Gualtieri04} are a case in point. Consider an $\SU{3,3}\subset\SO{6,6}$ structure defined by a pure spinor $\Phi^\pm$. To see that the differential conditions $\dd\Phi^\pm=0$ constrain the appropriate generalised intrinsic torsion, recall first that the generalised torsion of a generalised connection $D_MV^N=\partial_MV^N+\Omega_M{}^N{}_PV^P$, where $V\in\Gamma(E\otimes\det T^*M)$, is given by~\cite{Siegel93a,Gualtieri10,CSW11b}
\begin{equation}
\begin{aligned}
   (T_1)_{MNP} &= -3\Omega_{[MNP]} &&\in \Gamma(\ext^3E) , \\
   (T_2)_M &= - \Omega_P{}^P{}_M  &&\in \Gamma(E) , 
\end{aligned}
\end{equation}
where indices are raised and lowered using the $\Orth{d,d}$ metric. It is straightforward to show, decomposing under $\SU{3,3}$, that the intrinsic torsion is 
\begin{equation}
   \Tint^{\SU{3,3}}
      = \rep{6} + \rep{20} + \CC , 
\end{equation}
and that these are precisely the representations that appear in $\dd\Phi^\pm$. It is worth stressing that it is key that the $\bbR^+$ factor is included. Without it the $E^*\simeq E$ component of $W$ is missing~\cite{Gualtieri10} and the condition $\dd\Phi^\pm=0$ is \emph{not} equivalent to vanishing intrinsic torsion since there is then no $\rep{6}$ representation in $\Tint$. 

In what follows, we will simply look at the list of Wolf spaces and prehomogeneous vector spaces (PVS) that arise for split real forms of the simply laced groups, to see if any other potentially interesting examples of integrable structures exist. We will exclude $\Ex{8(8)}$ because the standard formalism of generalised geometry fails for this case~\cite{CSW11,BCKT13}; although other interesting approaches are possible~\cite{HS14,Rosabal14,CR15} we will not consider them here.  For the hypermultiplet structures we will see that the list is very simple, while for the PVS cases it is considerably richer. 


\subsection{Hypermultiplet structures and integrability}
\label{sec:H-struc-Odd}

From the list given in~\cite{AC05}, the only case based on simply laced groups that we have not yet considered is the hyper-K\"ahler cone $W=\SO{d,d}\times\bbR^+/(\SU{2}\times\SO{d-4,d})$, where the corresponding Wolf space is
\begin{equation}
   W/\bbH^* = \SO{d,d}/(\SO4\times\SO{d-4,d}) .
\end{equation}
Note that this case already arose for $d=5$, as the $\Ex{5(5)}\simeq\Spin{5,5}$ hypermultiplet structure discussed in section~\ref{sec:E6_structures} and where the generalised tangent space transformed in the spinor representation $\rep{16}$.\footnote{Note that such $\Spin{d,d}\times\bbR^+$ generalised geometries with $d=6,7,8$ can also appear as subsectors of exceptional generalised geometries~\cite{S-C13}.} Here, however, we are interested in the case of conventional $\Orth{d,d}\times\bbR^+$ generalised geometry, valid for any $d$, where the generalised tangent bundle transforms in the vector $\rep{2d}$ representation.

Recall that in conventional $\Orth{d,d}\times\bbR^+$ generalised geometry, the generalised tangent bundle and adjoint bundle take the forms
\begin{equation}
\begin{aligned}
   E &\simeq TM \oplus T^*M , \\
   \ad\tilde{F} &\simeq \bbR \oplus (TM\otimes T^*M) 
      \oplus \ext^2TM \oplus \ext^2 T^*M , 
\end{aligned}
\end{equation}
and one defines $(\det T^*M)^p$ to have weight $p$ under the $\bbR^+$ action~\cite{GMPW08,CSW11b}. We then define
\begin{defn}
An $\SO{d,d}$ \emph{hypermultiplet structure} is an $\SU2\times\SO{d-4,4}\subset\SO{d,d}\times\bbR^+$ generalised structure.
\end{defn} 
\noindent
It can be defined as usual by choosing a triplet of sections $J_\alpha$ of a weighted adjoint bundle as in~\eqref{eq:J-def} that give a highest weight $\su{2}$ subalgebra of $\so{d,d}$ satisfying~\eqref{eq:su2_algebra} and~\eqref{eq:J_a_norm}.

The integrability conditions again take the standard form~\eqref{eq:e7_mm}, namely
\begin{equation}
\label{eq:mm_Odd} 
   \mu_{\alpha}(V) = 0 
      \qquad\text{for all $V\in\Gamma(E)$} ,
\end{equation}
which is equivalent to the structure admitting a torsion-free, compatible generalised connection. It is straightforward to show, decomposing under $\SU2\times\SU2\times\SO{d-4,d}$ where the first factor is the $\SU2$ generated by $J_\alpha$, that the generalised intrinsic torsion is 
\begin{equation}
   \Tint^{\SU{2}\times\SO{d-4,d}}
      =(\rep{2},\rep{2},\rep{1})+(\rep{3},\rep{1},\rep{2d-4}) ,
\end{equation}
and that these are precisely the representations constrained by the moment map conditions. Note that for $d=5$ this intrinsic torsion is different from that in section~\ref{sec:E5_structures} precisely because the generalised tangent bundle transforms in a different representation.  


\subsection{Prehomogeneous vector spaces, vector-multiplet structures and integrability}
\label{sec:PVS}

In any reformulation of supergravity in terms of a lower-, $D$-dimensional gauged supergravity theory, the infinite set of vector multiplets will correspond to sections of the generalised tangent bundle since it is these elements that generate the symmetries. If the vector multiplets in $D$ dimensions contain scalars, then these define \VM{} structures given by sections of $E$. 

We have already seen how these appear for $\Ex{d(d)}\times\bbR^+$ generalised geometry. For $\SO{d,d}\times\bbR^+$ there is one remaining possibility noted in~\cite{Hitchin01}. A generalised vector is generically stabilised by $\SO{d-1,d}$ and defines a point in the flat homogeneous space $P=\SO{d,d}\times\bbR^+/\SO{d-1,d} \simeq \bbR^{d,d}$, which is a cone over the hyperboloid 
\begin{equation}
   P/\bbR^+ = \SO{d,d} / \SO{d-1,d} . 
\end{equation}
Thus we can define 
\begin{defn}
An $\SO{d,d}$ \emph{vector-multiplet structure} is an $\SO{d-1,d}\subset\SO{d,d}\times\bbR^+$ generalised structure.
\end{defn} 
\noindent
It is defined by a section of the \emph{weighted} generalised tangent bundle
\begin{equation}
   K\in\Gamma(E\otimes(\det T^*M)^{1/2}) \qquad
   \text{such that $\eta(K,K) > 0$} ,
\end{equation}
where $\eta$ is the $\Orth{d,d}$ metric so that $\eta(K,K)\in\Gamma(\det T^*M)$. Note that for $d=5$ the \TM{} structure $Q$ in $\Ex{5(5)}\simeq\Spin{5,5}$ generalised geometry, discussed in section~\ref{sec:E5_structures}, also defined a $\Spin{4,5}$ generalised structure. Again, the difference here is that the $\Orth{d,d}$ generalised tangent bundle $E$ is in the vector representation rather than the spinor representation, as was the case for $\Ex{5(5)}$.

The integrability conditions essentially take the standard form~\eqref{eq:e7_comp_int1}, except that since $K$ is a weighted section we cannot take the generalised Lie derivative along $K$. We need to rescale, so that the definition becomes
\begin{defn}
An \emph{integrable} or \emph{torsion-free} $\SO{d,d}$ vector-multiplet structure $K$ is one satisfying
\begin{equation}
\label{eq:V-Odd}
   \Dorf_{K/\epsilon} K = 0 , 
\end{equation}
where $\epsilon=\sqrt{\eta(K,K)}$ so that $K/\epsilon\in\Gamma(E)$.
\end{defn}
\noindent 
It is then straightforward to show, decomposing under $\SO{d-1,d}$, that the corresponding generalised intrinsic torsion is just a singlet corresponding to $\eta(K/\epsilon,T_2)$
\begin{equation}
   \Tint^{\SO{d-1,d}} = \rep{1} , 
\end{equation}
and it is precisely this singlet that is constrained by condition~\eqref{eq:V-Odd}. 

We can also consider the case of compatible $\SO{d,d}$ hyper- and vector-multiplet structures, defining as before
\begin{defn}
The two structures $J_\alpha$ and $K$ are \emph{compatible} if together they define an $\SU2\times\SO{d-5,d}\subset\SO{d,d}\times\bbR^+$ generalised structure. The necessary and sufficient conditions are
\begin{equation}
   J_\alpha \cdot K = 0 , \qqqq
   \tr (J_\alpha J_\beta) = - \eta(K,K)\delta_{\alpha\beta} ,
\end{equation}
where $\cdot$ is the adjoint action. 
\end{defn}
\noindent
The intrinsic torsion, decomposing under $\SU2\times\SU2\times\SO{d-5,d}$ where the first $\SU2$ factor is generated by $J_\alpha$,  is given by
\begin{equation}
   \Tint^{\SU{2}\times\SO{d-5,d}}
      =(\rep{2},\rep{1},\rep{1})
           + (\rep{3},\rep{1},\rep{2d-5}) 
           + (\rep{1},\rep{1},\rep{1}) ,
\end{equation}
and we see that the separate integrability of $J_\alpha$ and $K$ is enough to ensure the integrability of the $\{J_\alpha,K\}$ structure. 

It is interesting to note that in all our examples, \VM{} structures (and the \TM{} structure of section~\ref{sec:E6_structures}) were based on coset spaces that were prehomogeneous vector spaces (PVS), that is, were open orbits over the relevant generalised geometry structure group acting on a particular representation. This property will be true of any structure associated to a multiplet in $D$ dimensions that contains form-fields, such as vector- or tensor-multiplets, because the form-field degrees of freedom form vector spaces under the action of the structure group $G_\text{frame}$. It is interesting, therefore, following Hitchin~\cite{Hitchin05}, to look at the list of possible PVSs based on the split real forms of the simply laced groups to see which cases have been covered. 

A PVS can be one of two types~\cite{SK77}. Let $R$ be the representation space on which a group $\Gf$ acts. One then defines 
\begin{quote}
\begin{itemize}
\item[type 1:]
$\Gf\times\SL{n;\bbR}$ acting on $R\otimes\bbR^n$ gives an open dense orbit, 
\item[type 2:]
$\Gf\times\GL{n;\bbR}$ acting on $R\otimes\bbR^n$ gives an open dense orbit (and $(\Gf,R)$ is not type~1).  
\end{itemize}
\end{quote}
For type 2 there is always a polynomial in $R$ that is invariant under $\Gf$. To match the three basic generalised geometries we need $\Gf\in\{\SL{d;\bbR},\SO{d,d},\Ex{d(d)}\}$ and an $\bbR^+$ factor. This implies $(\Gf,R)$ is of type 2 with $n=1$. (In fact, in the next section we will see that a case with $n=2$ can also be physically interesting.) The list of possibilities, ignoring those with trivial stabiliser group $G$, is given in table~\ref{tab:PVS1}. We note that other than case~(a), which is simply a conventional metric, all the other cases of structures have already been considered in the literature. Note that the very special real geometry in case~(b) encodes $D=5$ vector multiplets from M-theory on an $\SU3$-structure manifold, the special K\"ahler geometry in case~(c) encodes $D=4$ vector multiplets for $\SU3$-structure reductions of type II, and similarly the special K\"ahler geometry in case~(g) encodes $D=4$ vector multiplets for $\SU{3,3}$ structures in type II. 

\begin{table}
\centering
\begin{tabular}{@{}rllll@{}} 
\toprule
& $\Gf/G$ & R  & geometry of $\Gf\times\bbR^+/G$ & \\
\midrule
(a) & $\GL{d;\bbR}/\SO{d;\bbR}$ & $S^2\bbR^n$ 
    & & \\
(b) & $\GL{2n;\bbR}/\Symp{2n;\bbR}$ & $\ext^2\bbR^{2n}$ 
    & very special real ($n=3$) & \cite{Hitchin01} \\
(c) & $\GL{6;\bbR}/\SL{3;\bbC}$ & $\ext^3\bbR^6$ 
    & special K\"ahler & \cite{Hitchin00}\\
(d) & $\GL{7;\bbR}/G_2$ & $\ext^3\bbR^6$ 
    & & \cite{Hitchin00a}\\
(e) & $\GL{8;\bbR}/\SU3$ & $\ext^3\bbR^6$ 
    & & \cite{Hitchin01} \\
(f) & $\SO{d,d}/\SO{d-1,d}$ & $\bbR^{2d}$ 
    & flat & \S\ref{sec:E6_structures}, \S\ref{sec:PVS}, \\
(g) & $\Spin{6,6}/\SU{3,3}$ & $\ext^\pm\bbR^6$ 
    & special K\"ahler & \cite{Hitchin02}, \cite{Gualtieri04} \\
(h) & $\Spin{7,7}/G_2\times G_2$ & $\ext^+\bbR^7$ 
    & & \cite{Witt06} \\
(i) & $\Ex{6(6)}/\Fx{4(4)}$ & $\rep{27}$ 
    & very special real & \S\ref{sec:E6_structures} \\
(j) & $\Ex{7(7)}/\Ex{6(2)}$ & $\rep{56}$ 
    & special K\"ahler &  \S\ref{sec:V-structures} \\
\bottomrule
\end{tabular}
\protect\caption{The PVSs that arise in $\SL{d;\bbR}$, $\SO{d,d}$ and $\Ex{d(d)}$ generalised geometry. The geometry of $\Gf\times\bbR^+/G$ is listed whenever it corresponds to that of a vector or tensor multiplet. We also reference the paper or section of this paper where the corresponding (generalised) structure is first discussed.}
\label{tab:PVS1}
\end{table}

It turns out that the integrability conditions for the corresponding (generalised) structures follow a standard pattern. As we have already noted for the cases discussed here, the invariant polynomial on the PVS can be used to define a Hitchin functional $H$, as first described in~\cite{Hitchin00,Hitchin00a,Hitchin02,Witt06,Hitchin05}. If we denote the structure $K\in\Gamma(N)$, where $N$ is the generalised tensor bundle corresponding to the representation $R$, then varying the functional one can always define a second invariant structure 
\begin{equation}
   K' = \delta H \in \Gamma(\det T^*M\otimes N^*), 
\end{equation}
where $\delta$ is the exterior derivative on the infinite-dimensional space of structures. (For example, from~\eqref{eq:hatK-def}, for $\Ex{7(7)}/\Ex{6(2)}$ we note that $K'$ is just $-\imath_{\hat{K}}\Omega$.) There are then two possibilities, depending on whether $N$ is the generalised tangent bundle $E$ or not, 
\begin{equation}
\label{eq:PVS-integrability}
\begin{aligned}
   \text{type A, $N\neq E$:}& & && &&
      \dd K=\dd K'=0 , \\
   \text{type B, $N=E$:}& & && &&
      \Dorf_KK=0 . 
\end{aligned}
\end{equation}
Recall that type~B are the vector-multiplet structures. Note that for all the examples where $N\neq E$ one can define an appropriate covariant generalised geometry notion of the action of the exterior derivative taking sections of $N$ to sections of some other generalised tensor bundle, something which is not possible for generic generalised tensor bundles $N$. Note also that in certain cases the $\dd K'=0$ condition is implied by the $\dd K$ condition. For example in the symplectic structures of case~(b), one has $K=\omega$ and $K'=\frac{1}{(n-1)!}\omega^{n-1}$, so that $\dd K=0$ implies $\dd K'=0$. 

For completeness, let us note that PVS structures have also been considered in the $\SO{d+1,d}\times\bbR^+$ generalised geometry~\cite{Baraglia12} that describes NS-NS degrees of freedom coupled to a single $\Uni1$ gauge field. These possibilities are listed in table~\ref{tab:PVS2}. Hitchin~\cite{Hitchin05} has also considered the type 2, $n=2$ case $\SO{5,5}\times\GL{2;\bbR}/\Gx{2(2)}\times\SL{2;\bbR}$ with $R=\ext^*\bbR^5\otimes\bbR^2$. 

\begin{table}[h]
\centering
\begin{tabular}{@{}rlll@{}} 
\toprule
& $\Gf/G$ & R  & \\
\midrule
(a)$'$ & $\SO{d+1,d}/\SO{d,d}$ & $\bbR^{2d+1}$ 
    & \\
(b)$'$ & $\SO{4,3}/\Gx{2(2)}$ & $\ext^*\bbR^3$ 
    & \cite{Rubio13} \\
(c)$'$ & $\Spin{5,4}/\Spin{4,3}$ & $\ext^*\bbR^4$ 
    &   \\
(d)$'$ & $\Spin{6,5}/\SU{3,2}$ & $\ext^*\bbR^5$ 
    & \cite{Rubio14} \\
\bottomrule
\end{tabular}
\protect\caption{The PVSs that arise in $\SO{d+1,d}$ generalised geometry.}
\label{tab:PVS2}
\end{table}


\subsection{Examples}
\label{sec:NS5}

Physically, we are interested in those structures that actually describe supersymmetric backgrounds in the NS-NS sector of type II supergravity. The presence of hypermultiplet structures means we are interested in backgrounds preserving eight supercharges. These have been analysed in~\cite{Strominger86,GMW03}. The ones of interest all preserve type II Killing spinors of only one type: they have $\nabla^+$ special holonomy in the language of~\cite{GMW03}. As such they are not described by generalised complex structures. 

The simplest case is the $D=6$, $\mathcal{N}=(1,0)$ background of~\cite{Strominger86}, corresponding to an NS5-brane at a point in a hyper-K\"ahler space, as described in section~\ref{sec:NS5-IIB}. In this case, the Killing spinor is stabilised by an $\SU2\times\SO4\subset\SO4\times\SO4\subset\SO{4,4}$ subgroup. However this corresponds to an $\SO{4,4}$ \HM{} structure $J_\alpha$. To see this, we can reduce the $\Ex{5(5)}$ description given in section~\ref{sec:NS5-IIB} to the $\SO{4,4}$ subgroup, generated by $(r,B^1,\beta^2)$ in~\eqref{eq:IIB_adjoint}. As an element of the $\SO{4,4}$ adjoint representation, the $\Ex{5(5)}$ hypermultiplet structure reduces to 
\begin{equation}
   \tilde{J}_{\alpha}
     = - \tfrac{1}{2}\kappa I_{\alpha}
        - \tfrac{1}{2}\kappa\omega_{\alpha}
        - \tfrac{1}{2}\kappa\omega_{\alpha}^{\sharp},
\end{equation}
with $\kappa^{2}=\ee^{-2\phi}\vol_{4}$. From the $\SO{4,4}$ truncation of the $\Ex{5(5)}$ adjoint algebra~\eqref{eq:IIB_adjoint} and Killing form~\eqref{eq:IIB_Killing}, it is easy to check the $\su{2}$ algebra~\eqref{eq:su2_algebra} and normalisation~\eqref{eq:J_a_norm}. Including the NS-NS two-form leads to the twisted object $J_\alpha=\ee^B\tilde{J}_\alpha\ee^{-B}$ and the integrability conditions are just the standard vanishing moment maps. No other structures are needed to define the supersymmetric background. This reduction to $\SO{4,4}$ explains why in section~\ref{sec:NS5-IIB} all the differential conditions on the $\SU2$ structure $\omega_\alpha$ arose from the moment maps, with the conditions on $Q$ being satisfied identically. 

For $\mathcal{N}=1$, $D=5$ the metric has the form 
\begin{equation}
\label{eq:NS3-5d} 
\dd s^{2}= \dd \tilde{s}^{2}(M_{\SU2}) + \zeta^2 , 
\end{equation}
where the metric $\dd \tilde{s}^{2}(M_{\SU2})$ admits an $\SU2$ structure defined by a triplet of two-forms $\omega_\alpha$ as above, and $\zeta$ is a one-form. The conditions for supersymmetry are given in~\cite{GMW03}. The Killing spinor is stabilised by an $\SU2\times\SO5\subset\SO{5,5}$ subgroup, which from the discussion above we see corresponds to a pair of compatible structures $\{J_\alpha,K\}$. The untwisted objects then take the form 
\begin{equation}
\begin{aligned}
   \tilde{J}_{\alpha}
     &= - \tfrac{1}{2}\kappa I_{\alpha}
        - \tfrac{1}{2}\kappa\omega_{\alpha}
        - \tfrac{1}{2}\kappa\omega_{\alpha}^{\sharp}, \\
   \tilde{K} &= \kappa \zeta^\sharp + \kappa \zeta , 
\end{aligned}
\end{equation}
with $\kappa^{2}=\ee^{-2\phi}\vol_{4}$, and twisted structures $J_\alpha=\ee^B\tilde{J}_\alpha\ee^{-B}$ and $K=\ee^B\tilde{K}$. The integrability conditions~\eqref{eq:mm_Odd} and~\eqref{eq:V-Odd} are equivalent to the Killing spinor equations. 

The $\mathcal{N}=2$, $D=4$ case includes an interesting new twist. We need to extend the $\SO{6,6}\times\bbR^+$ generalised tangent space to include the six-form potential $\tilde{B}$, dual to $B$ in $D=4$, and also the ``dual graviton''. This means taking 
\begin{equation}
   E \simeq TM \oplus T^*M \oplus \ext^5 T^*M \oplus (T^*M\otimes\ext^6 T^*M) , 
\end{equation}
which is a truncation of the $\Ex{7(7)}$ type II generalised tangent bundle~\eqref{eq:E-typeII} to the NS-NS sector, keeping $(v,\lambda^1,\sigma^2,\tau)$ in~\eqref{eq:V-IIB}. The subgroup of $\Ex{7(7)}$ acting on $E$ is then 
\begin{equation}
\label{eq:GframeSL2}
   G_\text{frame} = \SL{2;\bbR}\times\SO{6,6}\times\bbR^+ ,
\end{equation}
such that $E$ transforms in the $(\rep{2},\rep{12})_{\rep{1}}$ representation. The generalised metric defines an $\SO2\times\SO6\times\SO6$ structure, where the $\SL{2;\bbR}/\SO2$ coset encodes the NS-NS four-dimensional axion-dilaton. The adjoint representation is generated by $(r,B^1,\beta^2,\tilde{a}^2,\tilde{\alpha}^1)$ in~\eqref{eq:IIB-adj}. The $\Ex{7(7)}$ generalised geometry of M5-brane backgrounds was described in section~\ref{sec:Hypermultiplet-structures} and we can simply truncate the results there to $\SL{2;\bbR}\times\SO{6,6}$. The metric, truncated from~\eqref{eq:M5-metric}, takes the form 
\begin{equation}
\dd s^{2}= \dd \tilde{s}^{2}(M_{\SU2}) + \zeta_1^2 + \zeta_2^2 , 
\end{equation}
with a pair of one-forms $\zeta_i$. From~\eqref{eq:J-M5R2} and~\eqref{eq:T-M5R2}, we find the hyper- and vector-multiplet structures are
\begin{equation}
\begin{aligned}
   \tilde{J}_{\alpha}
     &= - \tfrac{1}{2}\kappa I_{\alpha}
        - \tfrac{1}{2}\kappa\omega_{\alpha}
        - \tfrac{1}{2}\kappa\omega_{\alpha}^{\sharp}, \\
   \tilde{K} &= \zeta_1^\sharp + \zeta_1 - \zeta_1 \wedge \vol_4 
       + \zeta_2\otimes \vol_6 , 
\end{aligned}
\end{equation}
with $\kappa^{2}=\ee^{-2\phi}\vol_{4}$, and twisted structures $J_\alpha=\ee^B\tilde{J}_\alpha\ee^{-B}$ and $K=\ee^B\tilde{K}$. Note that in deriving $\tilde{K}$ we take the real part of $\tilde{X}$ in~\eqref{eq:T-M5R2}. As in section~\ref{sec:Integrability}, the integrability conditions are simply
\begin{equation}
   \mu_{\alpha}(V) = 0 
      \qquad\text{for all $V\in\Gamma(E)$} , \qqq
   \Dorf_K K =0 , \qqq
   \Dorf_K J_\alpha = 0 .  
\end{equation}
One can again show that these conditions are equivalent to those in~\cite{GMW03}. 

The vector-multiplet structure $K$ is still associated to a PVS. However, from the form of the structure group~\eqref{eq:GframeSL2} we see that we are now interested in type 2 structures with $n=2$. The relevant stabiliser groups $G\subset\SL{2;\bbR}\times\SO{6,6}\times\bbR^+$ are now 
\begin{equation}
\begin{aligned}
   &\text{\textit{hypermultiplet structure}, $J_\alpha$} && &
      G &= \SL{2;\bbR}\times\SU2\times\SO{2,6} , \\
   &\text{\textit{vector-multiplet structure}, $K$} && & 
      G &= \SO2\times\SO{4,6} . 
\end{aligned}   
\end{equation}
with the common subgroup $\SO2\times\SU2\times\SO6$. Note also that the space of vector-multiplet structures admits a special K\"ahler metric. 


\section{Discussion}\label{sec:Discussion}

In this paper we have given a new geometrical interpretation of generic flux backgrounds in type II supergravity and M-theory, preserving eight supercharges in $D=4,5,6$ Minkowski spacetime, as integrable $G$-structures in $\Ex{d(d)}\times\mathbb{R}^{+}$ generalised geometry. As in conventional geometry, integrability was defined as the existence of a generalised torsion-free connection that is compatible with the structure, or equivalently as the vanishing of the generalised intrinsic torsion, defining what we called an ``exceptional Calabi--Yau'' (\HV{}) space. 

We found the differential conditions on the structures implied by integrability, and showed that they took a simple form in terms of the generalised Lie derivative or moment maps for the action of the generalised diffeomorphism group. As with Calabi--Yau backgrounds, supersymmetric solutions are described as the intersection of two separate structures that can be associated to hypermultiplet and vector-multiplet degrees of freedom in the corresponding gauged supergravity. We saw how the simple examples of Calabi--Yau, generalised Calabi--Yau and hyper-K\"ahler structures appear in our formalism, as well as various other simple supersymmetric flux backgrounds. We also discussed the corresponding structures in $\Orth{d,d}\times\bbR^+$ generalised geometry, which capture the geometry of NS5-brane backgrounds. Finally, using the classification of prehomogeneous vector spaces we identified all the possible generalised structures associated to vector and tensor multiplets. 

There are many directions for future study. The extension to AdS backgrounds will be considered in a companion paper~\cite{AW15b}. The other obvious extension is to find the analogous structures for backgrounds with different amounts of supersymmetry. In $\Ex{d(d)}\times\bbR^+$ generalised geometry the supersymmetry parameters transform under the maximal compact subgroup $\Hd{d}$. As conjectured in~\cite{CSW14b}, supersymmetric backgrounds preserving $\mathcal{N}$ supersymmetries  should be given by integrable $G$-structures where $G\subset H_d$ is the stabiliser group of the $\mathcal{N}$ Killing spinors. Thus, for example, $D=4$, $\mathcal{N}=1$ backgrounds define an $\SU7\subset\SU8$ structure~\cite{PW08}. We have seen here that the structures are naturally associated to multiplets in the $D$-dimensional theory, and furthermore that the integrability conditions can be deduced from the standard form of the $D$-dimensional gauged supergravity. This should provide a relatively simple prescription for deriving the conditions for other examples. 

A very important question both for phenomenology and the AdS/CFT correspondence is to identify the deformations of the structures. We discussed some general properties of the moduli spaces, notably that they arise as hyper-K\"ahler and symplectic quotients. However, there are several open questions, in particular to understand how the moduli space of \HV{} structures splits into a product of a hyper-K\"ahler space and a special K\"ahler space. On general grounds, one expects that the deformation problem is described by some underlying differential graded Lie algebra (DGLA) with cohomology classes capturing the first-order deformations and obstructions, as described for example in~\cite{Manetti04}. For example, for \HM{} structures in IIA, this would be some generalisation of the Dolbeault complex. Such extensions appear in generalised complex geometry~\cite{Gualtieri04,Gualtieri04a,Cavalcanti07}, but this would go further to include R-R degrees of freedom. In the generalised complex structure case, starting from a conventional complex structure, it is known that the extra deformations can be associated with gerbe and non-commutative deformations of the algebraic geometry~\cite{Gualtieri04,Kapustin04}. An open question is how to understand the corresponding R-R deformations when they exist. In the AdS context, solving the deformation problem gives a way of finding the exactly marginal deformations in the dual field theory. We will take some steps in this direction in a forthcoming paper~\cite{AGGPW15}.

As we have seen, the integrability of both the \HM{} and \VM{} structures are captured by moment maps. Typically this is closely allied to notions of stability (see for example~\cite{Thomas06}), which, if they exist, would here define integrable complex or symplectic structures (and their generalisations) under the action of some quaternionic version of the full  generalised diffeomorphism group. A natural question is also to understand the reduction of structures, as has been done for generalised complex geometry~\cite{BCG06}: how, given a generalised Killing symmetry, structures with eight supercharges on $M$ define structures on $\tilde{M}$ of one dimension lower. Physically this would realise the r-map of~\cite{WV91}. 

Conventional generalised complex geometry is known to capture the A and B topological string models on backgrounds with $H$-flux~\cite{Kapustin04,KL07,PW05}. The geometries defined here should encode some extension to M-theory or with the inclusion of R-R flux. It was previously proposed~\cite{DGNV05,Nekrasov04} that the relevant topological M-theory was related to Hitchin's formulation of $\Gx{2}$ structures~\cite{Hitchin00a}, which combines both the A and B model. Here we have a slightly different picture with two candidate structures in M-theory. Note that in principle either the \HM{} structures or the \VM{} structures could be viewed as generalisations of the A and B models, with mirror symmetry mapping \HM{} (or \VM{}) structures in IIA to \HM{} (or \VM{}) structures in IIB. However, the integrability conditions on the \VM{} structure are considerably weaker -- for example, for a generalised complex structure they do not imply that $\dd\Phi^\pm=0$. In this case it would appear one would need to choose a fixed background $J_\alpha$ and also impose the $\Dorf_X J_\alpha$ condition. The hypermultiplet structure integrability, on the other hand, does imply $\dd\Phi^\pm=0$, and hence these give the natural candidates for generalisations of the topological string models. It would be particularly interesting to consider the quantisation of these models, as in~\cite{PW05} though now with a hyper-K\"ahler rather than symplectic space of structures. 


\subsection*{Acknowledgements}

We would like to thank Charles Strickland-Constable, Michela Petrini and Mariana Gr\~{a}na for helpful discussions. AA is supported by an EPSRC PhD Studentship. DW is supported by the STFC grant ST/J000353/1 and the EPSRC Programme Grant EP/K034456/1 ``New Geometric Structures from String Theory''. This work has been a long time in the making and DW gratefully acknowledges support during various stages of this research from the Banff International Research Station; the Simons Center for Geometry and Physics, Stony Brook University; the Galileo Galilei Institute for Theoretical Physics, Florence; the Mainz Institute for Theoretical Physics, Johannes Gutenberg-University Mainz; the Fondation Math\'ematique Jacques Hadamard
and the Institut de Physique Th\'eorique, CEA-Saclay; and especially the Berkeley Center for Theoretical Physics at UC Berkeley. 


\appendix


\section{Notation}

Our notation follows \cite{CSW14}. Wedge products and contractions are given by
\begin{equation}
\begin{split}(v\wedge u)^{a_{1}\dots a_{p+p'}} & \coloneqq\frac{(p+p')!}{p!p'!}v^{[a_{1}\dots}u^{a_{p+1}\dots a_{p+p'}]},\\
(\lambda\wedge\rho)_{a_{1}\dots a_{q+q'}} & \coloneqq\frac{(q+q')!}{q!q'!}\lambda_{[a_{1}\dots a_{q}}\rho_{a_{q+1}\dots a_{q+q'}]},\\
(v\lrcorner\lambda)_{a_{1}\dots a_{q-p}} & \coloneqq\frac{1}{p!}v^{b_{1}\dots b_{p}}\lambda_{b_{1}\dots b_{p}a_{1}\dots a_{q-p}}\quad\text{if }p\leq q,\\
(v\lrcorner\lambda)^{a_{1}\dots a_{p-q}} & \coloneqq\frac{1}{q!}v^{a_{1}\dots a_{p-q}b_{1}\dots b_{q}}\lambda_{b_{1}\dots b_{q}}\quad\text{if }p\geq q,\\
(jv\lrcorner j\lambda)_{\phantom{a}b}^{a} & \coloneqq\frac{1}{(p-1)!}v^{ac_{1}\dots c_{p-1}}\lambda_{bc_{1}\dots c_{p-1}},\\
(j\lambda\wedge\rho)_{a,a_{1}\ldots a_{d}} &\coloneqq\frac{d!}{(q-1)!(d+1-q)!}\lambda_{a[a_{1}\ldots a_{q-1}}\rho_{a_{q}\ldots a_{d}]}.
\end{split}
\end{equation}
Given a basis $\{\hat{e}_{a}\}$ for $TM$ and a dual basis $\{e^{a}\}$ for $T^{*}M$, there is a natural $\gl{d}$ action on tensors. For example, the action on a vector and a three-form is
\begin{equation}
(r\cdot v)^{a}=r_{\phantom{a}b}^{a}v^{b},\qquad(r\cdot\lambda)_{abc}=-r_{\phantom{d}a}^{d}\lambda_{dbc}-r_{\phantom{d}b}^{d}\lambda_{adc}-r_{\phantom{d}c}^{d}\lambda_{abd}.
\end{equation}
When writing the components of generalised tensors, we sometimes use the notation that $(\ldots)_{(p)}$ and $(\ldots)^{(q)}$ denote $p$-form and $q$-vector components respectively. For a $p$-form $\rho$, we denote by $\rho^{\sharp}$ the $p$-vector obtained by raising the indices of $\rho$ using the conventional metric on the manifold.

We define the Hodge star as
\begin{equation}
\star e^{a_1 \ldots a_p}=\frac{1}{q!} \epsilon^{a_1 \ldots a_p}_{\phantom{a_1 \ldots a_p}b_1 \ldots b_q} e^{b_1 \ldots b_q}.
\end{equation}
With a Euclidean metric we have $\epsilon_{1\ldots d}=\epsilon^{1\ldots d}=1$, so that $\star 1 = \vol$ and $\star \vol =1$. With a mostly plus Minkowski metric we have $\epsilon_{0\ldots d-1}=-\epsilon^{0\ldots d-1}=1$, so that $\star 1 = \vol$ and $\star \vol =-1$. In particular this choice implies
\begin{equation}
(\lambda^\sharp\lrcorner\rho)\vol_d = \rho\wedge\star\lambda.
\end{equation}


\section{Examples of \texorpdfstring{$\mathcal{N}=2$}{N=2}, \texorpdfstring{$D=4$}{D=4} backgrounds}\label{sec:examples}

In this appendix, we shall summarise a number of simple $\mathcal{N}=2$ backgrounds in both type II and M-theory, with and without fluxes. We use these to provide concrete examples of $\Ex{7(7)}$ structures in section~\ref{sec:H-examples} and to show how the usual supersymmetry conditions are recovered from integrability conditions in section~\ref{sec:int_examples}.

\subsection{Calabi--Yau manifolds in type II and \texorpdfstring{$\SU{3}$}{SU(3)} structures}

Calabi--Yau manifolds admit a single covariantly constant spinor $\chi^+$ defining an $\SU3\subset\Spin6\simeq\SU4$ structure. Equivalently it admits a symplectic form $\omega$ and a holomorphic three-form $\Omega$ that are compatible. One can choose a frame $\{e^a\}$ for the metric on $M$ where these take the form 
\begin{equation}
\label{eq:SU3-structure}
\omega=e^{12}+e^{34}+e^{56},
\qqq
\Omega=(e^{1}+\ii e^{2})\wedge(e^{3}+\ii e^{4})\wedge(e^{5}+\ii e^{6}),
\end{equation}
where $e^{mn}=e^m\wedge e^n$. Raising an index on $\omega$ defines an almost complex structure $I$ on the six-dimensional space
\begin{equation}
\label{eq:acs}
I_{\phantom{m}n}^{m}=-\omega_{\phantom{m}n}^{m}
   =\tfrac{1}{8}\ii(\bar{\Omega}^{mpq}\Omega_{npq}
      -\Omega^{mpq}\bar{\Omega}_{npq}), \qqq
     I^q_{\phantom{q}m} \Omega_{qnp}= \ii \Omega_{mnp}.
\end{equation}
By construction, the forms satisfy the compatibility conditions~\eqref{eq:SU3-consistency}. In the language of $G$-structures, $\omega$ and $\Omega$ define $\Symp{6;\bbR}$ and $\SL{3;\bbC}$ structures respectively. The consistency conditions~\eqref{eq:SU3-consistency} imply that the common subgroup is given by $\Symp{6;\bbR}\cap\SL{3;\bbC}=\SU3$. The fact that $\chi$ is covariantly constant is equivalent to the integrability conditions 
\begin{equation}
   \dd\omega = 0 , \qqq \dd\Omega = 0 . 
\end{equation}


\subsection{\texorpdfstring{$\text{CY}_3 \times \text{S}^1$}{CY3 x S1} in M-theory}

Let us also briefly note the form of the M-theory lift of the type IIA Calabi--Yau background. The seven-dimensional internal space is a product $M=M_{\SU3}\times \text{S}^1$ with metric 
\begin{equation}
   \dd s^2(M) = \dd s^2(M_{\SU3}) + \zeta^2 ,
\end{equation}
where $\zeta=\dd y$, with $y$ the coordinate on the M-theory circle, and $\dd s^2(M_{\SU3})$ is the IIA Calabi--Yau metric on $M_{\SU3}$. The Killing spinors take the same form as \eqref{eq:CY-KS} but are now viewed as complex $\Spin7$ spinors. They again determine an $\SU3$ structure, which can equivalently be defined by the triplet of forms $\{\omega,\Omega,\zeta\}$. If we raise an index to define the vector $\zeta^\sharp=\partial_y$, we have the compatibility conditions
\begin{equation}
   \tfrac{1}{3!}\omega\wedge\omega\wedge\omega 
      = \tfrac{1}{8}\ii \Omega\wedge\bar{\Omega} , \qqq
   \omega\wedge\Omega = 0 , \qqq
   \imath_{\zeta^\sharp} \omega = 0 , \qqq
   \imath_{\zeta^\sharp} \Omega = 0 , 
\end{equation}
and the integrability conditions 
\begin{equation}
   \dd\omega = 0 , \qqq
   \dd\Omega = 0 , \qqq
   \dd \zeta = 0 . 
\end{equation}
Note that they imply $\zeta^\sharp$ is a Killing vector.


\subsection{Generalised Calabi--Yau metrics in type II and pure spinors}
\label{sec:GCY}

Returning to type II, we now consider the case where all NS-NS fields are non-zero, that is, we include non-trivial $H=\dd B$ flux and dilaton, and the Killing spinors take the form~\eqref{eq:KS-genCY}. The background can be characterised using $\Orth{d,d}\times\bbR^+$ generalised geometry as follows~\cite{GMPT05}. 

The generalised tangent bundle $E\simeq TM\oplus T^*M$ admits a natural $\Orth{d,d}$ metric $\eta$. The background is defined by two complex polyforms, taking $d=6$, 
\begin{equation}
   \Phi^\pm \in \Gamma(\ext^\pm T^*M) , 
\end{equation}
which can then be viewed as sections of the positive and negative helicity $\Spin{6,6}\times\bbR^+$ spinor bundles, where the $\bbR^+$ factor acts by a simple rescaling. The generalised spinors are not generic but are ``pure'' meaning they are stabilised by an $\SU{3,3}\subset\Spin{6,6}$ subgroup. They also satisfy the consistency conditions 
\begin{equation}
\label{eq:Phi-consistency}
   \langle\Phi^+,\bar{\Phi}^+ \rangle 
       = \langle\Phi^-,\bar{\Phi}^- \rangle ,
   \qqq
   \langle\Phi^{+},V\cdot\Phi^{-}\rangle = \langle\bar{\Phi}^{+},V\cdot\Phi^{-}\rangle = 0 \quad \forall V ,
\end{equation}
where, given $V=\xi+\lambda\in\Gamma(TM\oplus T^*M)$, one defines the Clifford action $V\cdot\Phi^\pm=V^A\Gamma_A\Phi^\pm
=\imath_\xi\Phi^\pm+\lambda\wedge\Phi^\pm$. In addition, $\langle\cdot,\cdot\rangle$ is the $\Spin{6,6}$-invariant spinor bilinear, or Mukai pairing, given by 
\begin{equation}
   \langle\Psi,\Sigma \rangle
      = \sum_p (-)^{[(p+1)/2]}\Psi_{(p)}\wedge\Sigma_{(6-p)} , 
\end{equation}
where $\Psi_{(p)}$ denotes the $p$-form component of $\Psi$ and $[p]$ is the integer part of $p$. 

Each pure spinor defines an (almost) generalised complex structure $\mathcal{J}^\pm\in\Gamma(\ad\tilde{F})$, where $\ad\tilde{F}\simeq\Gamma((TM\otimes T^{*}M) \oplus \ext^{2}T^*M \oplus \ext^{2}TM)$ is the principle $\Orth{6,6}$ frame bundle for $E$. The generalised complex structures are given by  
\begin{equation}
\label{eq:GCS-J-def}
\mathcal{J}_{\phantom{\pm A}B}^{\pm A} 
= \ii\frac{\langle\Phi^{\pm},\Gamma_{\phantom{A}B}^{A} 
	\bar{\Phi}^{\pm}\rangle}{\langle\Phi^{\pm}, 
	\bar{\Phi}^{\pm}\rangle},
\end{equation}
where $\Gamma^A$ are $\Orth{6,6}$ gamma matrices with $A=1,\ldots,12$, and indices are raised and lowered using the $\Orth{6,6}$ metric. Note that acting on pure spinors it has the property 
\begin{equation}
   \tfrac{1}{4}\mathcal{J}_{AB}^{\pm}\Gamma^{AB}\Phi^{\pm}
       = 3\ii\Phi^{\pm}.
\end{equation}

The integrability conditions are 
\begin{equation}
   \dd \Phi^+ = 0 , \qqq \dd \Phi^- = 0 ,
\end{equation}
which define what is known as a generalised Calabi--Yau metric~\cite{Gualtieri04}. These conditions imply that each almost generalised complex structure is separately integrable. Each is also equivalent to the existence of a torsion-free generalised connection compatible with the $\SU{3,3}_\pm$ structure defined by $\Phi^\pm$. 


\subsection{D3-branes on \texorpdfstring{$\text{HK} \times \bbR^2$}{HK x R2} in type IIB}
\label{sec:D3-bkgd}

Let us now turn to three further flux examples. The first corresponds to D3-branes in type IIB at a point in a space $M=M_{\SU2}\times\bbR^2$, where $M_{\SU2}$ is a four-dimensional hyper-K\"ahler space. This is in the class of the solutions first given in~\cite{GP01,Gubser00,GP02} and analysed in detail for $M=M_{\SU2}\times\bbR^2$ in~\cite{BCFFST01}. We have a conformal factor $\Delta$ and an R-R five-form flux $F$, and in general also an imaginary self-dual three-form flux. The metric on $M$ takes the form
\begin{equation}
\label{eq:D3-metric}
\dd s^{2}= \dd \tilde{s}^{2}(M_{\SU2}) + \zeta_1^2 + \zeta_2^2 , 
\end{equation}
where $\dd \tilde{s}^{2}(M_{\SU2})$ is an $\SU2$-structure metric on $M_{\SU2}$ and
\begin{equation}
   \zeta_1 = \ee^{-\Delta} \dd x , \qqq
   \zeta_2 = \ee^{-\Delta} \dd y . 
\end{equation}
The type IIB axion-dilaton $\tau=A+\ii\ee^\phi$ is constant, and for convenience we take $\tau=\ii$. 

The two $\SU8$ Killing spinors take the form
\begin{equation}
\label{eq:D3-KS}
   \epsilon_1 = \begin{pmatrix} \chi_1^+ \\ -\ii\chi_1^- \end{pmatrix} , 
   \qqq
   \epsilon_2 = \begin{pmatrix} \chi_2^+ \\ -\ii\chi_2^- \end{pmatrix} .
\end{equation}
The two spinors $\chi_i^+$ define a conventional $\SU2$ structure, which is simply the one defined by the hyper-K\"ahler geometry. It is determined by a triplet of symplectic forms $\omega_{\alpha}$ and the pair of one-forms $\{\zeta_1,\zeta_2\}$. One can always choose a frame $\{e^a\}$ for the metric on $M$ where these take the form 
\begin{equation}
\label{eq:SU2-structure}
\begin{gathered}
   \omega_{1}=e^{14}+e^{23}, \qqq
   \omega_{2}=e^{13}-e^{24}, \qqq
   \omega_{3}=e^{12}+e^{34}, \\
   \zeta_1 = e^5 , \qqq 
   \zeta_2 = e^6 . 
\end{gathered}
\end{equation}
The corresponding triplet of complex structures is given by  $(I_{\alpha})_{\phantom{m}n}^m=-(\omega_{\alpha})_{\phantom{m}n}^{m}$, such that for $\Omega=\omega_2 + \ii \omega_1$, we have $(I_3)^p_{\phantom{p}m}\Omega_{pn}=\ii \Omega_{mn}$. The volume form on $M$ is defined by
\begin{equation}
\tfrac{1}{2}\omega_{\alpha}\wedge\omega_{\beta}
    \wedge \zeta_{1}\wedge \zeta_{2} = \delta_{\alpha\beta} \vol_6 .
\end{equation}

If we include only five-form flux, the integrability conditions for the structure are
\begin{equation}
   \dd(\ee^{\Delta}\zeta_i) = 0 , \qqq
   \dd(\ee^{2\Delta}\omega_\alpha) = 0 , \qqq
   \dd\Delta = -\tfrac{1}{4}\star F ,
\end{equation}
where $F$ is the component of the five-form flux on $M$ and $\star$ is the six-dimensional Hodge duality operator, calculated using the metric~\eqref{eq:D3-metric}. If one also includes a non-zero three-form flux on $M$, it has to have the form~\cite{BCFFST01} 
\begin{equation}
H+\ii F_3 = \dd \gamma_{I}(z)\wedge\tau_I,
\end{equation}
where $\gamma_I(z)$ are analytic functions of $z=x+\ii y$ and $\tau_I$ are harmonic anti-self-dual two-forms on the hyper-K\"ahler space.  The functions $\gamma_I(z)$ are constrained by a differential equation arising from the Bianchi identity for $F$. 


\subsection{Wrapped M5-branes on \texorpdfstring{$\text{HK} \times \bbR^3$}{HK x R2} in M-theory}
\label{sec:M5-bkgd}

For our final two examples, we consider the M-theory backgrounds corresponding to wrapped M5-branes in a seven-dimensional geometry that is a product of a four-dimensional hyper-Kähler space with $\bbR^3$. There are two possibilities: the branes can either wrap a K\"ahler two-cycle in the four-dimensional hyper-Kähler space or wrap an $\bbR^2$ plane in $\bbR^3$. In each case, the spacetime is a product $M=M_{\SU2}\times\bbR^3$ with the metric 
\begin{equation}
\label{eq:M5-metric}
   \dd s^{2}= \dd \tilde{s}^{2}(M_{\SU2})
      + \zeta_1^2 + \zeta_2^2 + \zeta_3^2 ,
\end{equation}
where $M_{\SU{2}}$ admits an $\SU2$ structure, $\dd \tilde{s}^{2}(M_{\SU2})$ is the metric determined by the structure, $\zeta_i$ are one-forms, and there is a non-trivial four-form flux $F$. Crucially, because of the back-reaction of the wrapped brane, the $\SU2$ structure is not integrable, in other words the metric is no longer hyper-K\"ahler. 

One can choose a frame $\{e^a\}$ for the metric on $M$ such that the forms determining the $\SU2$ structure are given by 
\begin{equation}
\begin{gathered}
   \omega_{1}=e^{14}+e^{23}, \qqq
   \omega_{2}=e^{13}-e^{24}, \qqq
   \omega_{3}=e^{12}+e^{34}, \\
   \zeta_1 = e^5, \qqq 
   \zeta_2 = e^6, \qqq
   \zeta_3 = e^7 . 
\end{gathered}
\end{equation}
The corresponding triplet of complex structures is given by  $(I_{\alpha})_{\phantom{m}n}^m=-(\omega_{\alpha})_{\phantom{m}n}^{m}$, while the volume form on $M$ is defined by
\begin{equation}
\tfrac{1}{2}\omega_{\alpha}\wedge\omega_{\beta}
    \wedge \zeta_{1}\wedge \zeta_{2}\wedge \zeta_3 = \delta_{\alpha\beta} \vol_7 .
\end{equation}

The integrability conditions differ in the two cases. Consider first the case where the M5-brane wraps a K\"ahler cycle, calibrated by $\omega_3$, in the hyper-K\"ahler manifold. The metric takes the form~\cite{FHP06,FS99}
\begin{equation}
   \dd s^{2}= \dd \tilde{s}^{2}(M_{\SU2})
      + \ee^{-4\Delta} \bigl( \dd x^2 + \dd y^2 + \dd z^2 \bigr) ,
\end{equation}
so that 
\begin{equation}
   \zeta_1 = \ee^{-2\Delta}\dd x  , \qqq
   \zeta_2 = \ee^{-2\Delta}\dd y  , \qqq
   \zeta_3 = \ee^{-2\Delta}\dd z  .
\end{equation}
The remaining conditions can then be written as
\begin{equation}\label{K2_torsion}
\begin{gathered}
   \dd(\ee^{\Delta}\omega_1) = \dd(\ee^{\Delta}\omega_2) = 0 , \qqq
   \dd(\ee^{4\Delta}\omega_3) = \ee^{4\Delta}\star F, \\
   \dd(\ee^{4\Delta}\omega_3\wedge \zeta_1\wedge \zeta_2\wedge \zeta_3) = 0 , 
\end{gathered}
\end{equation}
where $\star$ is the Hodge duality operator calculated using the metric~\eqref{eq:M5-metric}. Note that the integrability conditions preserve the $\SO3$ symmetry between the $\zeta_\alpha$ but break the symmetry between the $\omega_\alpha$. 

For an M5-brane wrapping $\bbR^2$, the metric takes the form 
\begin{equation}
   \dd s^{2}= \ee^{-4\Delta}\dd \tilde{s}_{\text{HK}}^{2}(M_{\SU2})
       + \ee^{2\Delta} \bigl( \dd x^2 + \dd y^2)
       + \ee^{-4\Delta}\dd z^2 . 
\end{equation}
where $\dd \tilde{s}_{\text{HK}}^{2}(M_{\SU2})$ is a hyper-K\"ahler metric on $M_{\SU2}$, and
\begin{equation}
   \zeta_1 = \ee^{\Delta}\dd x  , \qqq
   \zeta_2 = \ee^{\Delta}\dd y  , \qqq
   \zeta_3 = \ee^{-2\Delta}\dd z .
\end{equation}
In addition
\begin{equation}\label{R2_torsion}
\begin{gathered}
   \dd(\ee^{4\Delta}\omega_1) = \dd(\ee^{4\Delta}\omega_2) 
      = \dd(\ee^{4\Delta}\omega_3) = 0, \\
   \dd(\ee^{4\Delta}\zeta_1\wedge \zeta_2) = \ee^{4\Delta}\star F, \qqq    \dd(\ee^{4\Delta}\zeta_3\wedge \vol_4) = 0,
\end{gathered}
\end{equation}
where $\frac12\omega_\alpha\wedge\omega_\beta=\delta_{\alpha\beta}\vol_4$. Now the symmetry between the $\zeta_\alpha$ is broken but that between the $\omega_\alpha$ is preserved. 

These examples are interesting because we have the same $\SU2$ structure in each case but very different integrability conditions. A seven-dimensional $\SU2$ structure in M-theory actually admits four independent globally defined spinors.\footnote{This counting is reflected in the fact that compactifying M-theory on $\text{K3}\times \text{T}^3$ breaks half the supersymmetry.} In the two examples, different pairs of spinors are picked out by the Killing spinor equations. When we turn to generalised geometry, we will see that these different choices give two very different embeddings of the $\SU2$ structure into the generalised structure. Note that, dimensionally reducing along $\zeta_3$, these solutions also correspond to wrapped NS5-branes in type IIA. In the first case of branes wrapped on a K\"ahler cycle, the ten-dimensional Killing spinors actually take the form~\eqref{eq:KS-genCY}, and so these geometries are included in the class of $\SU3\times\SU3$ NS-NS backgrounds described in section~\ref{sec:GCY}. However, when the brane is wrapped on $\bbR^2$, the Killing spinors take the form
\begin{equation}
\label{eq:NS5-KS}
   \epsilon_1 = \begin{pmatrix} \chi_1^+ \\ 0 \end{pmatrix} , 
   \qqq
   \epsilon_2 = \begin{pmatrix} \chi_2^- \\ 0 \end{pmatrix} ,
\end{equation}
and, although the background is purely NS-NS, we see that it is not described by an integrable $\SU3\times\SU3$ structure, an exceptional case first noted in~\cite{GMPT05}. 


\section{Hyper-K\texorpdfstring{\"a}{a}hler geometry of Wolf spaces}
\label{app:wolf}

A Wolf space is a symmetric quaternionic-K\"ahler space $W/\bbH^*=\Gf/(G\times\SU{2})$ (as always we are ignoring discrete factors). The Riemannian case was first studied by Wolf in~\cite{Wolf65} and classified by Alekseevsky in~\cite{Alekseevsky68}, while the pseudo-Riemannian case, of relevance here, was analysed by Alekseevsky and Cort\'es in~\cite{AC05}. It is known that every quaternionic-K\"{a}hler manifold admits a bundle over it whose structure group is $\SU{2}$~\cite{Sakamoto74}. More importantly, there exists a tri-Sasaki structure on this bundle~\cite{Konishi75} and the cone over the $\SU2$ bundle is hyper-K\"ahler~\cite{Swann91}. The geometry on this ``Swann bundle'' $W$ for Wolf spaces has been explicitly constructed in~\cite{KS01}.

We can construct the tri-Sasaki and hyper-K\"ahler structures as follows. The tri-Sasaki space over the Wolf space is simply the symmetric space $S=\Gf/G$. As for any symmetric space, given an element $k\in \Gf$ one can decompose the left-invariant one-form $\theta$ as
\begin{equation}
\theta=k^{-1}\dd k=\pi+A,
\end{equation}
where $\pi\in\gf\ominus\mathfrak{g}$ and $A\in\mathfrak{g}$. The  one-forms $\pi$ descend to one-forms on $S$, while $A$ transforms as a $G$-connection. Since $S$ is the tri-Sasaki space over the Wolf space, $\Gf$ contains an $\SU2$ factor that commutes with $G$. We can then define a triplet of maps $\hat{\jmath}_\alpha:G\to\gf$ as parametrising the orbit
\begin{equation}
    \hat{\jmath}_{\alpha}(k)=k\hat{\jmath}_{\alpha}^{(0)} k^{-1},
\end{equation}
where $\hat{\jmath}_{\alpha}^{(0)}$ is some fixed set of $\su2\subset\mathfrak{g}'$ generators, stabilised by $G$. We normalise such that $\hat{\jmath}_\alpha$ satisfy the algebra
\begin{equation}
[\hat{\jmath}_{\alpha},\hat{\jmath}_{\beta}]=2\epsilon_{\alpha\beta\gamma}\hat{\jmath}_{\gamma}.
\end{equation}
By definition $\hat{\jmath}_\alpha(kg)=\hat{\jmath}_\alpha(k)$ for all $g\in G$. Thus $\hat{\jmath}_\alpha$ descend to a triplet of $\gf$-valued functions on $S=\Gf/G$
\begin{equation}
   \hat{\jmath}_\alpha : S \to \gf ,
\end{equation}
where, by definition, there is a one-to-one correspondence between points in $S$ and points on the orbit in $\gf$. The exterior derivative of $\hat{\jmath}_{\alpha}$ on $S$ is 
\begin{equation}
\begin{split}
   \dd\hat{\jmath}_{\alpha} 
      &= (\dd k)k^{-1}\hat{\jmath}_{\alpha} + \hat{\jmath}_{\alpha}k\dd k^{-1} \\
      & =[\hat{\jmath}_{\alpha},\theta] \\
      & =[\hat{\jmath}_{\alpha},\pi],
\end{split}
\end{equation}
where we have used $[\hat{\jmath}_{\alpha},A]=0$ as the $\hat{\jmath}_{\alpha}$ are stabilised by $G$. 

The tri-Sasaki structure is defined by a triplet of one-forms whose derivatives give a triplet of symplectic forms on the base of the $\SU{2}$ fibration. Following the discussion in~\cite{Grandini07}, the one-forms are given by 
\begin{equation}
\begin{aligned}
   \hat{\eta}_{\alpha} 
     &= -\tfrac{1}{2}\epsilon_{\alpha\beta\gamma}
          \tr(\dd\hat{\jmath}_{\beta}\cdot\hat{\jmath}_{\gamma}) \\
     & = \tr(\pi\cdot\hat{\jmath}_{\alpha}),
\end{aligned}
\end{equation}
which are clearly the left-invariant forms projected onto the $\su{2}$ subalgebra. 

Now consider the metric cone over the tri-Sasaki space $W=\Gf\times\mathbb{R}^{+}/G$, with cone coordinate $r$. The one-forms on the cone are inherited from those on the base as~\cite{Grandini07} 
\begin{equation}
   \eta_{\alpha}=r^{2}\hat{\eta}_{\alpha}.
\end{equation}
From the definition of $\hat{\eta}_{\alpha}$ in terms of the $\hat{\jmath}_{\alpha}$, this can be viewed as taking the triplet of functions $j_\alpha:W\to\gf$ on the cone to be 
\begin{equation}
\label{eq:jr}
   j_{\alpha} = r\hat{\jmath}_{\alpha}.
\end{equation}
An exterior derivative gives the symplectic forms
\begin{equation}
\begin{split}
   \omega_{\alpha} 
      &= \tfrac{1}{2}\dd\eta_{\alpha} \\
      &= \tfrac{1}{2}\epsilon_{\alpha\beta\gamma}
           \tr(\dd j_{\beta}\wedge\dd j_{\gamma}).
\end{split}
\end{equation}
Note that the symplectic forms are manifestly closed. Given two vector fields $v,w\in\Gamma(W)$, if we define the triplet of functions $v_\alpha=\imath_v\dd j_\alpha$, then
\begin{equation}
   \omega_\alpha(v,w) = \epsilon_{\alpha\beta\gamma}
           \tr(v_{\beta}w_{\gamma}).
\end{equation}
Any change in the functions $j_\alpha$ defining a point in $W$ can be generated by the adjoint action of $a_v\in\gf$, so we can also view vector fields as $v_\alpha=[a_v,j_\alpha]$. We then have 
\begin{equation}
\begin{split}\omega_{\alpha}(v,w) 
 & =\epsilon_{\alpha\beta\gamma}\tr\left([a_v,j_{\beta}][a_w,j_{\gamma}]\right)\\
 & =2\tr([a_v,a_w]j_{\alpha}).
\end{split}
\end{equation}
This is the analogue of the Kirillov--Kostant--Souriau symplectic structure on coadjoint orbits, as discussed in~\cite{KS01}.


\section{Special K\texorpdfstring{\"a}{a}hler geometry}
\label{sec:SK}

There are a number of different ways to define rigid (or affine) special K\"ahler geometry~\cite{CRTV97,Freed98,Hitchin99}. The most appropriate to our needs follows~\cite{Freed98}, stating that it is a $2n$-dimensional K\"ahler manifold $\Mv$ with a \emph{flat}, torsion-free connection $\hat{\nabla}$ satisfying 
\begin{equation}
   \hat{\nabla}_m\Omega_{np} = 0 , \qquad
   \hat{\nabla}_{[m}\mathcal{I}^p{}_{n]} = 0 ,
\end{equation}
where $\Omega$ is the K\"ahler form and $\mathcal{I}$ is the complex structure. Note that $\hat{\nabla}$ is not the Levi-Civita connection, since these conditions do not imply $\hat{\nabla}$ is metric compatible. 

Locally, by the Poincar\'e Lemma, the condition on $\mathcal{I}$ can be integrated. The usual formulation is to note that, since $\hat{\nabla}$ is torsion-free, one also has $\hat{\nabla}_{[m}\delta^k{}_{n]}=0$, thus locally there exists a complex vector field $X$ such that 
\begin{equation}
\label{eq:X-def}
   \hat{\nabla}_n X^m 
       = \delta^m{}_n - \ii \mathcal{I}^m{}_n . 
\end{equation}
Writing the real and imaginary parts as 
\begin{equation}
\label{eq:X-x}
   X^m = x^m + \ii\hat{x}^m , 
\end{equation}
so that $\nabla_nx^m=\delta^m{}_n$ and $\nabla_n\hat{x}^m=-\mathcal{I}^m{}_n$, one notes that the metric is given by $g_{mn}=\Omega_{mp}\mathcal{I}^p{}_n=-\Omega_{mp}\hat{\nabla}_n\hat{x}^p
=-\hat{\nabla}_n(\Omega_{mp}\hat{x}^p)$. But since $g_{mn}$ is symmetric, this means there exists a local real function $H$ such that the metric is given by the Hessian
\begin{equation}
\label{eq:hessian}
   g_{mn} = - \hat{\nabla}_m\hat{\nabla}_n H , 
\end{equation}
and $\Omega_{mn}\hat{x}^n=\hat{\nabla}_m H=\partial_m H$. Note that in these conventions, following~\cite{Freed98}, $H$ is equal to minus the K\"ahler potential. 

The fact that $\hat{\nabla}$ is torsion-free and flat means one can always introduce real coordinates such that $\hat{\nabla}_m=\partial/\partial x^m$. This notation is consistent with~\eqref{eq:X-x} since the condition $\nabla_n\re X^m=\delta^m{}_n$ means that in flat coordinates we can always locally identify $\re X^m$ with the coordinate $x^m$. It is conventional to use a different index notation $x^\Sigma$ to distinguish flat coordinates (or equivalently $\Sigma$ is the index for a flat frame). If one requires that the symplectic structure takes a standard form in the flat coordinates, then the choice of $x^\Sigma$ is determined up to affine symplectic transformations
\begin{equation}
   x'^\Sigma = P^\Sigma{}_ \Xi x^\Xi + c^\Sigma , 
\end{equation}
where $P\in\Symp{2n;\bbR}$ and $c$ is constant. Note that in flat coordinates $g_{\Sigma\Xi}=-\partial_\Sigma\partial_\Xi H$. Since $\hat{\nabla}$ is not the Levi-Civita connection, one cannot introduce coordinates that are both flat and complex. However, one can go halfway and define so-called ``special coordinates'' $X^I$ such that
\begin{equation}
   X = X^\Sigma\frac{\partial}{\partial \lambda^\Sigma}
       = X^I\frac{\partial}{\partial x^I} 
         - F_I\frac{\partial}{\partial y_I} , 
\end{equation}
where $x^\Sigma=(x^I,y_I)$ are flat Darboux coordinates (that is ones where  $\Omega=\dd x^I\wedge \dd y_I$), implying that $x^I=\re X^I$ and $y_I=-\re F_I$. Furthermore, the condition~\eqref{eq:X-def} implies that there is a local holomorphic function $F(X^I)$, called the prepotential, such that $F_I=\partial F/\partial X^I$.   

Again following~\cite{Freed98}, one can define a local (or projective) special K\"ahler manifold in terms of the complex cone over it, in analogy to the way a quaternionic-K\"ahler manifold defines a hyper-K\"ahler cone. Suppose $\Mv$ is a rigid special K\"ahler manifold such that there is a globally defined holomorphic complex vector field $X$ satisfying~\eqref{eq:X-def} that generates a $\bbC^*$ action that preserves the structure. Then the rigid K\"ahler structure on $Y$ descends to a local special K\"ahler structure on the quotient space $\Mv/\bbC^*$.\footnote{Strictly, the fermions provide an additional integral condition on the cohomology of the K\"ahler form on the quotient~\cite{CRTV97}.} One can also show that, as a function of any set of flat coordinates, $H$ is homogeneous of degree two. Furthermore, the K\"ahler potential $K$ on $\Mv/\bbC^*$ is given by 
\begin{equation}
\label{eq:LSK-K}
   \ee^{-K}= H = \tfrac{1}{4}\ii\Omega(X,\bar{X}) ,
\end{equation}
where we use the homogeneity of $H$ to derive the last equality. 

In gauged $\mathcal{N}=2$ supergravity one identifies an action of a group $\mathcal{G}_\text{V}$ on $\Mv/\bbC^*$, which can be lifted to an action on $\Mv$ that commutes with the $\bbC^*$ action. Supersymmetry requires that the action of $\mathcal{G}_\text{V}$ preserves the special K\"ahler structure. If $\hat{k}_{\hat{\lambda}}\in \Gamma(T\Mv)$ is the vector field corresponding to the action of an element of the Lie algebra $\hat{\lambda}\in\mathfrak{g}_\text{V}$, then one first requires
\begin{equation}
   \mathcal{L}_{\hat{k}_{\hat{\lambda}}} \Omega = 0,   \qqqq
   \mathcal{L}_{\hat{k}_{\hat{\lambda}}} \mathcal{I} = 0 , 
\end{equation}
or, in other words, that $\hat{k}_{\hat{\lambda}}$ is a real holomorphic Killing vector. In addition, it must map flat coordinates to flat coordinates by a symplectic rotation, equivalent to the condition, in components, that it is linear in $x^\Sigma$, that is
\begin{equation}
\label{eq:k-flat}
   {\hat{k}_{\hat{\lambda}}}^\Sigma = p_{\hat{\lambda}}{}^\Sigma{}_\Xi x^\Xi , 
\end{equation}
where $p_{\hat{\lambda}}\in\symp{n}(\bbR)$. It is easy to see that the corresponding moment map is given by 
\begin{equation}
\label{eq:SK-mm}
   \mu_{\hat{\lambda}} = \tfrac{1}{2}p_{\hat{\lambda}\,\Sigma\Xi}x^\Sigma x^\Xi ,
\end{equation}
where $p_{\hat{\lambda}\Sigma\Xi}=p_{\hat{\lambda}}{}^\Lambda{}_\Xi
\Omega_{\Lambda\Sigma}$. The particular gauging of the $\mathcal{N}=2$ theory is encoded in an embedding tensor $\hat{\Theta}^{\hat{\lambda}}_\Lambda$~\cite{WST05b,WM11}. This can be used to define a set of (constant) generators in $\symp{n}(\bbR)$
\begin{equation}
\label{eq:X-k}
   \XX_{\Lambda\Xi}{}^\Sigma 
       = \hat{\Theta}^{\hat{\lambda}}_\Lambda p_{\hat{\lambda}}{}^\Sigma{}_\Xi ,
\end{equation}
so that, by definition, they must satisfy~\cite{WST05b}
\begin{equation}
\label{eq:X-idents}
   \XX_{\Lambda[\Xi\Sigma]} = 0 , \qqq
   \XX_{\Lambda_1\Gamma}{}^\Sigma \XX_{\Lambda_2\Xi}{}^\Gamma
       - \XX_{\Lambda_2\Gamma}{}^\Sigma \XX_{\Lambda_1\Xi}{}^\Gamma
       = \XX_{\Lambda_1\Lambda_2}{}^\Gamma \XX_{\Gamma\Xi}{}^\Sigma , 
\end{equation}
where $\XX_{\Lambda\Xi\Sigma}=\XX_{\Lambda\Xi}{}^\Gamma\Omega_{\Gamma\Sigma}$. They also satisfy a ``representation constraint''
\begin{equation}
\label{eq:X-ident-rep}
   \XX_{(\Lambda\Xi\Sigma)}=0. 
\end{equation}
Finally, we note that contracting the moment map~\eqref{eq:SK-mm} with the embedding tensor gives $\hat{\Theta}^{\hat{\lambda}}_\Lambda\mu_{\hat{\lambda}}=\tfrac{1}{2}\XX_{\Lambda\Xi\Sigma}x^\Xi x^\Sigma$. Using the condition $\XX_{\Lambda\Xi\Sigma}X^\Xi X^\Sigma=0$ given in~\cite{WM11}, which is a consequence of $\hat{k}_{\hat{\lambda}}$ being holomorphic, we have 
\begin{equation}
\label{eq:SK-mm-1}
   \hat{\Theta}^{\hat{\lambda}}_\Lambda\mu_{\hat{\lambda}}
       = \tfrac{1}{4}\XX_{\Lambda\Xi\Sigma}X^\Xi \bar{X}^\Sigma . 
\end{equation}


\section{\texorpdfstring{$\Ex{d(d)}$}{Ed(d)} generalised geometry}
\label{app:gen-geom}


\subsection{E\texorpdfstring{$_{d(d)}$}{d(d)} for M-theory}

We review from \cite{CSW14} a construction of $\Ex{d(d)}\times\mathbb{R}^{+}$ using the $\GL{d}$ subgroup appropriate to M-theory, including useful representations, tensor products and the generalised Lie derivative.

On a $d$-dimensional manifold $M$, the generalised tangent bundle is
\begin{equation}
E\simeq TM\oplus\ext^{2}T^{*}M\oplus\ext^{5}T^{*}M\oplus(T^{*}M\otimes\ext^{7}T^{*}M).
\end{equation}
We write sections of this bundle as
\begin{equation}
\label{eq:V-Mtheory}
V=v+\omega+\sigma+\tau,
\end{equation}
where $v\in\Gamma(TM)$, $\omega\in\Gamma(\ext^{2}T^{*}M)$, $\sigma\in\Gamma(\ext^{5}T^{*}M)$ and $\tau\in\Gamma(T^{*}M\otimes\ext^{7}T^{*}M)$. The adjoint bundle is
\begin{equation}
\ad\tilde{F}\simeq\mathbb{R}\oplus(TM\otimes T^{*}M)\oplus\ext^{3}T^{*}M\oplus\ext^{6}T^{*}M\oplus\ext^{3}TM\oplus\ext^{6}TM.
\end{equation}
We write sections of the adjoint bundle as
\begin{equation}
R=l+r+a+\tilde{a}+\alpha+\tilde{\alpha},
\end{equation}
where $l\in\mathbb{R}$, $r\in\Gamma(\End TM)$, $a\in\Gamma(\ext^{3}T^{*}M)$ etc. We take $\{\hat{e}_{a}\}$ to be a basis for $TM$ with a dual basis $\{e^{a}\}$ on $T^{*}M$ so there is a natural $\gl{d}$ action on tensors. The $\ex{d(d)}$ subalgebra is generated by setting $l=r_{\phantom{a}a}^{a}/(9-d)$. This relation fixes the weight of generalised tensors under the $\mathbb{R}^+$ factor, so that a scalar of weight $k$ is a section of $(\det T^{*}M)^{k/(9-d)}$
\begin{equation}\label{RplusMtheory}
\rep{1}_k \in \Gamma\Bigl((\det T^{*}M)^{k/(9-d)}\Bigr).
\end{equation}

We define the adjoint action of $R\in\Gamma(\ad\tilde{F})$ on $V\in \Gamma(E)$ to be $V'=R\cdot V$. The components of $V'$ are
\begin{equation}
\begin{aligned}
v' & =lv+r\cdot v+\alpha\lrcorner\omega-\tilde{\alpha}\lrcorner\sigma,\\
\omega' & =l\omega+r\cdot\omega+v\lrcorner a+\alpha\lrcorner\sigma+\tilde{\alpha}\lrcorner\tau,\\
\sigma' & =l\sigma+r\cdot\sigma+v\lrcorner\tilde{a}+a\wedge\omega+\alpha\lrcorner\tau,\\
\tau' & =l\tau+r\cdot\tau-j\tilde{a}\wedge\omega+ja\wedge\sigma.
\end{aligned}
\label{eq:M_adjoint}
\end{equation}
We define the adjoint action of $R$ on $R'$ to be $R''=[R,R']$. The components of $R^{\prime\prime}$ are
\begin{equation}
\label{eq:ad-ad-M}
\begin{aligned}
l^{\prime\prime} & =\tfrac{1}{3}(\alpha\lrcorner a'-\alpha^{\prime}\lrcorner a)+\tfrac{2}{3}(\tilde{\alpha}^{\prime}\lrcorner\tilde{a}-\tilde{\alpha}\lrcorner\tilde{a}^{\prime}),\\
r^{\prime\prime} & =[r,r^{\prime}]+j\alpha\lrcorner ja^{\prime}-j\alpha^{\prime}\lrcorner ja-\tfrac{1}{3}\id(\alpha\lrcorner a^{\prime}-\alpha^{\prime}\lrcorner a)\\
 & \eqspace+j\tilde{\alpha}^{\prime}\lrcorner j\tilde{a}-j\tilde{\alpha}\lrcorner j\tilde{a}^{\prime}-\tfrac{2}{3}\id(\tilde{\alpha}^{\prime}\lrcorner\tilde{a}-\tilde{\alpha}\lrcorner\tilde{a}^{\prime}),\\
a^{\prime\prime} & =r\cdot a^{\prime}-r^{\prime}\cdot a+\alpha^{\prime}\lrcorner\tilde{a}-\alpha\lrcorner\tilde{a}^{\prime},\\
\tilde{a}^{\prime\prime} & =r\cdot\tilde{a}^{\prime}-r^{\prime}\cdot\tilde{a}-a\wedge a^{\prime},\\
\alpha^{\prime\prime} & =r\cdot\alpha^{\prime}-r^{\prime}\cdot\alpha+\tilde{\alpha}^{\prime}\lrcorner a-\tilde{\alpha}\lrcorner a^{\prime},\\
\tilde{\alpha}^{\prime\prime} & =r\cdot\tilde{\alpha}^{\prime}-r^{\prime}\cdot\tilde{\alpha}-\alpha\wedge\alpha^{\prime}.
\end{aligned}
\end{equation}

The dual of the generalised tangent bundle is $E^{*}$. We embed the usual derivative operator in the one-form component of $E^{*}$ via the map $T^{*}M\rightarrow E^{*}$. In coordinate indices $M$, one defines
\begin{equation}
\partial_{M}=\begin{cases}
\partial_{m} & \text{for }M=m,\\
0 & \text{otherwise}.
\end{cases}
\end{equation}
We then define a projection to the adjoint as
\begin{equation}
\oadj\colon E^{*}\otimes E\rightarrow\ad\tilde{F}.
\end{equation}
Explicitly, as a section of $\ad \tilde{F}$ we have
\begin{equation}
\label{eq:del-ad-M}
\partial\oadj V=\partial\otimes v+\dd\omega+\dd\sigma.
\end{equation}

The generalised Lie (or Dorfman) derivative is defined as
\begin{equation}
\label{eq:Dorf-def-M}
\Dorf_{V}W=V^{B}\partial_{B}W^{A}-(\partial\oadj V)_{\phantom{A}B}^{A}W^{B}.
\end{equation}
This can be extended to act on tensors by using the adjoint action of $\partial\oadj V\in\Gamma(\ad\tilde{F})$ in the second term. We will need explicit expressions for the Dorfman derivative of sections of $E$ and $\ad\tilde{F}$. The Dorfman derivative acting on a generalised vector is
\begin{equation}
\begin{split}\Dorf_{V}V^{\prime} & =\mathcal{L}_{v}v^{\prime}+(\mathcal{L}_{v}\omega^{\prime}-\imath_{v^{\prime}}\dd\omega)+(\mathcal{L}_{v}\sigma^{\prime}-\imath_{v^{\prime}}\dd\sigma-\omega^{\prime}\wedge\dd\omega)\\
 & \eqspace+(\mathcal{L}_{v}\tau^{\prime}-j\sigma^{\prime}\wedge\dd\omega-j\omega^{\prime}\wedge\dd\sigma).
\end{split}
\label{eq:M_Dorf_vector}
\end{equation}
The Dorfman derivative acting on a section of the adjoint bundle is
\begin{equation}
\begin{split}\Dorf_{V}R & =(\mathcal{L}_{v}r+j\alpha\lrcorner j\dd\omega-\tfrac{1}{3}\id\alpha\lrcorner\dd\omega-j\tilde{\alpha}\lrcorner j\dd\sigma+\tfrac{2}{3}\id\tilde{\alpha}\lrcorner\dd\sigma)+(\mathcal{L}_{v}a+r\cdot\dd\omega-\alpha\lrcorner\dd\sigma)\\
 & \eqspace+(\mathcal{L}_{v}\tilde{a}+r\cdot\dd\sigma+\dd\omega\wedge a)+(\mathcal{L}_{v}\alpha-\tilde{\alpha}\lrcorner\dd\omega)+(\mathcal{L}_{v}\tilde{\alpha}).
\end{split}
\label{eq:M_Dorf_adjoint}
\end{equation}

For $\Ex{5(5)}$, we also need the vector bundle transforming in the $\rep{10_{2}}$ representation of $\Spin{5,5}\times\mathbb{R}^{+}$. We define this bundle as
\begin{equation}
N\simeq T^{*}M\oplus\ext^{4}T^{*}M.
\end{equation}
We write sections of this bundle as
\begin{equation}
Q=m+n,
\end{equation}
where $m\in\Gamma(T^{*}M)$ and $n\in\Gamma(\ext^{4}T^{*}M)$. We define the adjoint action of $R\in\Gamma(\ad\tilde{F})$ on $Q\in\Gamma(N)$ to be $Q'=R\cdot Q$, with components
\begin{equation}
\begin{split}
m' & =2lm+r\cdot m-\alpha\lrcorner n,\\
n' & =2ln+r\cdot n-a\wedge m.
\end{split}
\label{Q_adj_M}
\end{equation}
Using $\rep{16}^{c}\times\rep{10}\rightarrow\rep{16}$, we define a projection to $E$ as
\begin{equation}
\proj{E}:E^{*}\otimes N\rightarrow E.
\end{equation}
Explicitly, as a section of $E$, this allows us to define 
\begin{equation}\label{e5_proj_M}
\dd Q \coloneqq \partial\proj{E} Q =\dd m+\dd n.
\end{equation}

We define a patching of the bundle $E$ such that on the overlaps of local patches $U_{i}\cap U_{j}$ we have
\begin{equation}
V_{(i)}=\ee^{\dd\Lambda_{(ij)}+\dd\tilde{\Lambda}_{(ij)}}V_{(j)},
\end{equation}
where $\Lambda_{(ij)}$ and $\tilde{\Lambda}_{(ij)}$ are locally two- and five-forms respectively. This defines the gauge-invariant field strengths as 
\begin{equation}
F=\dd A,\qqq
\tilde{F}=\dd\tilde{A}-\tfrac{1}{2}A\wedge F .
\end{equation}

The \emph{twisted} Dorfman derivative $\hat{\Dorf}_{\tilde{V}}$ of an untwisted generalised tensor $\tilde{\mu}$ is defined as
\begin{equation}
\label{eq:M_twisted_dorf}
\hat{\Dorf}_{\tilde{V}}\tilde{\mu}=\ee^{-A-\tilde{A}}\Dorf_{\ee^{A+\tilde{A}}\tilde{V}}(\ee^{A+\tilde{A}}\tilde{\mu}).
\end{equation}
The twisted Dorfman derivative $\hat{\Dorf}_{\tilde{V}}$ is given by the same expression as the usual Dorfman derivative with the substitutions
\begin{equation}
\dd\omega\rightarrow\dd\tilde{\omega}-\imath_{\tilde{v}}F,\qqq\dd\sigma\rightarrow\dd\tilde{\sigma}-\imath_{\tilde{v}}\tilde{F}+\tilde{\omega}\wedge F.
\end{equation}
The projection $\partial\proj{E}Q$ also simplifies in a similar fashion allowing us to define 
\begin{equation}
\dd_{F}Q \coloneqq \ee^{-A}\bigl(\partial\proj{E}(\ee^{A}Q)\bigr)=\dd m+\dd n-F\wedge m .
\end{equation}

The quadratic invariant for $\Ex{5(5)}$ is
\begin{equation}
\eta(Q,Q)=-m\wedge n.\label{eq:M_quadratic}
\end{equation}
The cubic invariant for $\Ex{6(6)}$ is
\begin{equation}
c(V,V,V)=-(\imath_{v}\omega\wedge\sigma+\tfrac{1}{3!}\omega\wedge\omega\wedge\omega).\label{eq:M_cubic}
\end{equation}
The symplectic invariant for $\Ex{7(7)}$ is
\begin{equation}
s(V,V^{\prime})=-\tfrac{1}{4}(\imath_{v}\tau^{\prime}-\imath_{v^{\prime}}\tau+\sigma\wedge\omega^{\prime}-\sigma^{\prime}\wedge\omega).\label{eq:M_symplectic}
\end{equation}
The $\ex{d(d)}$ Killing form is
\begin{equation}
\tr(R,R^{\prime})=\tfrac{1}{2}\Bigl(\tfrac{1}{9-d}\tr(r)\tr(r^{\prime})+\tr(rr^{\prime})+\alpha\lrcorner a^{\prime}+\alpha^{\prime}\lrcorner a-\tilde{\alpha}\lrcorner\tilde{a}^{\prime}-\tilde{\alpha}^{\prime}\lrcorner\tilde{a}\Bigr).\label{eq:M_Killing}
\end{equation}

The form of the $\Ex{d(d)}$-invariant volume $\kappa^{2}$ depends on the compactification ansatz. For compactifications of the form
\begin{equation}
g_{11}=\ee^{2\Delta}g_{11-d}+g_{d},
\end{equation}
the invariant volume is
\begin{equation}
\kappa^{2}=\ee^{(9-d)\Delta}\sqrt{g_{d}}.\label{eq:M_inv_vol}
\end{equation}


\subsection{E\texorpdfstring{$_{d+1(d+1)}$}{d+1(d+1)} for type IIB}\label{rep_IIB}

We provide details of the construction of $\Ex{d+1(d+1)}\times\mathbb{R}^{+}$ using the $\GL{d}\times\SL{2}$ subgroup appropriate to type IIB supergravity, including useful representations, tensor products and the generalised Lie derivative.

On a $d$-dimensional manifold $M$, the generalised tangent bundle is
\begin{equation}
\begin{split}E & \simeq TM\oplus T^{*}M\oplus(T^{*}M\oplus\ext^{3}T^{*}M\oplus\ext^{5}T^{*}M)\oplus\ext^{5}T^{*}M\oplus(T^{*}M\otimes\ext^{6}T^{*}M)\\
 & \simeq TM\oplus(T^{*}M\otimes S)\oplus\ext^{3}T^{*}M\oplus(\ext^{5}T^{*}M\otimes S)\oplus(T^{*}M\otimes\ext^{6}T^{*}M),
\end{split}
\end{equation}
where $S$ transforms as a doublet of $\SL{2}$. We write sections of this bundle as
\begin{equation}
\label{eq:V-IIB}
V=v+\lambda^{i}+\rho+\sigma^{i}+\tau,
\end{equation}
where $v\in\Gamma(TM)$, $\lambda^{i}\in\Gamma(T^{*}M\otimes S)$, $\rho\in\Gamma(\ext^{3}T^{*}M)$, $\sigma\in\Gamma(\ext^{5}T^{*}M\otimes S)$ and $\tau\in\Gamma(T^{*}M\otimes\ext^{6}T^{*}M)$. The adjoint bundle is
\begin{equation}
\begin{split}
   \ad\tilde{F} & =
      \mathbb{R}\oplus(TM\otimes T^{*}M)
      \oplus(S\otimes S^{*})_0
      \oplus(S\otimes\ext^{2}TM)
      \oplus(S\otimes\ext^{2}T^{*}M)\\ & \eqspace\quad
      \oplus\ext^{4}TM\oplus\ext^{4}T^{*}M
      \oplus(S\otimes\ext^{6}TM)\oplus(S\otimes\ext^{6}T^{*}M),
\end{split}
\end{equation}
where the subscript on $(S\otimes S^*)_0$ denotes the traceless part. We write sections of the adjoint bundle as
\begin{equation}
\label{eq:IIB-adj}
R=l+r+a+\beta^{i}+B^{i}+\gamma+C+\tilde{\alpha}^{i}+\tilde{a}^{i},
\end{equation}
where $l\in\mathbb{R}$, $r\in\Gamma(\End TM)$, etc. We take $\{\hat{e}_{a}\}$ to be a basis for $TM$ with a dual basis $\{e^{a}\}$ on $T^{*}M$ so there is a natural $\gl{d}$ action on tensors. 

The $\ex{d+1(d+1)}$ subalgebra is generated by setting $l=r_{\phantom{a}a}^{a}/(8-d)$. This fixes the weight of generalised tensors under the $\mathbb{R}^+$ factor, so that a scalar of weight $k$ is a section of $(\det T^{*}M)^{k/(8-d)}$
\begin{equation}\label{RplusIIB}
\rep{1}_k \in \Gamma\Bigl((\det T^{*}M)^{k/(8-d)}\Bigr).
\end{equation}

We define the adjoint action of $R\in\Gamma(\ad\tilde{F})$ on $V\in \Gamma(E)$ to be $V^{\prime}=R\cdot V$. The components of $V^{\prime}$ are
\begin{equation}
\begin{aligned}
v^{\prime} & =lv+r\cdot v+\gamma\lrcorner\rho+\epsilon_{ij}\beta^{i}\lrcorner\lambda^{j}+\epsilon_{ij}\tilde{\alpha}^{i}\lrcorner\sigma^{j},\\
\lambda^{\prime i} & =l\lambda^{i}+r\cdot\lambda^{i}+a_{\phantom{i}j}^{i}\lambda^{j}-\gamma\lrcorner\sigma^{i}+v\lrcorner B^{i}+\beta^{i}\lrcorner\rho-\tilde{\alpha}^{i}\lrcorner\tau,\\
\rho^{\prime} & =l\rho+r\cdot\rho+v\lrcorner C+\epsilon_{ij}\beta^{i}\lrcorner\sigma^{j}+\epsilon_{ij}\lambda^{i}\wedge B^{j}+\gamma\lrcorner\tau,\\
\sigma^{\prime i} & =l\sigma^{i}+r\cdot\sigma^{i}+a_{\phantom{i}j}^{i}\sigma^{j}-C\wedge\lambda^{i}+\rho\wedge B^{i}+\beta^{i}\lrcorner\tau+v\lrcorner\tilde{a}^{i},\\
\tau^{\prime} & =l\tau+r\cdot\tau+\epsilon_{ij}j\lambda^{i}\wedge\tilde{a}^{j}-j\rho\wedge C-\epsilon_{ij}j\sigma^{i}\wedge B^{j}.
\end{aligned}
\label{eq:IIB_adjoint}
\end{equation}
We define the adjoint action of $R$ on $R'$ to be $R''=[R,R']$. The components of $R''$ are
\begin{equation}
\label{eq:ad-ad-IIB}
\begin{aligned}l^{\prime} & =\tfrac{1}{2}(\gamma\lrcorner C^{\prime}-\gamma^{\prime}\lrcorner C)+\tfrac{1}{4}\epsilon_{kl}(\beta^{k}\lrcorner B^{\prime l}-\beta^{\prime k}\lrcorner B^{l})+\tfrac{3}{4}\epsilon_{ij}(\tilde{\alpha}^{i}\lrcorner\tilde{a}^{\prime j}-\tilde{\alpha}^{\prime i}\lrcorner\tilde{a}^{j}),\\
r^{\prime\prime} & =(r\cdot r^{\prime}-r^{\prime}\cdot r)+\epsilon_{ij}(j\beta^{i}\lrcorner jB^{\prime j}-j\beta^{\prime i}\lrcorner jB^{j})-\tfrac{1}{4}\id\epsilon_{kl}(\beta^{k}\lrcorner B^{\prime l}-\beta^{\prime k}\lrcorner B^{l})\\
 & \eqspace+(j\gamma\lrcorner jC^{\prime}-j\gamma^{\prime}\lrcorner jC)-\tfrac{1}{2}\id(\gamma\lrcorner C^{\prime}-\gamma^{\prime}\lrcorner C)\\
 & \eqspace+\epsilon_{ij}(j\tilde{\alpha}^{i}\lrcorner j\tilde{a}^{\prime j}-j\tilde{\alpha}^{\prime i}\lrcorner j\tilde{a}^{j})-\tfrac{3}{4}\epsilon_{ij}(\tilde{\alpha}^{i}\lrcorner\tilde{a}^{\prime j}-\tilde{\alpha}^{\prime i}\lrcorner\tilde{a}^{j}),\\
a_{\phantom{\prime\prime i}j}^{\prime\prime i} & =(a\cdot a^{\prime}-a^{\prime}\cdot a)_{\phantom{i}j}^{i}+\epsilon_{jk}(\beta^{i}\lrcorner B^{\prime k}-\beta^{\prime i}\lrcorner B^{k})-\tfrac{1}{2}\delta_{\phantom{i}j}^{i}\epsilon_{kl}(\beta^{k}\lrcorner B^{\prime l}-\beta^{\prime k}\lrcorner B^{l})\\
 & \eqspace+\epsilon_{jk}(\tilde{\alpha}^{i}\lrcorner\tilde{a}^{\prime k}-\tilde{\alpha}^{\prime i}\lrcorner\tilde{a}^{k})-\tfrac{1}{2}\delta_{\phantom{i}j}^{i}\epsilon_{kl}(\tilde{\alpha}^{k}\lrcorner\tilde{a}^{\prime l}-\tilde{\alpha}^{\prime k}\lrcorner\tilde{a}^{l}),\\
\beta^{\prime\prime i} & =(r\cdot\beta^{\prime i}-r^{\prime}\cdot\beta^{i})+(a\cdot\beta^{\prime}-a^{\prime}\cdot\beta)^{i}-(\gamma\lrcorner B^{\prime i}-\gamma^{\prime}\lrcorner B^{i})-(\tilde{\alpha}^{i}\lrcorner C^{\prime}-\tilde{\alpha}^{\prime i}\lrcorner C),\\
B^{\prime\prime i} & =(r\cdot B^{\prime i}-r^{\prime}\cdot B^{i})+(a\cdot B^{\prime}-a^{\prime}\cdot B)^{i}+(\beta^{i}\lrcorner C^{\prime}-\beta^{\prime i}\lrcorner C)-(\gamma\lrcorner\tilde{a}^{\prime i}-\gamma^{\prime}\lrcorner\tilde{a}^{i}),\\
\gamma^{\prime\prime} & =(r\cdot\gamma^{\prime}-r^{\prime}\cdot\gamma)+\epsilon_{ij}\beta^{i}\wedge\beta^{\prime j}+\epsilon_{ij}(\tilde{\alpha}^{i}\lrcorner B^{\prime j}-\tilde{\alpha}^{\prime i}\lrcorner B^{j}),\\
C^{\prime\prime} & =(r\cdot C^{\prime}-r^{\prime}\cdot C)-\epsilon_{ij}B^{i}\wedge B^{\prime j}+\epsilon_{ij}(\beta^{i}\lrcorner\tilde{a}^{\prime j}-\beta^{\prime i}\lrcorner\tilde{a}^{j}),\\
\tilde{\alpha}^{\prime\prime i} & =(r\cdot\tilde{\alpha}^{\prime i}-r^{\prime}\cdot\tilde{\alpha}^{i})+(a\cdot\tilde{\alpha}^{\prime}-a^{\prime}\cdot\tilde{\alpha})^{i}-(\beta^{i}\wedge\gamma^{\prime}-\beta^{\prime i}\wedge\gamma),\\
\tilde{a}^{\prime\prime i} & =(r\cdot\tilde{a}^{\prime i}-r^{\prime}\cdot\tilde{a}^{i})+(a\cdot\tilde{a}^{\prime}-a^{\prime}\cdot\tilde{a})^{i}+(B^{i}\wedge C^{\prime}-B^{\prime i}\wedge C).
\end{aligned}
\end{equation}

The dual of the generalised tangent bundle is $E^{*}$. We embed the usual derivative operator in the one-form component of $E^{*}$ via the map $T^{*}M\rightarrow E^{*}$. In coordinate indices $M$, one defines
\begin{equation}
\partial_{M}=\begin{cases}
\partial_{m} & \text{for }M=m,\\
0 & \text{otherwise}.
\end{cases}
\end{equation}
We then define a projection to the adjoint as
\begin{equation}
\oadj\colon E^{*}\otimes E\rightarrow\ad\tilde{F}.
\end{equation}
Explicitly, as a section of $\ad \tilde{F}$ we have
\begin{equation}
\label{eq:del-ad-IIB}
\partial\oadj V=\partial\otimes v+\dd\lambda^{i}+\dd\rho+\dd\sigma^{i}.
\end{equation}
The generalised Lie (or Dorfman) derivative is defined as
\begin{equation}
\label{eq:Dorf-def-IIB}
\Dorf_{V}W=V^{B}\partial_{B}W^{A}-(\partial\oadj V)_{\phantom{A}B}^{A}W^{B}.
\end{equation}
This can be extended to act on tensors by using the adjoint action of $\partial\oadj V\in\Gamma(\ad\tilde{F})$ in the second term. We will need explicit expressions for the Dorfman derivative of sections of $E$ and $\ad\tilde{F}$. The Dorfman derivative acting on a generalised vector is
\begin{equation}
\begin{split}\Dorf_{V}V^{\prime} & =\mathcal{L}_{v}v^{\prime}+(\mathcal{L}_{v}\lambda^{\prime i}-\imath_{v^{\prime}}\dd\lambda^{i})+(\mathcal{L}_{v}\rho^{\prime}-\imath_{v^{\prime}}\dd\rho+\epsilon_{ij}\dd\lambda^{i}\wedge\lambda^{\prime j})\\
 & \eqspace+(\mathcal{L}_{v}\sigma^{\prime i}-\imath_{v^{\prime}}\dd\sigma^{i}+\dd\rho\wedge\lambda^{\prime i}-\dd\lambda^{i}\wedge\rho^{\prime})\\
 & \eqspace+(\mathcal{L}_{v}\tau^{\prime}-\epsilon_{ij}j\lambda^{\prime i}\wedge\dd\sigma^{j}+j\rho^{\prime}\wedge\dd\rho+\epsilon_{ij}j\sigma^{\prime i}\wedge\dd\lambda^{j}).
\end{split}
\label{eq:IIB_Dorf_vector}
\end{equation}
The Dorfman derivative acting on a section of the adjoint bundle is
\begin{equation}
\begin{split}\Dorf_{V}R & =(\mathcal{L}_{v}l+\tfrac{1}{2}\gamma\lrcorner\dd\rho+\tfrac{1}{4}\epsilon_{kl}\beta^{k}\lrcorner\dd\lambda^{l}+\tfrac{3}{4}\epsilon_{kl}\tilde{\alpha}^{k}\lrcorner\dd\sigma^{l})\\
 & \eqspace+(\mathcal{L}_{v}r+j\gamma\lrcorner j\dd\rho-\tfrac{1}{2}\id\gamma\lrcorner\dd\rho+\epsilon_{ij}j\beta^{i}\lrcorner j\dd\lambda^{j}-\tfrac{1}{4}\id\epsilon_{kl}\beta^{k}\lrcorner\dd\lambda^{l}\\
 & \eqspace\phantom{{}+({}}+\epsilon_{ij}j\tilde{\alpha}^{i}\lrcorner j\dd\sigma^{j}-\tfrac{3}{4}\id\epsilon_{kl}\tilde{\alpha}^{k}\lrcorner\dd\sigma^{l})\\
 & \eqspace+(\mathcal{L}_{v}a_{\phantom{i}j}^{i}+\epsilon_{jk}\beta^{i}\lrcorner\dd\lambda^{k}-\tfrac{1}{2}\delta_{\phantom{i}j}^{i}\epsilon_{kl}\beta^{k}\lrcorner\dd\lambda^{l}+\epsilon_{jk}\tilde{\alpha}^{i}\lrcorner\dd\sigma^{k}-\tfrac{1}{2}\delta_{\phantom{i}j}^{i}\epsilon_{kl}\tilde{\alpha}^{k}\lrcorner\dd\sigma^{l})\\
 & \eqspace+(\mathcal{L}_{v}\beta^{i}-\gamma\lrcorner\dd\lambda^{i}-\tilde{\alpha}^{i}\lrcorner\dd\rho)\\
 & \eqspace+(\mathcal{L}_{v}B^{i}+r\cdot\dd\lambda^{i}+a_{\phantom{i}j}^{i}\dd\lambda^{j}+\beta^{i}\lrcorner\dd\rho-\gamma\lrcorner\dd\sigma^{i})\\
 & \eqspace+(\mathcal{L}_{v}\gamma+\epsilon_{ij}\tilde{\alpha}^{i}\lrcorner\dd\lambda^{j})\\
 & \eqspace+(\mathcal{L}_{v}C+r\cdot\dd\rho+\epsilon_{ij}\dd\lambda^{i}\wedge B^{j}+\epsilon_{ij}\beta^{i}\lrcorner\dd\sigma^{j})+(\mathcal{L}_{v}\tilde{\alpha}^{i})\\
 & \eqspace+(\mathcal{L}_{v}\tilde{a}^{i}+r\cdot\dd\sigma^{i}+a_{\phantom{i}j}^{i}\dd\sigma^{j}-\dd\lambda^{i}\wedge C+B^{i}\wedge\dd\rho).
\end{split}
\label{eq:IIB_Dorf_adjoint}
\end{equation}

For $\Ex{5(5)}$, we also need the vector bundle transforming in the $\rep{10_{2}}$ representation of $\Spin{5,5}\times\mathbb{R}^{+}$. We define this bundle as
\begin{equation}
N\simeq S\oplus\ext^{2}T^{*}M\oplus S\otimes\ext^{4}T^{*}M.
\end{equation}
We write sections of this bundle as
\begin{equation}
Q=m^{i}+n+p^{i},
\end{equation}
where $m^{i}\in\Gamma(S)$, $n\in\Gamma(\ext^{2}T^{*}M)$ and $p^{i}\in\Gamma(S\otimes\ext^{4}T^{*}M)$. We define the adjoint action of $R\in\Gamma(\ad\tilde{F})$ on $Q\in\Gamma(N)$ to be $Q'=R\cdot Q$, with components
\begin{equation}
\begin{split}
m'^{i} & =2lm^{i}+a_{\phantom{i}j}^{i}m^j+\beta^{i}\lrcorner n-\gamma\lrcorner p^{i},\\
n' & =2ln+r\cdot n+\epsilon_{ij}\beta^{i}\lrcorner p^{j}+\epsilon_{ij}m^{i}B^{j},\\
p'^{i} & =2lp^{i}+r\cdot p^{i}+a_{\phantom{i}j}^{i}p^{j}+B^{i}\wedge n-m^{i}C.
\end{split}
\label{Q_adj_IIB}
\end{equation}
Using $\rep{16}^{c}\times\rep{10}\rightarrow\rep{16}$, we define a projection to $E$ as
\begin{equation}
\proj{E}:E^{*}\otimes N\rightarrow E.
\end{equation}
Explicitly, as a section of $E$, this allows us to define 
\begin{equation}\label{e5_proj_IIB}
\dd Q \coloneqq \partial\proj{E}Q = \dd m^{i}+\dd n.
\end{equation}

We define a patching of the bundle such that on the overlaps of local patches $U_{i}\cap U_{j}$ we have
\begin{equation}
V_{(i)}=\ee^{\dd\Lambda_{(ij)}^{i}+\dd\tilde{\Lambda}_{(ij)}}V_{(j)},
\end{equation}
where $\Lambda_{(ij)}^{i}$ and $\tilde{\Lambda}_{(ij)}$ are locally one- and three-forms respectively. This defines the gauge-invariant field strengths as 
\begin{equation}
F^{i}=\dd B^{i},\qqq 
F=\dd C-\tfrac{1}{2}\epsilon_{ij}F^{i}\wedge B^{j}.
\label{eq:IIB_flux}
\end{equation}
We embed the NS-NS and R-R three-form fluxes as $F_{3}^{1}=H$ and $F_{3}^{2}=F_{3}$.

The \emph{twisted} Dorfman derivative $\hat{\Dorf}_{V}$ of an untwisted generalised tensor $\tilde{\mu}$ is defined by
\begin{equation}
\label{eq:IIB_twisted_dorf}
\hat{\Dorf}_{\tilde{V}}\tilde{\mu}=\ee^{-B^{i}-C}\Dorf_{\ee^{B^{i}+C}\tilde{V}}(\ee^{B^{i}+C}\tilde{\mu}).
\end{equation}
 The twisted Dorfman derivative $\hat{\Dorf}_{V}$ is given by the same expression as the usual Dorfman derivative but with the substitutions
\begin{equation}
\begin{aligned}
\dd\lambda^{i}&\rightarrow\dd\tilde{\lambda}^{i}-\imath_{\tilde{v}}F^{i}, \\
\qqq\dd\rho&\rightarrow\dd\tilde{\rho}-\imath_{\tilde{v}}F-\epsilon_{ij}\tilde{\lambda}^{i}\wedge F^{j}, \\
\qqq\dd\sigma^{i}&\rightarrow\dd\tilde{\sigma}^{i}+\tilde{\lambda}^{i}\wedge F-\tilde{\rho}\wedge F^{i}.
\end{aligned}
\end{equation}
The projection $\partial\proj{E}Q$ also simplifies in a similar fashion allowing us to define 
\begin{equation}\label{dQ_twisted}
\dd_{F^{i}}Q \coloneqq 
   \ee^{-B^{i}}\bigl(\partial\proj{E}(\ee^{B^{i}}Q)\bigr)
   =\dd m^{i}+\dd n+\epsilon_{ij}m^{i}F^{j}.
\end{equation}

The quadratic invariant for $\Ex{5(5)}$ is
\begin{equation}
\eta(Q,Q)=\epsilon_{ij}m^i p^j -\tfrac{1}{2} n\wedge n.\label{eq:IIB_quadratic}
\end{equation}
The cubic invariant for $\Ex{6(6)}$ is
\begin{equation}
c(V,V,V)=-\tfrac{1}{2}(\imath_{v}\rho\wedge\rho+\epsilon_{ij}\rho\wedge\lambda^{i}\wedge\lambda^{j}-2\epsilon_{ij}\imath_{v}\lambda^{i}\sigma^{j}).\label{eq:IIB_cubic}
\end{equation}
The symplectic invariant for $\Ex{7(7)}$ is
\begin{equation}
s(V,V^{\prime})=-\tfrac{1}{4}\bigl((\imath_{v}\tau^{\prime}-\imath_{v^{\prime}}\tau)+\epsilon_{ij}(\lambda^{i}\wedge\sigma^{\prime j}-\lambda^{\prime i}\wedge\sigma^{j})-\rho\wedge\rho^{\prime}\bigr).\label{eq:IIB_symplectic}
\end{equation}
The $\ex{d+1(d+1)}$ Killing form is
\begin{equation}
\begin{split}\tr(R,R^{\prime}) & =\tfrac{1}{2}\Bigl(\tfrac{1}{8-d}\tr(r)\tr(r^{\prime})+\tr(rr^{\prime})+\tr(aa^{\prime})+\gamma\lrcorner C^{\prime}+\gamma^{\prime}\lrcorner C+\epsilon_{ij}(\beta^{i}\lrcorner B^{\prime j}+\beta^{\prime i}\lrcorner B^{j})\\
 & \eqspace\phantom{\frac{1}{2}\biggl[}+\epsilon_{ij}(\tilde{\alpha}^{i}\lrcorner\tilde{a}^{\prime j}+\tilde{\alpha}^{\prime i}\lrcorner\tilde{a}^{j})\Bigr).
\end{split}
\label{eq:IIB_Killing}
\end{equation}

The form of the $\Ex{d+1(d+1)}$-invariant volume $\kappa^{2}$ depends on the compactification ansatz. For compactifications of the form
\begin{equation}
g_{10}=\ee^{2\Delta}g_{10-d}+g_{d},
\end{equation}
the invariant volume includes a dilaton dependence and is given by
\begin{equation}
\kappa^{2}=\ee^{-2\phi}\ee^{(8-d)\Delta}\sqrt{g_{d}}.\label{eq:IIB_inv_vol}
\end{equation}

We can include non-zero axion and dilaton in our formalism using the $\SL{2}$ frame given in \cite{LSW14}. Let $\hat{f}_{\phantom{i}\hat{i}}^{i}$ be an $\SL{2}$ frame written in terms of a parametrisation of $\SL{2}/\SO{2}$ as
\begin{equation}
\hat{f}_{\phantom{i}\hat{i}}^{i}=\begin{pmatrix}\ee^{\phi/2} & 0\\
A\ee^{\phi/2} & \ee^{-\phi/2}
\end{pmatrix}.
\end{equation}
Comparing with the split frame of \cite{LSW14}, we see we can write a generalised vector as
\begin{equation}
V=v + \ee^{-\phi/2} \lambda^i + \ee^{-\phi}\rho + \ee^{-3\phi/2} \sigma^i,
\end{equation}
where $\lambda^i=\hat{f}_{\phantom{i}\hat{i}}^{i} \lambda^{\hat{i}}$ etc., and $\lambda^{\hat{i}}$ contains no explicit axion-dilaton dependence. Using this we can determine where the dilaton appears in the adjoint for $\Ex{d(d)}$ and $Q$ for $\Ex{5(5)}$
\begin{equation}
\begin{split}
R&=l+r+a^i_{\phantom{i}j}+\ee^{\phi/2}\beta^i+\ee^{-\phi/2}B^i+\ee^\phi \gamma + \ee^{-\phi}C+\ee^{3\phi/2}\tilde{\alpha}^i+\ee^{-3\phi/2}\tilde{a}^i , \\
Q&=\ee^{-\phi/2}m^i+\ee^{-\phi}n+\ee^{-3\phi/2}p^i .
\end{split}
\end{equation}
Looking back to $\tilde{J}_\alpha$ and $\tilde{Q}$ for the NS5-brane solution in \eqref{NS5_J} and \eqref{NS5_Q}, we see they are indeed of this form. The various powers of the dilaton correspond to the exponentiated action of the adjoint element given by
\begin{equation}
l+r = \frac{\phi}{4}(-1+\id).
\end{equation}


\section{Moment maps and quotients}
\label{app:mm}

In this appendix, we briefly review the notion of moment maps, and symplectic and hyper-K\"ahler quotients, including the infinite-dimensional example of flat gauge connections on a Riemann surface due to Atiyah and Bott~\cite{AB83}. 

Consider a manifold $Y$ with a symplectic form $\Omega$ that is closed, $\dd\Omega=0$. Suppose there is an action of a Lie group $\mathcal{G}$ on $Y$ that preserves the symplectic structure -- that is $\mathcal{G}$ acts on $Y$ via symplectomorphisms. An element $g$ in the Lie algebra $\mathfrak{g}$ of $\mathcal{G}$ induces a vector field $\rho_g$ on $Y$. As the group $\mathcal{G}$ acts via symplectomorphisms, the Lie derivative of $\Omega$ with respect to $\rho_g$ vanishes. Together with $\dd\Omega=0$, this implies $\dd\imath_{\rho_g}\Omega=0$ and so $\imath_{\rho_g}\Omega$ is closed. A moment map for the action of the group $\mathcal{G}$ on the manifold $Y$ is a map $\mu\colon Y \times\mathfrak{g}\rightarrow \bbR$ such that, for all $g\in\mathfrak{g}$,
\begin{equation}
\dd\mu(g)=\imath_{\rho_g}\Omega.
\end{equation}
The moment map is defined up to an additive constant of integration. If $\mathfrak{g}^{*}$ is the dual of the Lie algebra $\mathfrak{g}$, one can also view $\mu$ as a map from $Y$ to $\mathfrak{g}^*$. If $\mathcal{G}$ is non-Abelian one can fix the constant by requiring that the map is equivariant, that is, that $\mu$ commutes with the action of $\mathcal{G}$ on $Y$. Still viewing $\mu$ as a map from $Y$ to $\mathfrak{g}^*$, one can then form the \emph{symplectic quotient}
\begin{equation}
Y \qquotient \mathcal{G} = \mu^{-1}(0) \quotient \mathcal{G}.
\end{equation}
This quotient space inherits a symplectic structure from $Y$ and is a manifold if $\mathcal{G}$ acts freely on $Y$. (Generically the reduced space is not a manifold, but is a ``stratified space''.) 

On a hyper-K\"ahler manifold $Y$, one can consider an action of $\mathcal{G}$ that preserves all three symplectic forms $\Omega_\alpha$. Instead of a single moment map, one can then consider a triplet of maps $\mu_\alpha:Y\to\mathfrak{g}^*$ satisfying 
\begin{equation}
\dd\mu_\alpha(g)=\imath_{\rho_g}\Omega_\alpha .
\end{equation}
Choosing them to be equivariant, one can then define the \emph{hyper-K\"ahler quotient}~\cite{HKLR87}
\begin{equation}
   Y \qqquotient \mathcal{G}  
      =\mu_{1}^{-1}(0)\cap\mu_{2}^{-1}(0)\cap\mu_{3}^{-1}(0) \quotient \mathcal{G}.
\end{equation}
This space inherits a hyper-K\"ahler structure from $Y$, and the quotient is a manifold if $\mathcal{G}$ acts freely.  

We can also consider the case where both the group and the symplectic space are infinite dimensional. A well-known example is the work of Atiyah--Bott~\cite{AB83}. Let $G$ be a compact Lie group and $P$ be a principal $G$-bundle over a compact Riemann surface $\Sigma$. The group of gauge transformations $\mathcal{G}$ is the set of $G$-equivariant diffeomorphisms of $P$. Infinitesimally it is generated by sections of the adjoint bundle $\ad P$, that is $\Lie(\mathcal{G})=\Gamma(\ad P)$. Let $Y$ be the infinite-dimensional space of connections on $P$.  The curvature of a connection $A\in Y$ is
\begin{equation}
F=\dd A + \tfrac{1}{2}[A,A].
\end{equation}
One can associate the tangent space $T_AY$ at $A\in Y$ with the space of $\ad P$-valued one-forms $\Omega^{1}(\Sigma,\ad P)$. Given two elements $\alpha,\beta\in T_{A}Y$, one can define a symplectic product
\begin{equation}
   \Omega(\alpha,\beta)
    = \int_{\Sigma}\tr (\alpha\wedge\beta),
\end{equation}
where $\tr$ is a gauge-invariant inner product on $\mathfrak{g}$, for example the Killing form if $\mathfrak{g}$ is semi-simple. To see that $\Omega$ is non-degenerate note that, given a metric on $\Sigma$, we have 
\begin{equation}
   \Omega(\alpha,\star\alpha)
      = \int_{\Sigma} \tr(\alpha\wedge\star\alpha)
      = \left\Vert\alpha\right\Vert^{2}
      \geq 0,
\end{equation}
and so $\Omega(\alpha,\star\alpha)=0$ if and only if $\alpha=0$. Furthermore, any connection $A$ can be written as $A=A^{(0)}+\alpha$ for some fixed connection $A^{(0)}$ and $\alpha\in\Omega^{1}(\Sigma,\ad P)$ (in other words $Y$ is an affine space modelled on $\Omega^{1}(\Sigma,\ad P)$), meaning that in this parametrisation $\Omega$ is independent of $A$ and hence, in particular, $\Omega$ is a closed two-form on $Y$. 

The moment map for the $\mathcal{G}$-action on $Y$ is $\mu=F$. To see this note that, given an element $\Lambda$ of $\Lie(\mathcal{G})\simeq\Gamma(\ad{P})$, the induced vector field on $Y$ is just the gauge transformation of $A$, namely 
\begin{equation}
   \rho_{\Lambda} = \dd\Lambda+[A,\Lambda] . 
\end{equation}
Thus we have, for any $\alpha\in\Gamma(TY)$, 
\begin{equation}
\begin{split}
   \imath_{\rho_{\Lambda}}\Omega(\alpha) = \Omega(\rho_{\Lambda},\alpha) 
      &= \int_\Sigma \tr \bigl[(\dd\Lambda+[A,\Lambda])\wedge \alpha\bigr] 
      = \int_\Sigma \tr \bigl[\Lambda\wedge(\dd\alpha+[A,\alpha])\bigr] \\
      &= \imath_\alpha \biggl(\delta \int_\Sigma \tr \Lambda F \biggr) , 
\end{split}
\end{equation}
where $\delta$ is the exterior derivative on $Y$, that is, in coordinates, the functional derivative $\delta/\delta A_m(x)$. Viewed as a map $\mu:Y\to\Lie(\mathcal{G})^*$, we see that $\mu=F$. 

This map is equivariant, and so we may form the symplectic reduction by quotienting by the space of gauge transformations $\mathcal{G}$
\begin{equation}
   Y \qquotient \mathcal{G}=\mu^{-1}(0) \quotient \mathcal{G}.
\end{equation}
This is the moduli space of flat connections, that is $A\in Y$ such that $F=0$ modulo gauge equivalence. The space of connections $Y$ and the group of gauge transformations $\mathcal{G}$ are infinite dimensional, but the moduli space is actually finite dimensional. 


\section{Intrinsic torsion and \texorpdfstring{$\SU6$}{SU(6)} structures}
\label{sec:int-tor-su6}

Following~\cite{CSW14b}, we first calculate the intrinsic torsion space $\Tint^{\SU6}$ for generalised $\SU6$ structures. Decomposing under $\SU2\times\SU6$ the space of generalised torsions decomposes as 
\begin{equation}
\begin{split}
   W = \rep{56} + \rep{912}
      &=  (\rep{1},\rep{1}) + 2(\rep{1},\rep{15}) + (\rep{1},\rep{21})
         + (\rep{1},\rep{35}) + (\rep{1},\rep{105})
         \\ & \eqspace
         + 3(\rep{2},\rep{6}) + (\rep{2},\rep{20}) + (\rep{2},\rep{84})
         + (\rep{3},\rep{1}) + (\rep{3},\rep{15}) + \CC
\end{split}
\end{equation}
The space of $\SU6$ connections is given by 
\begin{equation}
\begin{split}
   K_{\SU6} 
      &= \bigl( \repp{1}{1} + \repp{2}{6} + \repp{1}{15} + \CC \bigr)
          \times \repp{1}{35} \\
      &= \repp{1}{15} + \repp{1}{21} + \repp{1}{35} + \repp{1}{105} 
          \\ & \eqspace
          + \repp{1}{384} 
          + \repp{2}{6} + \repp{2}{84} + \repp{2}{120} + \CC
\end{split}
\end{equation}
Thus we have 
\begin{equation}
\label{eq:SU6Tint}
   \Tint^{\SU6} \supseteq (\rep{1},\rep{1}) + (\rep{1},\rep{15})
         + 2(\rep{2},\rep{6}) + (\rep{2},\rep{20}) 
         + (\rep{3},\rep{1}) + (\rep{3},\rep{15}) + \CC ,
\end{equation}
where equality holds if there are no unexpected kernels in the map $\tau:K_{\SU{6}}\to W$. To see that this is indeed the case, we need the explicit map. In $\SU8$ indices, sections of $K_{\SU8}$ are given by
\begin{equation}
   \hat\Sigma = (\hat\Sigma_{\alpha\beta\phantom{\gamma}\delta}^{\smash{\phantom{\alpha\beta}\gamma}} , 
        \bar{\hat\Sigma}^{\alpha\beta}{}^\gamma{}_\delta , )
        \in (\rep{28}+\rep{\overline{28}})\times\rep{63} , 
\end{equation}
where the elements are antisymmetric on $\alpha$ and $\beta$ and traceless on contracting $\gamma$ with $\delta$. The space $W$ decomposes as 
\begin{equation}
   W = \rep{56} + \rep{912} = \rep{28} + \rep{36} + \rep{420} + \CC , 
\end{equation}
and the map $\tau$ is
\begin{equation}
\begin{aligned}
   \tau(\hat\Sigma)_{\alpha\beta} 
      &= \hat\Sigma_{\alpha\gamma\phantom{\gamma}\beta}^{\smash{\phantom{\alpha\beta}\gamma}}
      && \in \rep{36} + \rep{28} , \\
   \tau(\hat\Sigma)_{\alpha\beta\gamma}{}^\delta 
      &= 3\hat\Sigma_{[\alpha\beta\phantom{\delta}\gamma]}^{0\phantom{\alpha}\,\,\,\delta}
      && \in \rep{420} , 
\end{aligned}
\end{equation}
where the ``0'' superscript on $\hat\Sigma_{[\alpha\beta\phantom{\delta}\gamma]}^{0\phantom{\alpha}\,\,\,\delta}$ means it is completely traceless. The $\rep{28}$ and $\rep{36}$ representations correspond to the symmetric and antisymmetric parts of $\tau(\hat\Sigma)_{\alpha\beta}$. There are similar expressions for the conjugate representations in terms of $\bar{\hat\Sigma}$. 

Turning to $\SU6$ connections, let $\Sigma$ be a section of $K_{\SU6}$. We can split the spinor indices $\alpha$ into $a=1,\ldots,6$ and $i=7,8$ so that the non-zero components are
\begin{equation}
\begin{aligned}
   \Sigma_{ab}{}^c{}_d &\in \repp{1}{15}\times\repp{1}{35} , \\
   \Sigma_{ai}{}^c{}_d = -\Sigma_{ia}{}^c{}_d 
       &\in \repp{2}{6}\times\repp{1}{35} , \\ 
   \Sigma_{ij}{}^c{}_d &\in \repp{1}{35} , 
\end{aligned}
\end{equation}
and similarly for the conjugate $\bar{\Sigma}$. We then find the non-zero components of $\tau(\Sigma)$ are
\begin{equation}
\label{eq:map}
\begin{aligned}
   \tau(\Sigma)_{ab} &= \Sigma_{ac}{}^c{}_b  
      && \in \repp{1}{15}+\repp{1}{21} , \\
   \tau(\Sigma)_{ib} &= \Sigma_{ic}{}^c{}_b  
      && \in \repp{2}{6} , \\
   \tau(\Sigma)_{abc}{}^d &= 3\Sigma^0_{[ab}{}^d{}_{c]} 
      - \tfrac{1}{2}\Sigma_{[a|e|}{}^e{}_{b}\delta^d_{c]}  
      && \in \repp{1}{105} + \repp{1}{15} , \\
   \tau(\Sigma)_{abi}{}^c &= 2\Sigma^0_{i[a}{}^c{}_{b]} 
      - \tfrac{1}{15}\Sigma_{ie}{}^e{}_{[a}\delta^c_{b]}
      &&\in \repp{2}{84} + \repp{2}{6} , \\ 
   \tau(\Sigma)_{aij}{}^c &= \Sigma_{ij}{}^c{}_{a}  
      &&\in \repp{1}{35} , \\
   \tau(\Sigma)_{abi}{}^j &= \tfrac{1}{3}\Sigma_{[a|c|}{}^c{}_{b]}\delta^j_i  
      &&\in \repp{1}{15} , \\ 
   \tau(\Sigma)_{aij}{}^k &= 
       - \tfrac{1}{3}\Sigma_{[i|c}{}^c{}_{a}\delta^k_{j]}  
      &&\in \repp{2}{6} ,  
\end{aligned}
\end{equation}
and hence $\Tint$ is indeed given by an equality in~\eqref{eq:SU6Tint}. Note in addition that 
\begin{equation}
\label{eq:SU6-zeros}
   \tau(\Sigma)_{abi}{}^i - \tfrac{2}{3}\tau(\Sigma)_{[ab]} = 0 , \qqqq 
   \tau(\Sigma)_{aij}{}^j + \tfrac{1}{6}\tau(\Sigma)_{ia} = 0 . 
\end{equation}

We now turn to showing which components of the intrinsic torsion enter each of the integrability conditions on the pair $\{J_\alpha,X\}$. For this it is useful to have an expression for $T(V)$ for $\SU6$ connections. We first note that the compatible $\SU6$ connection $\hat{D}$ must also be an $\SU8$ connection and hence can be written as 
\begin{equation}
   \hat{D} = D + \hat{\Sigma} ,
\end{equation}
where $\hat{\Sigma}\in K_{\SU8}$ and $D$ is a torsion-free $\SU8$ connection. (That such connections exist is central to the formulation of supergravity in terms of generalised geometry: they are the analogues of the Levi-Civita connection of conventional gravity~\cite{CSW11,CSW14}.) Since $D$ is torsion-free, the torsion of $\hat{D}$ is given by 
\begin{equation}
\label{eq:TS}
   T = \tau(\hat{\Sigma}). 
\end{equation}
We can then calculate $T(V)$. Writing  $V=(V^{\alpha\beta},\bar{V}_{\alpha\beta})$ for the decomposition $\rep{56}=\rep{28}+\rep{\overline{28}}$ and $T(V)=(T(V)_0,T(V)^\alpha{}_\beta,T(V)^{\alpha\beta\gamma\delta})$ for the decomposition of the adjoint $\rep{1}+\rep{133}=\rep{1}+\rep{63}+\rep{70}$, we define the adjoint action on a generalised vector $W$ as
\begin{equation}
\begin{aligned}
   \bigl[T(V)\cdot W \bigr]^{\alpha\beta} 
      &= T(V)_0 W^{\alpha\beta} 
         + T(V)^{\alpha}{}_\gamma W^{\gamma\beta}
         + T(V)^{\beta}{}_\gamma W^{\alpha\gamma}
         + T(V)^{\alpha\beta\gamma\delta} \bar{W}_{\gamma\delta} , \\
   \bigl[\overline{T(V)\cdot W} \bigr]_{\alpha\beta} 
      &= T(V)_0 \bar{W}_{\alpha\beta}
         - T(V)^{\gamma}{}_\alpha \bar{W}_{\gamma\beta}
         - T(V)^{\gamma}{}_\beta \bar{W}_{\gamma\beta}
         + \bar{T}(V)_{\alpha\beta\gamma\delta} W^{\gamma\delta} . 
\end{aligned}
\end{equation}
From the form of the generalised Lie derivative in $\SU8$ indices given in appendix~D of~\cite{CSW14b}, we find 
\begin{equation}
\begin{aligned}
   T(V)_0 &= \tfrac{1}{32}V^{\alpha\beta} \tau(\hat{\Sigma})_{\alpha\beta} 
        + \CC , \\
   T(V)^\alpha{}_\beta
      &= \tfrac{1}{32}V^{\gamma\gamma'}\Bigl(
           \tau(\hat{\Sigma})_{\gamma\gamma'\beta}{}^\alpha
           + \tfrac{5}{3} \tau(\hat{\Sigma})_{\beta\gamma} 
                \delta^\alpha_{\gamma'}
           + \tfrac{1}{3} \tau(\hat{\Sigma})_{\gamma\beta} 
                \delta^\alpha_{\gamma'}
           + \tfrac{1}{6}\tau(\hat{\Sigma})_{\gamma\gamma'}
                \delta^\alpha_\beta \Bigr) + \CC , \\
   T(V)^{\alpha\beta\gamma\delta}
      &= - \tfrac{1}{8} V^{\epsilon\epsilon'}\Bigl(
           \bar{\tau}(\hat{\Sigma}){}^{[\alpha\beta\gamma}{}_\epsilon
               \delta_{\epsilon'}^{\delta]} 
           - \bar{\tau}(\hat{\Sigma}){}^{[\alpha\beta}
               \delta_\epsilon^\gamma
               \delta_{\epsilon'}^{\delta]} \Bigr) - \star(\CC) , 
\end{aligned}
\end{equation}
where $\star(\CC)$ is the Hodge dual of the conjugate expression.  

We also have expressions for the structures $X$ and $J_\alpha$ in terms of the spinor indices. For $X$ the non-zero component is the singlet in the  $\rep{28}=\repp{1}{1}+\repp{2}{6}+\repp{1}{15}$ representation 
\begin{equation}
\label{eq:Tdecomp}
   X^{\alpha\beta} = (T^{ij},T^{ia},T^{ab})
      \propto (\epsilon^{ij},0,0) , 
\end{equation}
while for $J_\alpha$ it is the triplet in the $\rep{63}=\repp{1}{1}+\repp{3}{1}+\repp{2}{6}+\repp{2}{\overline{6}}+\repp{1}{35}$ representation 
\begin{equation}
   (J_\alpha)^\alpha{}_\beta 
     = \bigl((J_\alpha)_0\delta^i{}_j,(J_\alpha)^i{}_j,
              (J_\alpha)^i{}_a,(J_\alpha)^{ia},(J_\alpha)^a{}_b\bigr) 
     \propto (0,(\sigma_\alpha)^i_{\phantom{i}j},0,0,0) ,
\end{equation}
where $\sigma_\alpha$ are the Pauli matrices. The first thing to notice is that, substituting into the generalised Lie derivative in $\SU8$ indices given in appendix~D of~\cite{CSW14b}, we find
\begin{equation}
   \Dorf_X X \equiv 0 \qquad \text{identically,}
\end{equation}
simply from the form of the $X$ given in~\eqref{eq:Tdecomp}. 

For the moment maps, since $\kappa^2$ has weight two, the condition~\eqref{eq:mu-ex} on the intrinsic torsion can be written as 
\begin{equation}
\begin{split}
   &\tr \bigl(J_\alpha T(V) \bigr) + T(J_\alpha\cdot V)_0 \\
      & \qqqq \propto \tfrac{1}{2}V^{\gamma\gamma'}\sigma_\alpha{}^j{}_i\,
           \tau_{\gamma\gamma'j}{}^i
       + \tfrac{1}{6} V^{\gamma i}\sigma_\alpha{}^j{}_i
            (5\tau_{j\gamma} + \tau_{\gamma j}) 
       + V^{\gamma i} \sigma_\alpha{}^j{}_i (\tau_{\gamma j}-\tau_{j\gamma})
       + \CC ,
\end{split}
\end{equation}
where we abbreviate $\tau(\hat{\Sigma})_{\alpha\beta}$ and $\tau(\hat{\Sigma})_{\alpha\beta\gamma}{}^\delta$ as $\tau_{\alpha\beta}$ and $\tau_{\alpha\beta\gamma}{}^\delta$. This vanishes for all $V$ if and only if
\begin{equation}
\label{eq:mu-int-tor}
\begin{aligned}
   \sigma_\alpha{}^j{}_i\,\tau_{abj}{}^i &= 0 
           && \in \repp{3}{15} , \\
   (\tau_{aij}{}^j + \tfrac{1}{6}\tau_{ia}) - \tfrac{7}{6}\tau_{ai} &= 0 
           && \in \repp{2}{6} , \\
   \tau_{(ij)} &= 0 
           && \in \repp{3}{1} .
\end{aligned}
\end{equation}
Note, comparing with~\eqref{eq:SU6-zeros}, that the $\repp{2}{6}$ representation appearing in the second line is indeed independent of the $\repp{2}{6}$ component of the torsion generated by an $\SU6$ generalised connection. 

The non-zero components of $T(X)$ are 
\begin{equation}
\begin{aligned}
   T(X)_0 &\propto \epsilon^{ij}\tau_{ij} , & && && 
   T(X)^i{}_j
      &\propto - 2\epsilon^{ik} \tau_{(jk)}
          - \tfrac{1}{2}(\epsilon^{kl}\tau_{kl}) \delta^i_j , \\
   T(X)^i{}_a
      &\propto \epsilon^{kl}\tau_{akl}{}^i 
          - \tfrac{1}{3}\epsilon^{ik}(5\tau_{ak}+\tau_{ka}) , & && && 
   T(X)^a{}_b 
      &\propto \epsilon^{kl}\tau_{bkl}{}^a 
          + \tfrac{1}{6}(\epsilon^{kl}\tau_{kl})\delta^a_b , \\
   \bar{T}(X)^{ab ij}
      &\propto \epsilon^{ik}\bar{\tau}^{ab j}{}_k 
          - \epsilon^{jk}\bar{\tau}^{ab i}{}_k 
          + \tfrac{2}{3}\epsilon^{ij} \bar{\tau}^{[ab]} ,   & && &&    
   \bar{T}(X)^{abc i}
      &\propto \epsilon^{ik}\bar{\tau}^{abc}{}_k , 
\end{aligned}
\end{equation}
so the non-zero components of $T(X)\cdot\bar{X}$ are 
\begin{equation}
\label{eq:LTT-int-tor}
\begin{aligned}
   \left( T(X) \cdot \bar{X}\right)_{ij}
      & \propto 4\tau_{[ij]} && \in \repp{1}{1}, \\
   \left( T(X) \cdot \bar{X}\right)_{ia}
      & \propto 2 (\tau_{aij}{}^j + \tfrac{1}{6}\tau_{ia}) 
         + \tfrac{5}{3}\tau_{ai} && \in \repp{2}{6}' , \\
   \big( \overline{T(X) \cdot \bar{X}}\big)^{ab}
      & \propto - 2(\bar{\tau}^{abi}{}_i - \tfrac{2}{3}\bar{\tau}^{ab} )
      && \in (\rep{1},\overline{\rep{15}}) . 
\end{aligned}
\end{equation}
Note again that the linear combination of torsions in the second and third lines are independent of those appearing in an $\SU6$ generalised connection, and further that the combination in the second line is different from the one in the second line of~\eqref{eq:mu-int-tor}, and hence we denote it $\repp{2}{6}'$. Similarly, the non-zero components of $[T(X),J_\alpha]$ are 
\begin{equation}
\label{eq:LTJ-int-tor}
\begin{aligned}
   \left[ T(X), J_\alpha \right]^i{}_j
      & \propto (\epsilon^{kl}\tau_{kl}) \sigma_\alpha{}^i{}_j
          - 2\epsilon^{ik}\tau_{(lk)}\sigma_\alpha{}^l{}_j
          + 2\epsilon^{lk}\tau_{(jk)}\sigma_\alpha{}^i{}_l
          && \in \repp{1}{1} + \repp{3}{1} , \\
   \left[ T(X), J_\alpha \right]^i{}_a
      & \propto \bigl( 2 (\tau_{aij}{}^j 
          + \tfrac{1}{6}\tau_{ia}) 
          + \tfrac{5}{3}\tau_{ai}\bigr) \sigma_\alpha{}^i{}_j \epsilon^{jk}
          && \in \repp{2}{6}' , \\
   \left[ T(X), J_\alpha \right]^{abci}
      & \propto - \epsilon^{jk}\bar{\tau}^{abc}{}_k\sigma_\alpha{}^i{}_j
          && \in (\overline{\rep{2}},\overline{\rep{20}}) , \\
   \left[ T(X), J_\alpha \right]^{abij}
      & \propto 2 \bar{\tau}^{abj}{}_k\epsilon^{lk}\sigma_\alpha{}^i{}_l
          && \in (\rep{3},\overline{\rep{15}}) . 
\end{aligned}
\end{equation}
Note that the combination of torsions appearing in the second line is the same as the combination appearing in the second line of~\eqref{eq:LTT-int-tor}.



\end{document}